\documentstyle[axodraw,12pt]{article}

\newcommand{\be}{\begin{eqnarray}}
\newcommand{\ee}{\end{eqnarray}}

\newcommand{\vecg}[1]{\mbox{\boldmath $#1$}}
\renewcommand{\vec}[1]{{\bf #1}}

\input epsf
\newcommand{\grpicture}[1]
{
    \begin{center}
        \epsfxsize=200pt
        \epsfysize=0pt
        \vspace{-5mm}
        \parbox{\epsfxsize}{\epsffile{#1.ps}}
        \vspace{5mm}
    \end{center}
}

\begin{document}
\begin{flushright}
TPI -- MINN -- 96/23\\
NUC--MINN--96/21--T
\end{flushright}
\begin{center}
{\Large PHYSICS OF THERMAL QCD}
\vspace{1cm}

A.V. Smilga\\
\vspace{0.5cm}

{\it TPI, School of Physics and Astronomy, University of
Minnesota,
Minneapolis, MN 55455, USA}
\footnote
{Permanent address: ITEP, B. Cheremushkinskaya 25, Moscow
117259, Russia.}
\vspace{1cm}
\end{center}

\abstract{We give a review of modern theoretical
understanding of the physics of $QCD$ at finite temperature.
Three temperature regions are studied in details.
When the
temperature is low, the system presents a rarefied pion gas.
Its
thermodynamic and kinetic properties
are adequately described by chiral perturbation theory.

When the temperature is increased, other than pion degrees
of freedom are
excited, the interaction between the particles in the heat
bath becomes
strong, and the chiral theory is not applicable anymore. At
some point $T =
T_c$ a phase transition is believed to occur. The physics of
the
transitional region is discussed in details. The dynamics
and the very
existence of this phase transition strongly depends on the
nature of the
gauge group, the number of light quark flavors, and on the
value of quark
masses. If the quarks are very heavy, the order parameter
associated with
the phase transition is the correlator of Polyakov loops
$<P^*(\vec{x}) P(0)
>_T$ related to the static potential between heavy colored
sources. When the
quark masses are small, the proper order parameter is the
quark condensate
$<\bar q q>_T$ and the phase transition is associated with
the
restoration of the chiral symmetry. Its dynamics is best
understood in the
framework of the instanton liquid model. Theoretical
estimates and {\it
some}
numerical lattice measurements indicate that the phase
transition probably
 does {\it not} occur for the experimentally observed values
of quark masses
. We have instead a very sharp crossover, "almost" a second
order phase
transition.

At high temperatures $T \gg \mu_{hadr}$ the system is
adequately described
in terms of quark and gluon degrees of freedom and presents
the {\it
quark--gluon plasma} ($QGP$). Static and kinetic properties
of $QGP$ are
discussed. A particular attention is payed to the problem of
physical
observability, i.e. the physical meaningfulness of various
characteristics
of $QGP$ discussed in the literature.}

\vspace{1cm}

\newpage

\centerline{CONTENTS}

1. INTRODUCTION. \dotfill 3

\vspace{.5cm}

2. FINITE $T$ DIAGRAM TECHNIQUE. \dotfill 8

\bigskip 2.1 {\it Euclidean (Matsubara) technique.} \dotfill
8

\bigskip 2.2 {\it Real time (Keldysh) technique.} \dotfill 9

\vspace{.5cm}

3. PURE YANG--MILLS THEORY: DECONFINEMENT PHASE TRANSITION
\dotfill 19

\bigskip 3.1 {\it Preliminary remarks. } \dotfill 19

\bigskip 3.2 {\it Bubble confusion.} \dotfill 21

\bigskip 3.3 {\it More on phase transition.} \dotfill 26
\vspace{.5cm}

4. LUKEWARM PION GAS. \dotfill 31

\bigskip 4.1 {\it Chiral symmetry and its breaking} \dotfill
31

\bigskip 4.2 {\it Thermodynamics} \dotfill 38

\bigskip 4.3 {\it Pion collective excitations} \dotfill 39

\bigskip 4.4 {\it Nucleons} \dotfill 48

\bigskip 4.5 {\it Vector mesons. Experiment} \dotfill 53

\vspace{.5cm}

5. CHIRAL SYMMETRY RESTORATION \dotfill 55

\bigskip 5.1 {\it General considerations.\\  Order of phase
transition and critical behavior} \dotfill  55

\bigskip 5.2 {\it Insights from soft pion physics. Large
$N_c$} \dotfill 60

\bigskip 5.3 {\it The real world} \dotfill 66

\bigskip 5.4 {\it Instantons and percolation} \dotfill 70

\bigskip 5.5 {\it Disoriented chiral condensate} \dotfill 75

\vspace{.5cm}

6. QUARK--GLUON PLASMA \dotfill 76

\bigskip 6.1 {\it Static properties of $QGP$: a bird eye's
view} \dotfill 77

\bigskip 6.2 {\it Static properties of $QGP$: perturbative
corrections}
 \dotfill 83

\bigskip 6.3 {\it Collective excitations} \dotfill 90

\bigskip 6.4 {\it Damping mayhem and transport paradise}
\dotfill 97

\bigskip 6.5 {\it Chirality drift} \dotfill 105

7. AKNOWLEDGEMENTS \dotfill 116

REFERENCES \dotfill 116

\section{Introduction.}
The properties of $QCD$ medium at finite temperature have
been
the subject of intense study during the last 15--20 years.
It was
realized that the properties of the medium undergo a drastic
change as the temperature increases. At low temperatures,
the
system presents a gas of colorless hadron states:  the
eigenstates of the $QCD$ hamiltonian at zero temperature.
When
the temperature is small, this gas is composed mainly of
pions
--- other mesons and baryons have higher mass and their
admixture in the medium is exponentially small $\sim
\exp\{-M/T\}$. At small temperature, also the pion density
is
small --- the gas is rarefied and pions practically do not
interact with each other.

 However, when the temperature increases, pion density
grows,
the interaction becomes strong, and also other strongly
interacting hadrons appear in the medium. For temperatures
of
order $T \sim$ 150 Mev and higher, the interaction is so
strong that the hadron states do not present a convenient
basis
to describes the properties of the medium anymore, and no
analytic calculation is possible.

 On the other hand, when the temperature is very high, much
higher than the characteristic hadron scale $\mu_{hadr}
\sim$
0.5 Gev, theoretical analysis becomes possible again. Only
in
this range, the proper basis are not hadron states but
quarks
and gluons --- the elementary fields entering the $QCD$
lagrangian. One can say that, at high temperatures, hadrons
get
"ionized" to their basic compounds.
In the $0^{\underline{th}}$ approximation, the
system presents the heat bath of freely propagating colored
particles. For sure, quarks and gluons interact with each
other, but at high temperatures the effective coupling
constant is small $\alpha_s(T) \ll 1$ and the effects due to
interaction can be taken into account perturbatively
\footnote{We state right now, not to astonish the experts,
that there are limits of applicability of perturbation
theory even at very high temperatures, and we {\it are}
going to discuss them later on.}.
This interaction has the long--distance Coulomb nature,
and the properties of the system are
in
many respects very similar to the properties of the usual
non-relativistic plasma involving charged particles with
weak
Coulomb interaction. The only difference is that quarks and
gluons carry not the electric, but color charge. Hence the
name: {\it Quark-Gluon Plasma} ($QGP$).

 Thus the properties of the system at low and at high
temperatures have nothing in common. A natural question
arises:
What is the nature of the transition from low-temperature
hadron
gas to high- temperature quark-gluon plasma ? Is it a {\it
phase} transition ? If yes, what is its order ?  I want to
emphasize that this question is highly non-trivial. A
drastic
change in the properties of the system in a certain
temperature
range does not guarantee the presence of the phase
transition
{\it point} where free energy of the system or its specific
heat
is discontinuous. Recall that there is no phase transition
between  ordinary gas and ordinary plasma.

Whether or not the phase transition occurs in the real $QCD$
with particular values of quark masses is the question under
discussion
now. Our personal feeling is that the answer
is probably negative and what really happens is not the
phase
transition but a sharp crossover --- "almost" a second-order
phase transition.  Another discussed possibility is that a
weak first
order phase
transition (with small latent heat) still occurs. A point of
consensus
now is that
the real phase transition {\it does}
occur in some relative theories --- in pure Yang- Mills
theory
(when the quark masses are sent to infinity) and in $QCD$
with 2
or 3 exactly massless quark flavors.

 There are at least 4 reasons why this question is
interesting
to study:
\begin{enumerate}
 \item It is just an amusing
theoretical question.
 \item Theoretic conclusions can be
checked in lattice numerical experiments.  Scores of papers
devoted to lattice study of thermal properties of QCD have
been
published.  \item Perhaps, a direct experimental study would
be
possible on RHIC --- high-energy ion collider which is now
under
construction.
\item During the first
second of its evolution, our Universe passed through the
stage
of high-$T$ quark-gluon plasma which later cooled down to
hadron
gas (and eventually to dust and stars, of course). It is
essential to understand whether the phase transition did
occur
at that time. A {\it strong} first-order phase transition
would lead
to observable effects. In particular, it could have created
inhomogeneities
in baryon number density in early Universe which would later
affected
nucleosynthesis and therefore leave a signature in the
primordial nuclear
abundances \cite{strong}. We know (or almost know --- the
discussion of this question has not yet completely died
away)
that there were no such transition. But it is important to
understand why.
\end{enumerate}

 Note that there is also a related but {\it different}
question
--- what are the properties of relatively cold but very
dense
matter and whether there is a phase transition when the
chemical
potential corresponding to the baryon charge rather than the
temperature is increased.  This review will be devoted
exclusively to the thermal properties of QCD, and we shall
assume zero baryon charge density.

At present, we are able to describe the physics of thermal
phase transition
mainly in qualitative terms (we will address, however, the
issue of
critical exponents and the critical amplitudes where certain
theoretical
predictions can be made). As was already mentioned, there
are, however,
two regions where many {\it quantitative} theoretical
results can be
obtained. These are the small temperature region where the
system presents
a weakly interacting rarefied pion gas and the high
temperature region where
the system presents a weakly interacting (and also rarefied
in some sense)
quark--gluon plasma.

In both cases, we are facing a reasonably clean and  rather
interesting
problem of theoretical physics lying on cross-roads between
the relativistic field theory and condensed matter physics.
Personally, I had a great fun studying it. Unfortunately, at
present only a theoretical study of the problem is possible.
That concerns
especially the high temperature phase.
Hot hadron medium with the temperature above phase
transition can be produced for tiny fractions of a moment in
heavy ion collisions but:
 \begin{itemize}
\item It is not clear at all whether a real thermal
equilibrium is achieved.
\item A hot system created in the collision of heavy
nuclei rapidly expands and cools down emitting pions and
other particles. It is
not possible to probe the properties of the system directly,
but only indirectly via the characteristics of the final
hadron state.
\item Anyway,  the temperature
achieved at existing accelerators is not high enough for the
perturbation
theory to
work and there are no {\it quantitative} theoretical
predictions with which experimental data can be compared.
 RHIC would be somewhat better in this respect, but even
there the maximal
temperature one could expect to reach is of order 0.5 GeV
\cite{RHICT}.
This is above
the expected phase transition temperature $\sim 200$ MeV
(see a detailed
discussion in Chapters 4 and 5), but is still probably not
high enough for the effective coupling constant to be small
so that
perturbative calculations would be justified.
  \end{itemize}

Thus at present, there are no experimental tests of
non-trivial theoretical predictions for $QGP$ properties.
The
effects observed in experiment such as the famous
$J/\psi$ -- suppression (for a recent review see
\cite{Muller})
just indicate that a
hot and dense medium is created but says little on whether
it is $QGP$ or something else.

The absence of proper feedback between theory and experiment
is a
sad and unfortunate reality of our time: generally, what is
interesting theoretically is not possible to measure and
what is possible to measure is not interesting theoretically
\footnote{One of the possible exceptions of this general
rule
in the field of thermal QCD is  a fascinating perspective to
observe the phenomenon of disoriented chiral condensate at
RHIC and we will discuss it.}.

The situation is somewhat better with numerical lattice
experiments.
One of the advantages of the lattice approach is that the
question of
whether thermal equilibrium is
achieved does not arise --- path integrals on an Euclidean
cylinder with
imaginary time size $\beta = 1/T$ describe the partition
function and
other characteristics  of the
equilibrated system just by definition. A lot of numerical
results at the
temperatures below or slightly above critical temperature
exist and
in some cases they provide very useful insights which help
to understand
better the physics of the phase transition.
In principle, one can also measure on the lattice
the characteristics of the
high temperature phase.
Due to a finite size of the
lattice, it is more difficult, however, and such studies
just started. A limitation here is that
 we can calculate numerically only Euclidean
path integrals which means that we can study  only static
characteristics of the system. Kinetic properties (spectrum
of collective
excitations in $QCD$ heat bath, the transport phenomena such
as viscosity and
electric conductivity, etc) cannot be probed in this way.
(May be, it {\it is} possible, however, to study a
nontrivial kinetic
characteristics, the rate of axial charge non-conservation
in high
temperature phase, studying numerically {\it classical}
Yang--Mills
theory on a {\it hamiltonian} lattice. We will address this
question
at the end of the review.)

Thus we will not attempt (or almost will not attempt)  to
establish
relation of the results of theoretical calculations with
realistic accelerator experiments. Comparison with the
lattice data will be
done when the latter are available, i.e. in the low and
moderate
temperature region. What we {\it will} do,
however, is discussing the relation of theoretical results
with {\it gedanken} experiments. Suppose, we have a thermos
bottle with hot hadron matter or with $QGP$ on a laboratory
table
and are studying it
from {\it any} possible experimental angle.
We call a quantity physical if it can in principle be
measured in such a study and non-physical otherwise. We
shall see later that many quantities discussed by
 theorists may be called physical only with serious
reservations, and some are  not physical at all.

Many reviews on the subject have already been written (see
e.g.
\cite{Muller},   \cite{ShurPR} --
\cite{ITEP}). However, all of them were either written a
considerable time
ago and do not cover very important recent developments or
have a limited scope.
I will try here to fill up this gap and present a modern
discussion which covers
a reasonably broad set of questions. Of course, I could not
cover everything.
The issues which interested me more and which I am more
familiar with are
discussed  at greater length and in greater details. In
cases when the question
is still controversial and no general consensus exists, I
will present my own
viewpoint. A bias of this kind is, however, inevitable.

I had some problems in arranging the material and found no
way not to break
the causality: cross--references both back in text and {\it
forward} in text
will be abundant. The physics of {\it all} finite
temperature systems has
something in common, and the parallels will be drawn all the
time.
Moreover, some basic facts about quark--gluon plasma (the
subject of
Chapter 6) will be {\it used} in Chapter 3 where the physics
of pure
glue systems is discussed, etc.

But, anyway, some plan should be chosen, and it is the
following.
The next chapter presents technical preliminaries.
I will give a review of finite $T$ diagram technique. A
particular emphasis
is given to the real time formalism which is most convenient
when studying
kinetic properties of the system. It is much less known than
Euclidean Matsubara
technique (well suited to study static characteristics) and
has entered the
standard tool kit of field thenrists studying thermal field
theories only
recently.

Chapter 3 is devoted to physics of pure Yang--Mills theory
at finite
$T$. The system displays {\it deconfinement} phase
transition in temperature
associated with changing the behavior of the potential
between static quark
sources. In particular, we discuss the structure of the high
temperature
phase and show that there is only {\it one} such phase, not
$N_c$ different
phases with domain walls between them as people believed for
a long time
and some continue to believe up to
now. There is no such physical phenomenon as spontaneous
breaking of $Z_N$
symmetry at high temperature in the pure Yang--Mills system.

In Chapter 4 we go over to the theories involving light
dynamical quarks and
discuss first the properties of low temperature phase. We
discuss at length
the temperature dependence of chiral condensate (it
decreases with temperature)
and also explore the fate
of pions and massive hadron states at small nonzero
temperature.
We show that the leading
effect of the pion heat bath on massive hadrons is that the
latter acquire
finite width. Real part of the poles of corresponding Green
functions
 is also somewhat shifted. For nucleons, this shift is
rather tiny.
For vector mesons, the shift is also tiny in the temperature
region
up to $\sim 100$ MeV. What happens at higher temperatures is
not
quite clear by now and is a subject of intense discussions.

Chapter 5 is devoted to the physics of thermal phase
transition in $QCD$. It
is associated with restoration of chiral symmetry which is
broken spontaneously
at zero temperature. We show how the physics of the phase
transition depends
on the number of light quark flavors (For two massless
flavors it is the second
order
while for three flavors it is the first order. We argue in
particular that,
in the theory with 4 or may be 5 massless flavors, chiral
symmetry
would probably not broken  at all and, correspondingly,
there
would be no phase transition ). We discuss in details the
physical mechanism
of the phase transition which, in our opinion, is best
understood in the
framework
of the instanton-antiinstanton liquid model. In this model,
characteristic
vacuum
fields contributing in the Euclidean functional integral for
the partition
function present
a collections of quasi-classical objects: instantons and
antiinstantons. Each
(anti)instanton supports  fermion zero modes (one fermion and one 
antifermion zero mode for each flavor). When interaction between
quasi-particles is taken into account, the modes shift from
zero, but, as the
interaction turns out to be strong enough, the
characteristic eigenfunctions
become delocalized and characteristic distance between
neighboring eigenvalues
is very small $\sim 1/V^{Eucl}$. This brings about nonzero
fermion condensate.
When temperature is increased, the density of quasi-
particles decreases and,
which is even more important, their interaction decreases.
As a result,
instantons and antiinstantons tend to form "molecules" with
localized fermion
eigenfunctions. Characteristic eigenvalues are far from zero
in this case and
the chiral condensate is zero. Remarkably, it is basically
the same mechanism
which brings about the so called "percolation phase
transition in doped
semiconductors". When the density of impurity is high,
electrons can jump
between adjacent impurity atoms, acquire  mobility, and the
material
becomes a conductor.
 In the end of the chapter we discuss the phenomenon
of disoriented chiral condensate.

Physics of $QGP$ is discussed in Chapter 6. We discuss the
limits of
applicability of perturbation theory due to so called
"magnetic screening" - the
effect which is specific for nonabelian gauge theories. We
discuss also the
spectrum of plasmons and plasminos --- the collective
excitations with quantum
numbers of quarks and gluons in $QGP$. We address the
controversial issue of
the plasmon damping. We show that damping depends on the
gauge convention and
has as such little physical meaning. On the contrary, {\it
transport phenomena}
(such as viscosity, electric conductivity, energy losses of
a heavy
energetic particle passing through $QGP$, etc.) are quite
physical and can be
measured in a {\it gedanken} experiment. Finally, we discuss
a pure
non-perturbative effect of chirality non-conservation in
quark-gluon plasma
(or, if you will baryon number non-conservation in hot
electroweak plasma).
 The rate of non-conservation
is proportional to $T^4$ times a power of coupling constant.
What is this
power and whether a numerical algorithm exists where this
quantity can be
determined is still a question under discussion now.

\section{ Finite $T$ diagram technique.}
\setcounter{equation}0
 The main point of interest for us are the physical
phenomena in hot QCD system. However, as the main
theoretical tool to study them is the perturbation theory
and we want in some cases not only to quote the results, but
also to explain how they are obtained, we are in a position
to spell out  how the perturbative calculations at finite
$T$ are performed.

There are two ways of doing this --- in imaginary or in real
time. These techniques are completely equivalent and  which
one to use is mainly a matter of taste. Generally, however,
the Euclidean technique is more handy when one is interested
in pure static properties of the system (thermodynamic
properties and static correlators) where no real time
dependence is involved. On the other hand, when one is
interested in kinetic properties (spectrum of collective
excitations, transport phenomena, etc.), it is much more
convenient
to calculate directly in real time.

\subsection{Euclidean (Matsubara) technique.}
Many good reviews of Matsubara technique are available in
the literature (see
e.g. \cite{kapbook}) and we will describe it here only
briefly.
 Consider a theory of real scalar field described by the
hamiltonian $H[\phi(\vec{x}),\  \Pi(\vec{x})]$. The
partition
function of this theory at temperature $T$ can be written as

\be
\label{Zdef}
Z = {\rm Tr} \left\{e^{-\beta H} \right\} =
\int \prod_\vec{x} d\phi(\vec{x}) {\cal K} [\phi(\vec{x}),
\phi(\vec{x}) ; \beta]
\ee
where ${\cal K}$ is the quantum evolution operator in the
imaginary time $\beta = 1/T$ :
 \be
\label{K}
 {\cal K} [\phi'(\vec{x}), \phi(\vec{x}) ; \beta]
= \sum_n   \Psi_n^*[\phi'(\vec{x})]  \Psi_n[\phi(\vec{x})]
e^{-\beta E_n}
\ee
$\Psi_n$ are the eigenstates of the hamiltonian. One can
express the integral in RHS of Eq.(\ref{Zdef}) as an
Euclidean path integral:

\be
\label{Zpath}
Z = \int \prod_{\vec{x}, \tau} d\phi(\vec{x}, \tau)
\exp \left\{ - \int_0^\beta d\tau \int d\vec{x} \ {\cal L}
[\phi(\vec{x}, \tau] \right\}
\ee
where the periodic boundary conditions are imposed
  \be
  \label{bc}
 \phi(\vec{x}, \tau + \beta) = \phi(\vec{x}, \tau)
 \ee
A thermal average $<{\cal O}>_T$ of any operator ${\cal O}
[\phi(\vec{x}, \tau)]$ has the form \cite{Mats}

\be
\label{OT}
<{\cal O}>_T = Z^{-1}
\int \prod_{\vec{x}, \tau} d\phi(\vec{x}, \tau) {\cal O}
[\phi(\vec{x}, \tau)]
\exp \left\{ - \int_0^\beta d\tau \int d\vec{x} \ {\cal L}
[\phi(\vec{x}, \tau] \right\}
\ee

One can develop now the diagram technique in a usual way.
The only difference with the zero temperature case is that
the Euclidean frequencies of the field $\phi(\vec{x}, \tau)$
are now quantized due to periodic boundary conditions
(\ref{bc}):
\be
\label{p0bos}
p_0^n = 2\pi i nT
\ee
with integer $n$. To calculate something, one should draw
the same graphs as at zero temperature and go over into
Euclidean space where the integrals over Euclidean
frequencies are substituted by sums:
 \be
 \label{intT}
\int \frac {d^4p}{(2\pi)^4} f(p) \longrightarrow
T \sum_n \int \frac {d^3p}{(2\pi)^3} f(2\pi i nT, \vec{p})
\ee
The same recipe holds in any theory involving bosonic
fields. In theories with fermions, one should impose
antiperiodic boundary conditions on the fermion fields
$\psi(\vec{x}, \tau)$ (see e.g. \cite{Brown} for  detailed
pedagogical explanations), and the frequencies are quantized
to
\be
\label{p0ferm}
p_0^n = i\pi (2n+1) T
\ee
An important heuristic remark is that, when the temperature
is very high, in many cases only the bosons with zero
Matsubara frequencies $p_0 = 0$ contribute in $<{\cal
O}>_T$. The contribution of higher Matsubara frequencies and
also the contribution of fermions in $<{\cal O}>_T$ become
irrelevant and, effectively,  we are dealing with a 3-
dimensional theory. As was just mentioned, it is true in
many, but not in {\it all} cases. For example, it makes no
sense to neglect fermion fields when one is interested in
the properties of collective excitations with fermion
quantum numbers. For any particular problem of interest a
special study is required.

\subsection{Real time (Keldysh) technique.}
Matsubara technique is well suited to find thermal averages
of {\it static} operators ${\cal O}(\vec{x})$. If we are
interested in a time-dependent quantity, there are two
options: {\bf i)} To find first the thermal average $<{\cal
O}(\vec{x}, \tau)>$ for Euclidean $\tau$ and perform then
an analytic continuation onto the real time axis. It is
possible, but quite often rather cumbersome.
{\bf ii)} To work in the real time right from the beginning.
The corresponding technique was first developed in little
known papers \cite{Bakshi} and independently by Keldysh
\cite{Keld} who applied it to condensed matter problems with
a particular emphasis on the systems out of
equilibrium.
For systems at thermal equilibrium, it was effectively
reinvented
in a slightly different approach
in the seventies by the name ``thermo field dynamics''
\cite{TFD}.
It was fully apprehended by experts in
relativistic field theory only in the beginning of the
nineties.

There is no good review on real time diagram technique
addressed to field
theorists. The existing review \cite{Landsman} is very deep
and
extensive, but
is written rather formally and is hard to read. What we will
do here
is in a sense complementary to the review \cite{Landsman}.
We will not
derive the real time technique in a {\it quite} accurate and
regular way, but
rather
elucidate its physical foundations, show how multicomponent
Green's functions
appear, formulate and discuss the real time Feynman rules,
and discuss also
a simplistic but rather useful version of
the real time technique due to Dolan and Jackiw \cite{DJ}.

Consider a quantum mechanical system in thermal equilibrium.
Suppose a
thermal average (\ref{OT}) of a Heisenberg operator
${A}_0(t)$ is zero.
Let us perturb the system adding to the hamiltonian the term
  \be
  \label{pertV}
V(t) \ =\ V_0 \hat{B}_0 \delta(t)
  \ee
where $B_0$ is some  operator and we assume the constant
$V_0$ to be small.
The Heisenberg operator $A_H(t)$ of the full hamiltonian $H
\ = \ H_0 + V(t)$
is related to the Heisenberg operator ${A}_0(t)$ of
unperturbed hamiltonian
(or the operator in interaction representation) as
  \be
  \label{HeisA}
A_H(t) \ = \ \exp \left\{ i \int_{-\infty}^t  V(\tau) d\tau
\right\}
A_0(t)  \exp \left\{ -i \int_{-\infty}^t  V(\tau) d\tau
\right\}
  \ee
(Generally, $T$--ordered exponentials enter, but for an
instantaneous
perturbation $T$--ordering is irrelevant).
In the first order in $V_0$
  \be
 \label{HeisAV}
A_H(t) \ = \ A_0(t) - iV_0 \theta(t) \left[ A_0(t), B_0
\right]
  \ee
The thermal average is
 \be
  \label{AHav}
<A_H(t)>_T \ =\ - iV_0 <\theta(t) [A_0(t), B_0]>_T
  \ee
The expression in the RHS of Eq. (\ref{AHav}) is called the
response function
--- it determines the response of the system at some time $t
> 0$ at the
instantaneous small perturbation applied at $t = 0$. Let us
make a Fourier
transform and define the {\it generalized susceptibility}
  \be
\label{DABOm}
D^{AB} (\omega) \ = \ -i \int_{-\infty}^\infty e^{i\omega t}
\theta(t)
<\left[ A_0(t), B_0 \right]>_T dt \nonumber \\
 = \  \sum_{nm} e^{-E_n/T}
\left[ \frac {A_{nm}B_{mn}} {\omega + E_n - E_m + i0}
- \frac {B_{nm}A_{mn}} {\omega + E_m - E_n  + i0} \right]
 \ee
We see that $D^{AB} (\omega) $ is analytic in the upper
half--plane.
It is just the corollary of the fact that $D^{AB}(t < 0) \
=\ 0$. The relation
(\ref{DABOm}) is well known in statistical mechanics and is
called
the Kubo formula.
\footnote{ We have changed the sign convention compared to
the original Kubo
convention (see e.g. \cite{staty2}) to make a generalization
to quantum
field theory more transparent.}

Consider now a thermal field theory. Let it be first a
theory of real scalar
field with possible nonlinear interactions. Suppose the
perturbation is coupled
to $\phi(\vec{x}, 0)$ and we measure the response of the
system at some later
time
$t$ in terms of  $<\phi(\vec{x}, t)>_T$. The corresponding
response
function is called the retarded Green's function
  \be
\label{DRdef}
D_R(t, \vec{x}) \ =\ -i\theta(t) <[\phi(t, \vec{x}),
\phi(0)]>_T
\ee
Its Fourier image $D_R(\omega, \vec{p})$ is analytic in the
upper
 $\omega$ half-plane. One can show that the retarded Green's
function describes
a natural analytic continuation of the Matsubara Green's
function (defined at
a discrete set of points on the imaginary $\omega$ axis) on
the complex $\omega$
plane. We will be mainly interested with real $\omega$.

On the tree level
 \be
\label{DR0}
D^0_R(\omega, \vec{p}) \ =\ \frac 1{(\omega + i0)^2 -
\vec{p}^2 - m^2}
\ee

Let us separate now the free hamiltonian from the
interaction part and build
up the
perturbation theory for $D_R(t, \vec{x})$ and other
physically observable
quantities
in interaction representation. It is very well known that at
zero temperature
the
retarded propagators like (\ref{DR0}) are not quite
convenient for this
purpose. The
matter is that  to calculate, say, the exact retarded
Green's function, we
have to
draw the loops involving virtual particles which do not
necessarily go
forward in time.
In other words, exact retarded Green's function cannot be
expressed into
integrals
of only retarded propagators. Advanced components also play
a role and, as  a
result,
we are arriving at the "old diagram technique" which is
somewhat clumsy.

 Feynman showed that the calculations can be greatly
simplified using $T$ -
ordered
Green's functions. In that case, everything is expressed
into the integrals of
$T$ - ordered propagators. We know that the Feynman
technique for field theories
at zero temperature is explicitly Lorentz -- invariant.
However, at finite
temperature, Lorentz--invariance is lost --- the reference
frame where thermal
medium
is globally at rest is singled out. That gives an indication
that Feynman
trick probably
would not work at finite temperature and we are bound to use
a version of the
old
diagram technique which is not Lorentz--invariant.

Another indication of the trouble is that the notion of $S$-
-matrix which is
basic
at $T = 0$ and for which the Feynman diagram technique has
been constructed has
absolutely
no meaning at finite temperature (see \cite{Smatr} for a
formal proof).
Particles interact
 with the heat
bath all the time, scatter on the real particles there and
just have no chance
to arrive from infinity or to escape to infinity from
interaction point (see
\cite{Persik}
for a related physical discussion).

What can be directly measured in the heat bath are {\it
classical}
 response functions (see a detailed
discussion of this issue in Chapter 6). Suppose we disturb
normal
plasma (or pion gas or $QGP$) with a concentrated laser beam
at $t=0$ and
measure the
distribution of charges and corresponding electromagnetic
fields at later
time. Such
a gedanken experiment is quite feasible. Thus retarded
Green's functions (in
contrast
to the Feynman Green's functions) have a direct physical
meaning and
we should be
able to calculate them.

Let us show that the Feynman program fails indeed at $T \neq
0$ \cite{LP}.
 Consider
the exact Feynman propagator $D_F(x) \ =\ -i<T\{\phi(x)
\phi(0)\}>_T$. Heisenberg
field operators are
 \be
 \label{phiHeis}
\phi(\vec{x}, t) \ =\ S(-\infty, t) \phi_0(\vec{x}, t) S(t,
-\infty)
\ee
where $\phi_0(\vec{x}, t)$ are the operators in the
interaction representation,
 \be
 \label{S12}
S(t_2, t_1) \ =\ T \exp\left\{-i\int_{t_1}^{t_2} V(t') dt'
\right\},
 \ee
and $V(t)$ here is a nonlinear part of the hamiltonian and
has nothing to do
with the instantaneous perturbation (\ref{pertV}). Simple
transformations give
(see e.g. \cite{BLP})
  \be
\label{S-1S}
D_F(x) \ =\ -i<S^{-1}T\{\phi_0(x) \phi_0(0)S\}>_T
 \ee
where $S \ =\ S(\infty, -\infty)$ is the $S$- matrix
operator. At zero
temperature
we were interested with the vacuum average, $S$-matrix gave
just a phase factor
when acting on the vacuum state (this factor was  anyway
cancelled out
with the vacuum loops
in perturbative expansion of $T\{\phi(x) \phi(0)S\}$) and,
expanding $S$ in
powers
of $V$ , using Wick theorem etc., we obtained a standard
Feynman loop expansion
for the exact propagator.

Thermal average (\ref{OT}) involves, however, also averaging
over excited
states on which
the operator $S$ acts in a non-trivial way. We are bound to
take the operator
  \be
 \label{S-1}
S^{-1} \ =\ \tilde{T} \exp\left\{i\int_{-\infty}^{\infty}
V(t') dt' \right\}
  \ee
into account when performing Wick contraction of the field
operators
($\tilde{T}$ stands for anti-chronological ordering where
the operators
at later times stand on the right)

As a result, three types of Green's functions appear:
\begin {itemize}
\item The usual $T$ - ordered ones coming from the
contractions inside
$T\{\phi_0(x) \phi_0(0)S\}$;
\item Anti-$T$-ordered Green's functions coming from the
expansion of $S^{-1}$;
\item Finally, there are also {\it not} ordered Green's
functions
 $<\phi_0(x) \phi_0(0)>_T$ coming
from contracting the operators in $S^{-1}$ with the
operators in
$T\{\phi_0(x) \phi_0(0)S\}$.
\end{itemize}

Let us first deal with the latter and calculate
  \be
  \label{propdef}
-i<\phi_0(x) \phi_0(0)>_T\  = \ -i\sum_n e^{- \beta E_n}
<n|\phi_0(x)
\phi_0(0) |n>
  \ee
where $|n>$ present the eigenstates of unperturbed
hamiltonian so that
$E_n \ =\ \sum_{\vec{p}} N_{\vec{p}}\epsilon_{\vec{p}}$ with
integer
$\{N_{\vec{p}}\}$
, and $\epsilon_{\vec{p}} \ =\ \sqrt{\vec{p}^2 + m^2}$.
Introduce as usual a finite spatial volume $V$ and decompose
\be
\label{second}
\phi_0(x) = \sum_\vec{p} \frac 1{\sqrt{2\epsilon_{\vec{p}}
V}}
\left[ a_\vec{p} e^{-i \epsilon_{\vec{p}} t + i
\vec{p}\vec{x}}
+  a^+_{\vec{p}} e^{i \epsilon_{\vec{p}} t - i
\vec{p}\vec{x}} \right]
\ee

where $a^+_\vec{p}$ and  $a_\vec{p}$ are the creation and
annihilation operators. Substitute it in (\ref{propdef}).
At zero temperature only the vacuum
average contributes and we can use the fact $a_\vec{p}|0> =
0$ to obtain a usual expression. At
finite temperature, the excited states $|n>$ with
$a_\vec{p}|n> \neq 0$
contribute in the sum
(\ref{propdef}), and an
additional contribution  arises: $<n|a^+_{\vec{p}}
a_{\vec{p}}|n> \ =\
 N_{\vec{p}},
\ \ \ \ <n|a_{\vec{p}} a^+_{\vec{p}}|n> \ =\ 1 + N_{\vec{p}}
$. Trading
as usual the
sum for the integral and taking Fourier image , we obtain
  \be
\label{D12}
D^0_{21}(\omega, \vec{p}) = -i \int <\phi_0(x) \phi_0(0)>_T
e^{ipx} d^4x \
= \nonumber \\
-\frac {i\pi}{\epsilon_{\vec{p}}} \left[ (1 +
n_B(\epsilon_{\vec{p}}) )
\delta (\omega - \epsilon_{\vec{p}}) +
n_B(\epsilon_{\vec{p}})
\delta (\omega + \epsilon_{\vec{p}}) \right]
  \ee
where
  \be
  \label{nB}
n_B(\epsilon_{\vec{p}}) \ =\ <N_{\vec{p}}>_T \ =\
\frac 1{e^{\beta \epsilon_{\vec{p}}} -1}
  \ee
is the Bose distribution function. The indices 21 appear as
a reminder that
 the operator  standing on the left in  in (\ref{propdef})
and corresponding to
the ``final point''
   appeared
from the factor $S^{-1}$ in (\ref{S-1S}) while the operator
standing
on the right   --- from the factor $T\{\phi_0(x) \phi_0(0)
S\}$.
Similarly, a function
  \be
 \label{1221}
D^0_{12}(\omega, \vec{p}) = \
-i \int <\phi_0(0) \phi_0(x)>_T e^{ipx} d^4x \ = \
 D^0_{21}(-\omega, -\vec{p}) \ = \   D^0_{21}(-\omega,
\vec{p})
 \ee
can be defined.

$T$ -- ordered and anti-$T$ -- ordered tree propagators will
also differ from
their
zero temperature expressions. Proceeding in the same way, we
obtain
  \be
\label{D1122}
D^0_{11}(\omega, \vec{p})  \ =\ -i \int <T\{\phi_0(x)
\phi_0(0)\}>_T e^{ipx} \ =
\nonumber \\
= \frac 1{\omega^2 - \epsilon_{\vec{p}}^2 + i0}
- 2\pi i  n_B(\epsilon_{\vec{p}}) \delta (\omega^2 -
\epsilon_{\vec{p}}^2)
\nonumber \\
D^0_{22}(\omega, \vec{p})  \ =\ -i \int
<\tilde{T}\{\phi_0(x) \phi_0(0)\}>_T
e^{ipx} \ = \nonumber \\
= - \frac 1{\omega^2 - \epsilon_{\vec{p}}^2 - i0}
- 2\pi i n_B(\epsilon_{\vec{p}}) \delta (\omega^2 -
\epsilon_{\vec{p}}^2)
 \ee
Various kinds of thermal propagators in (\ref{D12},
\ref{1221}, \ref{D1122}) are
conveniently ``organized'' in a $2 \times
2$ matrix:
  \be
  \label{D0matr}
D^0(\omega, \vec{p}) \ =\ \left( \begin{array}{cc}
D_{11}^0(\omega, \vec{p}), D_{12}^0(\omega, \vec{p}) \\
D_{21}^0(\omega,
\vec{p}),
D_{22}^0(\omega, \vec{p}) \end{array} \right)
  \ee
To derive diagram technique, suppose first that the
interaction $V(t)$
 corresponds
to scattering on a static classical field $F(\vec{x})$:
  \be
 \label{fiscat}
  V(t) \ =\ \int \phi^2(\vec{x}, t) F(\vec{x}) d\vec{x}
  \ee
Let us consider first the Feynman propagator (\ref{S-1S}).
Expand both $S$ and
 $S^{-1}$ in $V(t)$ and compose various products of field
operators
contracting them according to Wick rules. $\phi(x)$
(corresponding to
 the outgoing external leg) can be paired with an operator
from the expansion
of $S$ or with the operator from the expansion of $S^{-1}$.
That gives us
either the 11 -- component of the Green's function
(\ref{D0matr}) or the
12-- component. Suppose the pairing occurred within $S^{-
1}$. The remaining
field operator in (\ref{fiscat}) can be further paired
either with an
operator in $S$ or with an operator in $S^{-1}$. In the
first case, we get
21 -- component and in the second --- 22 - component. The
last pairing occurs
with $\phi(0)$ and we have either 21 or 11 -- component
depending on where
the line came from.
 
\newpage

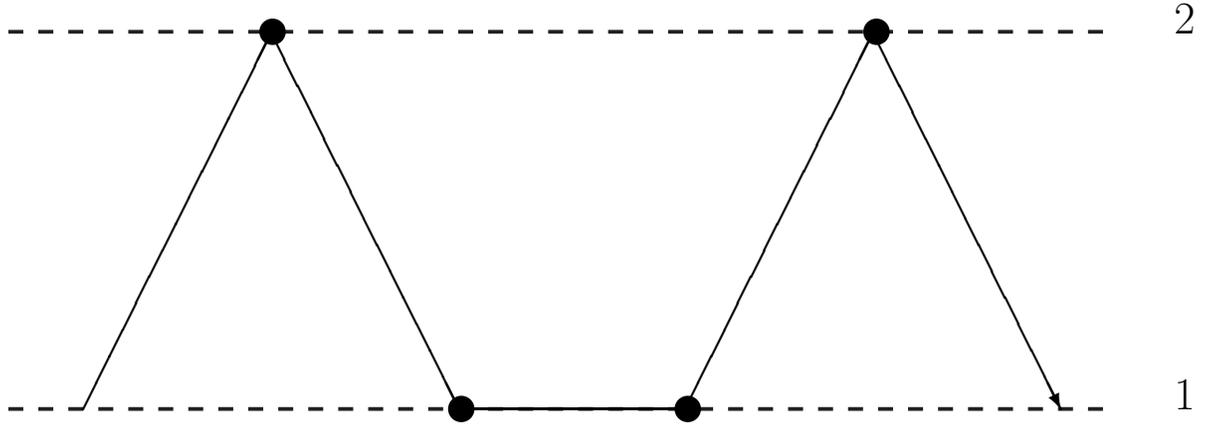
\begin{figure}
\begin{picture}(200,70)
\SetScale{2.845}
\DashLine(10,10)(155,10){2}
\DashLine(10,60)(155,60){2}
\put(165,10){\Large 1}
\put(165,60){\Large 2}
\thicklines
\put(20,10){\line(1,2){25}}
\put(45,60){\line(1,-2){25}}
\GCirc(45,60){1.5}{0}
\GCirc(70,10){1.5}{0}
\GCirc(100,10){1.5}{0}
\GCirc(125,60){1.5}{0}
\put(70,10){\line(1,0){30}}
\put(100,10){\line(1,2){25}}
\put(125,60){\vector(1,-2){25}}
\thinlines

\end{picture}
\caption{Keldysh propagator in external field.}

\label{wand}
\end{figure}

\newpage

The resultive ``wandering'' of the virtual particle is
depicted in
Fig. \ref{wand} ( Note that two lines in Fig. \ref{wand}
which, in our approach,
 correspond to the structures $T\{\phi(x)\phi(0)S\}$ (the
lower line)
and $S^{-1}$ (the upper line) have a direct
correspondence in the path integral approach
\cite{Semenoff,Landsman} which we
will
not discuss here.). When writing a corresponding analytical
expression, we
have to have in mind that the interaction vertex on the
upper line
corresponding to
$S^{-1}$ has an opposite sign compared to usual one --- just
because the sign
of $V(t)$ in $S^{-1}$ is reversed.

Likewise, we can calculate in any order the 22 - component
of the exact Green's
function in which case the virtual particle line starts and
ends in the
anti-chronological-ordered domain. For the mixed component
component $D_{21}$,
the wandering starts on the lower line in  Fig. \ref{wand}
and ends up on the  upper line. For the mixed component
component $D_{12}$ ---
the other way round. Two last statements follow from the
simple fact
  \be
<\phi(x) \phi(0)>\ =\ <S(-\infty, t) \phi_0(x) S(t, -\infty)
S(-\infty, 0) \phi_0(0) S(0, -\infty)>\ =\ \nonumber \\
<S(-\infty, t) \phi_0(x) S(t, \infty)
S(\infty, 0) \phi_0(0) S(0, -\infty)>\ =\ <\tilde{T}
\{\phi_0(x) S^{-1} \}
{T} \{\phi_0(0) S \}>
 \ee
and similarly
 \be
<\phi(0) \phi(x)>\ =\  <\tilde{T} \{\phi_0(0) S^{-1} \}
{T} \{\phi_0(x) S \}>
 \ee

We are ready now to formulate general Feynman rules in the
Keldysh technique.
 Suppose we want to find the exact thermal Green's function
in the theory
with interaction $\lambda \phi^3/6$
\footnote{Never mind that the theory does not exist
due to vacuum  instability, this is only an illustrative
example.}
 at finite $T$.
Present in in the matrix form like in (\ref{D0matr}). Then
11 -- component would
stand for thermal average of $T$ -- ordered product of exact
Heisenberg
operators
etc. Draw the same graphs as at $T = 0$. Each line
corresponds now not to a
scalar
function, but to a matrix (\ref{D0matr}). The vertices are
now tensors
$\Gamma_{abc}$
where $a,b,c = 1,2$. The tree vertices (to be substituted in
the graphs) are,
however,
very simple:
  \be
\label{Gam12}
\Gamma^0_{111} = \lambda,\ \ \Gamma^0_{222} = -\lambda
  \ee
and all other components are zero.  $\Gamma^0_{222}$ has the
opposite
sign compared to $\Gamma^0_{111}$ due to the reversed sign
of $V(t)$ in
(\ref{S-1}).

\newpage

\begin{figure}
\begin{center}
\begin{picture}(120,70)
\SetScale{2.845}
\thicklines

\put(10,35){\line(1,0){20}}
\put(90,35){\line(1,0){20}}

\CArc(60,35)(30,0,360)
\GCirc(30,35){1.5}{0}
\GCirc(90,35){1.5}{0}

\thinlines
\put(0,35){u}
\put(35,35){z}
\put(83,35){y}
\put(115,35){x}
\end{picture}
\end{center}
\caption{One--loop Green's function in $\phi^3$ theory.}

\label{FigD}
\end{figure}
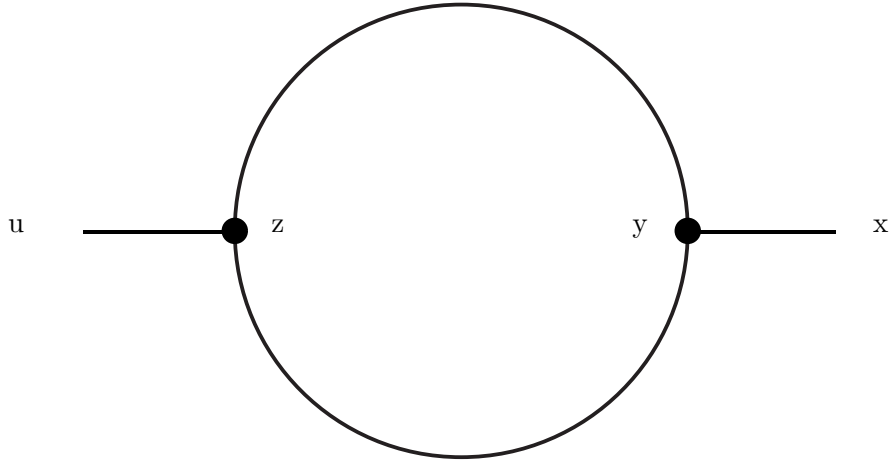

\newpage

Take the graph in Fig.\ref{FigD} as an example. The
corresponding
analytic expression  reads
  \be
\label{D2gr}
D_{ab}^{Fig.\ref{FigD}}(u-x) \ \sim  \nonumber \\
D^0_{ac}(u-z) \int d^4z \int d^4y \
\Gamma^0_{ced} D^0_{dd'}(z-y)   D^0_{ee'}(z-y)
\Gamma^0_{c'e'd'}\ \
  D^0_{c'b}(y-x)
 \ee
Summation over the indices which occur twice is assumed.

Similarly, exact Green's functions with arbitrary number of
legs can be found.
Generally, exact vertices have non-vanishing mixed
components like
$\Gamma_{121}$
etc.

Look again now at (\ref{D0matr}). Note that not all
components in the matrix
Green's function are independent. The relation
  \be
\label{svaz}
 D_{11}^0 \ +  D_{22}^0 \ =\  D_{12}^0 \ +\  D_{21}^0\
\equiv D_P^0
 \ee
holds. This relation is true in a {\it general} system, not
necessarily
in thermal
equilibrium. One can show that the same relation holds also
for the
exact Green's functions.

One can present (and it makes a lot of sense) the thermal
propagator in the
form which
involves only 3 independent components. To this end, we make
an orthogonal
 transformation
\be
\label{orttr}
D_\Omega \ =\ \Omega^{-1} D \Omega
 \ee
where
  \be
\label{Omega}
\Omega \ =\ \frac 1{\sqrt{2}}
\left( \begin{array}{cc} 1,1 \\-1,1 \end{array} \right)
 \ee
We have
 \be
\label{DOm}
D_\Omega \ = \ \left( \begin{array}{cc} 0, D_A \\ D_R, D_P
\end{array} \right)
 \ee
where $D_R = D_{11} - D_{12}$ and $D_A = D_{11} - D_{21}$
are nothing else as
retarded and advanced components of the Green's function.
Indeed, subtracting a
 straight
product from the $T$ -- product, one gets the commutator
(\ref{DRdef}) for $D_R$
and the commutator
  \be
\label{DAdef}
D_A(t, \vec{x}) \ =\ i\theta(-t) <[\phi(t, \vec{x}),
\phi(0)]>_T
\ee
for $D_A$. For the Fourier components the relation
$D_A(\omega, \vec{p}) \ = \ D_R^*(\omega, \vec{p})$ holds so
that
 $D_A(\omega, \vec{p})$ is analytic in the lower $\omega$
half-plane.

 $D_P$ is defined in (\ref{svaz}).
It is symmetric in $\omega$ due to (\ref{1221}).
Like $D_R$ and $D_A$, it also has a direct physical meaning.
For
example, the rate of photon production in quark-gluon plasma
is defined
is given by the imaginary part of the $P$ -- component of
the correlator
of electromagnetic currents. As far as the 2-point Green's
function is
concerned, $P$--component generally provides the information
about particle
distribution on the energy levels.

For future references, let us write down explicitly the
inverse relations
  \be
\label{invers}
D_{11} \ =\ \frac 12 (D_R + D_A + D_P), \ \ \
D_{12} \ =\ \frac 12 (-D_R + D_A + D_P),
\nonumber \\
D_{21} \ =\ \frac 12 (D_R - D_A + D_P), \ \ \
D_{22} \ =\ \frac 12 (-D_R - D_A + D_P),
  \ee

The matrix Green's functions (\ref{D0matr}), (\ref{DOm})
were invented in the
first place to describe the processes out of equilibrium.
For example, studying
P -- component of the 2-point Green's function in the state
which is
slightly out of thermal equilibrium, one can easily derive
Boltzmann kinetic
equation \cite{Keld,LP}. In the general case, $D_R = D_A^*$
and $D_P$ are
 completely independent
functions.

We are interested here in the system at thermal equilibrium.
Then $D_P$
is related to $D_R$ and $D_A$. It is not difficult to check
that, at the
tree level, the relation
  \be
\label{PbRA}
D_P^0(\omega, \vec{p}) \ =\ \coth \frac \omega {2T} (D_R^0 -
D_A^0)
  \ee
 holds. It can be proven that the same relation holds also
for the exact
Green's functions.

As the relation (\ref{svaz}) holds also
for the exact Green's function, the rotated exact Green's
function $D_\Omega$
retains
the form (\ref{DOm}) with zero in the upper left corner and
involving only
physical components.

All the steps of our derivation are easily generalized on
the fermion case. The
fermion Green's function has a matrix form. On the tree
level,
  \be
G^0_{11}(\omega, \vec{p})  \ =\ -i \int <T\{\psi_0(x) \bar
\psi_0(0)\}>_T
e^{ipx} \ d^4x
\nonumber \\
= (\omega \gamma^0 - \vec{p} \vecg{\gamma} + m)
\left[ \frac 1{\omega^2 - \epsilon_{\vec{p}}^2 + i0}
+ 2\pi i  n_F(\epsilon_{\vec{p}})  \delta (\omega^2 -
\epsilon_{\vec{p}}^2)
\right]  \nonumber \\
  G^0_{22}(\omega, \vec{p})  \ =\ -i \int
<\tilde{T}\{\psi_0(x)
\bar \psi_0(0)\}>_T e^{ipx}\ d^4x  \nonumber \\
= (\omega \gamma^0 - \vec{p} \vecg{\gamma} + m)\left[ -
\frac 1{\omega^2 - \epsilon_{\vec{p}}^2 - i0}
+ 2\pi i n_F(\epsilon_{\vec{p}})  \delta (\omega^2 -
\epsilon_{\vec{p}}^2)
\right]  \nonumber \\
G^0_{12}(\omega, \vec{p}) = i \int <\bar \psi_0(0) \psi_0(x)
>_T e^{ipx} d^4x \
= \nonumber \\
- (\omega \gamma^0 - \vec{p} \vecg{\gamma} + m)
 \frac {i\pi}{\epsilon_{\vec{p}}} \left[ (1 -
n_F(\epsilon_{\vec{p}}) )
\delta (\omega + \epsilon_{\vec{p}}) -
n_F(\epsilon_{\vec{p}})
\delta (\omega - \epsilon_{\vec{p}}) \right] \nonumber \\
G^0_{21}(\omega, \vec{p}) = -i \int <\psi_0(x) \bar
\psi_0(0)>_T e^{ipx} d^4x \
= \nonumber \\
- (\omega \gamma^0 - \vec{p} \vecg{\gamma} + m)
 \frac {i\pi}{\epsilon_{\vec{p}}} \left[ (1 -
n_F(\epsilon_{\vec{p}}) )
\delta (\omega - \epsilon_{\vec{p}}) -
n_F(\epsilon_{\vec{p}})
\delta (\omega + \epsilon_{\vec{p}}) \right]
  \ee
where
  \be
 \label{nF}
 n_F(\epsilon_{\vec{p}}) \ = \ \frac 1{e^{\beta
\epsilon_{\vec{p}} } + 1}
  \ee
is the Fermi distribution function.

Again, four components of the matrix Green's function
satisfy the relation
(\ref{svaz}) (which holds also in non-equilibrium case)
 and the orthogonal transformation (\ref{orttr}) brings the
Green's
function in the triangle form
 \be
\label{GOm}
G_\Omega \ = \ \left( \begin{array}{cc} 0, G_A \\ G_R, G_P
\end{array} \right)
 \ee
where $G_R \ = G_A^*$ and $G_P$ are the physical components.
In thermal
equilibrium, the relation
  \be
  \label{PfRA}
G_P(\omega, \vec{p})\  =\ \tanh \frac \omega {2T} (G_R -
G_A)
  \ee
holds.

The exact matrix Green's functions satisfy Dyson equations
  \be
\label {Dymatr}
D \ =\ D^0 + D^0 \Pi D, \ \ \ \ \ G \ =\ G^0 + G^0 \Sigma G
\ee
where $\Pi$ and $\Sigma$ are boson and fermion polarization
operators. Writing
$\Pi \ =\ (D^0)^{-1} - D^{-1},\ \ \Sigma\ =\ (G^0)^{-1} -
G^{-1}$ and using the
relation (\ref{svaz}), one can easily prove that
   \be
\label{svpol}
 \sum_{ab} \Pi_{ab} \ = \ 0, \ \ \sum_{ab} \Sigma_{ab} \ =\
0
  \ee
Also for a general multileg one-particle irreducible vertex
the relation
  \be
\label{svgam}
\sum_{\{a_i = 1,2\}} \Gamma_{\{a_i\}} \ = \ 0
  \ee
holds.

Rotating $\Pi$ and $\Sigma$ with the matrix (\ref{Omega}),
 one can present polarization operators in the triangle form
 \be
\label{PSOm}
\Pi_\Omega \ = \ \left( \begin{array}{cc} \Pi_P, \Pi_R \\
\Pi_A, 0
 \end{array} \right), \ \ \  \Sigma_\Omega \ = \ \left(
\begin{array}{cc}
\Sigma_P, \Sigma_R \\ \Sigma_A, 0  \end{array} \right)
 \ee
where $\Pi_R \ =\ \Pi_{11} + \Pi_{12},\ \Pi_A \ =\ \Pi_{11}
+ \Pi_{21},\
\Pi_P\ =\ \Pi_{11} + \Pi_{22}$ and similarly for $\Sigma$.
The remarkable fact is that the Dyson equations
(\ref{Dymatr}) for the
 retarded and advanced Green's function components have a
simple form
  \be
\label{DyRA}
D_R \ =\ D^0_R + D^0_R \Pi_R D_R, \ \ \ \ D_A \ =\ D^0_A +
D^0_A \Pi_A D_A
  \ee
and similarly for fermions. This is best seen using the
triangle form
(\ref{DOm}), (\ref{GOm}), (\ref{PSOm}). That means, in
particular, that to
find the poles
of the exact retarded Green's functions which determine
dispersion laws of
 collective excitations in thermal medium (see Chapter 6 for
more details),
one need to know only retarded components of the
corresponding polarization
operators.

There is, however, one subtle point which we are in a
position
to discuss. In our derivation, extensively used the notion
of $S$--
matrix in spite of the fact that, as was mentioned earlier,
it does
not exist generally speaking at finite $T$. This principal
difficulty
corresponds to a certain {\it technical} difficulty.
Consider e.g.
the scattering process in Fig. \ref{wand}. The amplitude
constructed
according to Keldysh rules is well defined as long as
scattering
at individual vertices occurs at nonzero angle. If the
external field
momentum is zero, we cannot define {\it individual} terms in
the elements of
the matrix product of adjacent Green's functions
$D^0(\vec{p}) D^0(\vec{p})$.
 Really, say $D^0_{12}(\vec{p})   D^0_{21}(\vec{p})$
involves a product of
two $\delta$- functions with the same argument and is not
defined.

Still, a consistent real time technique can be defined, and
the recipe is
the following: {\it i)} Let us assume that the {\it tree}
retarded and
advanced Green's
functions involve a small but nonzero {\it damping}. That
means that
$i\epsilon$
with small but finite $\epsilon$ should be substituted for
$i0$ in
(\ref{DR0}) etc.
{\it ii)} Calculate the graphs
(better in triangle basis) with such $D_R$ and $G_R$
assuming
$D_A = D_R^*, G_A =
G_R^*$ and substituting for $D_P$, $G_P$ the expressions
(\ref{PbRA})
and (\ref{PfRA}).
{\it iii)} Send $\epsilon$ to zero in the very end.
One can show that the amplitudes defined in such a way
involve no pathologies
and
are well defined \cite{Landsman}. Anyway, this recipe is
very physical: a
nonzero damping
$\epsilon$ means that no asymptotic states really exist.

$\epsilon$ as defined above is a technical parameter and is
eventually sent to
zero. However, there is also a physical damping due to
scattering processes
which displays itself as the imaginary part of the pole in
the exact Green's
functions. A natural physical estimate for damping is the
inverse free path time
$\tau_{free \ path}$ (this is not always so --- see a
detailed discussion of the
fascinating issue of so called {\it anomalous damping} in
Chapter 6 --- but
let us neglect these subtleties here). When one calculates
real time Green's
functions,
the scale $\tau_{free\ path}$ shows up as a barrier beyond
which the
perturbative series
explodes and higher order corrections are  not under good
control.

It suffices often to use a simplified version of the real
time technique due to Dolan and Jackiw \cite{DJ}. It amounts
to
neglecting all other components of the matrix Green's
function except
$D_{11}$, $G_{11}$. Generally, it is wrong --- in
particular, the amplitudes
involving
zero angle scattering would be pathological. But, for many
simple problems,
other matrix components do not contribute, indeed, and using
Dolan--Jackiw
technique is quite justified. Then the extra term $\sim
n_B(\epsilon)
\delta(\omega^2
- \epsilon^2)$ in $D^0_{11}$ (and similarly for fermions)
reflects the
presenbe of real particles in the heat bath with Bose
(Fermi) distribution.
Sometimes it is called  a ``temperature insertion''.

Suppose e.g. we are interested in the {\it real part} of the
retarded polarization
operator in the $\phi^3$ theory calculated by the one--loop
graph in
Fig.\ref{FigD}.
We have $\Pi_R = \Pi_{11} + \Pi_{12}$. $\Pi_{12}$ involves a
product of two
$\delta$ --
functions. That means that both virtual lines are put on
mass shell and hence
$\Pi_{12}$ is purely imaginary and does not contribute in
the real part. As for
$\Pi_{11}$ --- it depends on the one--loop level only on
$T$--
ordered components
of the Green's function. We have an ordinary $T = 0$ graph
and a graph with one
temperature insertion (a graph with temperature insertions
in both
lines contributes
in the imaginary part).

Such a calculation is quite meaningful. Not in $\phi^3$
theory which does not
exist,
of course, but for hot Yang--Mills theory where we also have
triple interaction
vertices. Real parts of dispersion laws of plasmon and
plasmino excitations are
calculated just in this way.

 \section{Pure Yang-Mills theory: deconfinement phase
transition.}
\setcounter{equation}0
\subsection{Preliminary remarks.}
We will start our discussion of the physics of hot
nonabelian gauge
theories with the theory not involving dynamical quarks ---
a pure
Yang--Mills theory. Like $QCD$, this theory is believed to
be
confining (there is no, of course, direct experimental
evidence as in
$QCD$, but it follows from theoretical arguments and from
lattice
measurements). Its physics is, however, rather different.
The spectrum
does not involve here mesons and baryons we are used to, but
only
glueball states. The lowest glueball state has a mass of
order $1.5\  GeV$.
It is not seen in the real world due to large mixing to
meson states, but in
the pure glue theory it should be just a stable particle.

As the temperature is increased, glueball states appear in
the heat
bath. When the temperature is very high, the system is
better described
in terms of original gluon degrees of freedom. The
interaction of gluons
at very high temperatures is weak and the system presents
pure gluon
plasma.

We will show now that there are serious reasons to believe
that, when
we increase the temperature and go over from the low
temperature glueball
phase to the high temperature gluon plasma phase, a phase
transition (the {\it deconfinement} phase transition) occurs
on the way.
  That was argued long ago by Polyakov \cite{Pol} and
Susskind
\cite{Sus}. On the heuristic level, the reasoning is the
following:

 At zero temperature the theory  enjoys confinement. That
means
that the potential between the test
heavy quark and antiquark grows linearly at large distances:
\be
\label{conf}
T=0:\ \ \ V_{Q\bar{Q}}(r) \sim \sigma r,\ \ \ r
\rightarrow
\infty
 \ee
 On the other hand, at high temperature when the system
presents
a weakly interacting plasma of gluons, the behavior of the
potential is quite different:

\be \label{Deb}
T \gg
\mu_{hadr}: \ \ \ V_{Q\bar{Q}}(r) \sim \frac {g^2(T)}{r}
e^{-m_D
r}
\ee
Here $m_D \sim gT$ is the Debye mass (see Chapter 6 for
detailed
discussion), and the potential is the
Debye screened potential much similar to the usual Debye
potential between static quarks in non-relativistic plasma.
There is no confinement at large $T$. There should be some
point
$T_c$ (the critical temperature) where the large $r$
asymptotics
of the potential changes and the phase transition from the
confinement phase to the Debye screening phase occurs.

   These simple arguments can be formulated in a rigorous
way.
Consider the partition function of the system written as the
Euclidean Matsubara path integral. Gluon fields satisfy
 periodic boundary conditions in imaginary time
\be \label{bc1}
A_\mu^a(\vec{x}, \beta) =
A_\mu^a(\vec{x}, 0)
\ee
where $\beta = 1/T$.
Let us choose a gauge where $A_0^a$ is
time-independent.  Introduce the quantity called the
Polyakov
loop
\be \label{P}
P(\vec{x}) = \frac 1{N_c} {\rm Tr} \exp\{
ig\beta A_0^a(\vec{x}) t^a\}
\ee
It is just a Wilson loop on the
contour which winds around the cylinder.  Consider the
correlator
\be \label{PP}
C_T(\vec{x}) = <P(\vec{x}) P^*(0)>_T
\ee
One can show \cite{Nad} that the correlator (\ref{PP}) is
related to the free energy of the test heavy quark-antiquark
pair immersed in the plasma.
\be \label{FQQ}
 C_T(\vec{x}) =
\frac 34 \exp\{-\beta F_{Q\bar{Q}}^{(3)}(r)\} + \frac 14
\exp \{
-\beta F_{Q\bar{Q}}^{(0)}(r)\}
\ee
where $r = |\vec{x}|$.
$F_{Q\bar{Q}}^{(3)}(r)$ and $F_{Q\bar{Q}}^{(0)}(r)$ are free
energies of test quark-antiquark pairs (alias static
potentials)
in the triplet and, correspondingly, the singlet net color
state. Let us take now the limit $r \rightarrow \infty$. The
quantity
\be \label{Cinf}
C_T(\infty) = \lim_{r \rightarrow
\infty} C_T(r)
\ee
plays the role of the {\it order parameter}
of the deconfinement phase transition. At small $T$,
$F_{Q\bar{Q}}(r)$  grow linearly at  $r \rightarrow
\infty$ and $C_T(\infty) = 0$. At large $T$, free energies
do not grow and $C_T(\infty)$ is some nonzero constant (if
one
would naively substitute in Eq.(\ref{FQQ}) the Debye form of
the
potentials (\ref{Deb}), one would get $C_{T \gg
\mu_{hadr}}(\infty) = 1$, but it is not quite true because
$F_{Q\bar{Q}}(r)$ involve also a constant depending on the
ultraviolet cutoff of the theory. See \cite{Jengo,bub} for
detailed discussion). There is a phase transition in
between.

What are the properties of this phase transition ? There are
not
quite rigorous but suggestive theoretical arguments based on
the
notion of ``universality class'' \cite{Svet,SvetPR} which
predict
different properties for different gauge groups. The main
idea
is that the pure YM theory based on $SU(2)$ color group has
some
common features with the Ising model (with global symmetry
$Z_2$), the theory with $SU(3)$ gauge group --- with a
generalized Ising model (the Potts model) with the global
symmetry $Z_3$ etc.  The Ising model has the second order
phase
transition, and the same should be true for pure $SU(2)$
gauge
theory. A system with $Z_3$ symmetry display, however, the
first
order phase transition, and the same should be true for pure
$SU( 3)$ theory. The lattice data
\cite{purlat}
are in a nice agreement with this prediction. Also, the
critical
exponents of the second order phase transition were measured
\cite{2ind}.
Their numerical values are close to the numerical values of
critical indices in the Ising model.

 The arguments presented are rather suggestive, and the
deconfinement
phase transition is probably there, indeed. However, there
is no
{\it absolute} clarity here yet. We will return to
discussion of
this question later on and now let us dwell on a very
confusing issue
of domain walls and bubbles in the high - $T$ phase.
\subsection{Bubble confusion.}
There was a long-standing confusion concerning the nature of
deconfinement phase transition in pure YM theory. It has
been
clarified only recently.

In scores of papers published since 1978, it was explicitly
or
implicitly assumed that one can use the cluster
decomposition
for the correlator (\ref{PP}) at large $T$ and attribute the
meaning to the temperature average $<P>_T$. Under this
assumption, the phase of this average can acquire $N_c$
different values: $<P>_T = C\exp\{2\pi ik/N_c\}, \ \ k =
0,1,\ldots, N_c-1$ which would correspond to $N_c$ distinct
physical phases and to the spontaneous breaking of the
discrete
$Z_N$-symmetry. Assuming that , the surface energy
density of the domain walls separating these phases has been
evaluated in \cite{pis}. Then the ``bubbles'' of one of the
high temperature phases inside another would be abundant in
early
Universe. These bubbles would be surrounded by the walls
carrying significant surface energy. That could influence
the
 evolution of the Universe and could lead
 to some observable effects.

However, the standard interpretation is wrong. In
particular:
\begin{enumerate}
\item Only the correlator (\ref{PP}) has the physical
meaning. The phase of the expectation value $<P>_T$ is not a
physically measurable quantity. There is only {\it one}
physical
phase in the hot YM system.
\item The ``walls'' found in \cite{pis} should not be
interpreted as physical objects living in Minkowski space,
but
rather as Euclidean field configurations, kind of ``planar
instantons'' appearing due to non-trivial $\pi_1({\cal G}) =
Z_N$
where ${\cal G} = SU(N)/Z_N$ is the true gauge symmetry
group of
the {\it pure} Yang-Mills system.
\item The whole bunch of arguments which is usually applied
to nonabelian gauge theories can be transferred with a
little
change to hot $QED$. The latter also involves planar
instantons
appearing due to non-trivial $\pi_1[U(1)] = Z$. These
instantons should {\it not} be interpreted as Minkowski
space
walls.
\end{enumerate}

 We refer the reader to our paper \cite{bub} where a
detailed argumentation
of these statements   was given. Here we restrict ourselves
by
outlining the  main  physical points in the arguments.

A. {\it Continuum theory.}

A preliminary remark is that the situation when the symmetry
is broken at high
temperatures and restores at low temperatures is very
strange and unusual.
The opposite is much more common in physics. We are aware of
only one model
example where spontaneous symmetry breaking survives and can
even be induced
at high temperatures \cite{Moh}. But the mechanism of this
breaking is
completely different from what could possibly occur in the
pure Yang-Mills
theory.

Speaking of the latter, we note first that
there is no much sense to speak about the spontaneous
breaking of
$Z_N$ - symmetry because such a symmetry is just not there
in the theory.
As was already mentioned, the true gauge group of pure YM
theory is
$SU(N)/Z_N$ rather than $SU(N)$. This is so because the
gluon fields belong
to the adjoint color representation and are not transformed
at all under the
action of the elements of the center $Z_N$ of the gauge
group $SU(N)$.

$<P>_T$ as such is not physical because it corresponds to
introducing a
single fundamental source in the system: $<P>_T\  = \exp\{-
\beta F_T\}$
where
$F_T$ is the shift in free energy of the heat bath where a
single static
 fundamental source source is immersed \cite{Larry}. But one
{\em can}not put a single fundamental source
in a finite spatial box with
periodic boundary conditions \cite{Hift} (Such a box should
be introduced
to regularize the theory in infrared). This is due to the
Gauss law
constraint: the total color charge of the system "source +
gluons in the
heat bath" should be zero, and adjoint gluons cannot screen
the fundamental
source. This observation resolves the troubling paradox:
complex $<P>_T$
would mean the complex free energy $F_T$ which is
meaningless.

The "states" with different $<P>_T$ could be associated with
different minima
of the effective potential \cite{Weiss}
 \begin{equation}
 \label{pot}
V_T^{eff}(A_0^3) = \frac{\pi^2T^4}{12}\left\{ 1 - \left[
\left(
\frac{gA_0^3}{\pi T} \right)_{mod.2} - 1 \right]^2
\right\}^2
 \end{equation}
For simplicity, we restrict ourselves here and in the
following
with the $SU(2)$ case.

This potential
is periodic in $A_0^3$ with the period $2\pi T/g$.
The minima at $A_0^3 = 4\pi nT/g$ correspond to $P=1$
while the minima at $A_0^3 = 2\pi (2n+1)T/g$ correspond to
$P = -1$. There
{\em are} also planar (independent of y and z)
configurations which interpolate
between $A_0^3 = 0$ at $x = -\infty$ and $A_0^3 = 2\pi T/g$
at $x = \infty$.
These configurations contribute to Euclidean path integral
and are
topologically
nonequivalent to the trivial configuration $A_0^3 = 0$
(Note that the configuration interpolating
between $A_0^3 = 0$ and $A_0^3 = 4\pi T/g$ {\it is}
topologically equivalent
to the trivial one. Such a configuration corresponds
to the equator on $S^3 \equiv
SU(2)$ which can be  easily slipped off. A topologically
non-trivial
configuration corresponds to a meridian going from the north
pole of the sphere
to its south pole and presents a noncontractible loop on
$SU(2)/Z_2$ ).
Actually, such configurations were known for a long time by
the nickname of
't {Hooft}
 fluxes \cite{flux}.

 Minimizing the surface
action density in a non-trivial topological class, we arrive
at the
configuration
which is rather narrow (its width is of order $(gT)^{-1}$)
and has the action
density
  \begin{equation}
  \label{act}
\sigma^{SU(2)} = \frac{4\pi^2 T^2}{3\sqrt{6} g} + CgT^2
  \end{equation}
(as was shown in \cite{bub}, the constant $C$ cannot be
determined analytically
in contrast to the claim
of \cite{pis} due to infrared singularities characteristic
for thermal gauge
theories \cite{Linde}). These topologically non-trivial
Euclidean configurations
are quite analogous to instantons. Only here they are
delocalized in two
transverse directions and thereby the relevant topology is
determined by
$\pi_1[{\rm {\cal G}}]$ rather than $\pi_3[{\rm {\cal G}}]$
as was the case
for usual localized instantons. But, by the same token as
the instantons
cannot be interpreted as real objects in the Minkowski space
even if they
are static (and, at high $T$, the instantons with the size
$\rho \gg T^{-1}$
become static), these planar configurations cannot be
interpreted as real
Minkowski space domain walls.

I want to elucidate here the analogy between nonabelian and
abelian theories.
The effective potential for standard QED at high temperature
has essentially
the same form as (\ref{pot}):
   \begin{equation}
 \label{potab}
V_T^{eff}(A_0) = -\frac{\pi^2T^4}{12}\left\{ 1 - \left[
\left(
\frac{eA_0}{\pi T} + 1 \right)_{mod.2} - 1 \right]^2
\right\}^2
 \end{equation}
It is periodic in $A_0$ and acquires minima at $A_0 = 2\pi
nT/e$. Here
different
minima correspond to the same value of the standard Polyakov
loop
$P_1( {\bf x}) = \exp\{ie\beta A_0( {\bf x})\} \ =\ 1$. One
can introduce
, however, the quantity $P_{1/N}( {\bf x}) =
\exp\{ie\beta A_0({\bf x})/N\}$ which corresponds to probing
the system
with a fractionally charged heavy source : $e_{{\rm source}}
= e/N$. Note that
 a fractional heavy source in a system involving
only the fermions with charge $e$ plays exactly the same
role
 as  a fundamental heavy source in the pure YM system
involving
only the adjoint color fields. A single fractional source
would distinguish
between different minima of the effective potential. If $N
\rightarrow
\infty$ , all minima would be distinguished, and we would
get infinitely many
distinct "phases".

But this is wrong. One cannot introduce a {\em single}
fractional source  and measure $<P_{1/N}>_T$ as such due to
the Gauss
law constraint --- the total electric charge should be zero
and integer--charged
electrons cannot screen a fractionally charged source.
What can be done is to introduce a pair of fractional
charges with opposite
signs and measure the correlator $<P_{1/N}({\bf x})
P_{1/N}^*(0)>_T$.
The latter is
a physical quantity but is not sensitive to the phase of
$P$. The same concerns
the correlator $<P_{1/N}({\bf x}_1) \ldots P_{1/N}({\bf
x}_N)>_T$ which
corresponds to putting $N$ fractional same-sign charges at
different spatial
points.

Finally, one can consider the configurations $A_0({\bf x})$
interpolating
between different minima of (\ref{potab}). They are
topologically inequivalent
to trivial configurations and also have the meaning of
planar instantons
\footnote{In the abelian case, there are infinitely many
topological classes:
 $\pi_1[U(1)] = Z$.}.
But not the meaning of the walls separating distinct
physical phases.
The profile and the surface action density of these abelian
planar instantons
can be found in the same way as it has been done in
Ref.\cite{pis} for
the nonabelian case. For configurations interpolating
between adjacent
minima, one gets
  \begin{equation}
  \label{sab}
\sigma^{u(1)} = \frac{2\pi^2 (2\sqrt{2} - 1)T^2}{3\sqrt{6}e}
+ CeT^2 \ln(e)
  \end{equation}
where $C$ is a numerical constant which {\it can}
in principle be analytically evaluated.

There is a very fruitful and
instructive analogy with the Schwinger model.  Schwinger
model is
the two-dimensional $QED$ with one massless fermion.
Consider this theory
at high temperature $T \gg g$ where $g$ is the coupling
constant (in
two dimensions it carries the dimension of mass). The
effective potential in
the constant $A_0$ background has the form which is very
much analogous to
(\ref{pot},\ref{potab}):
  \begin{equation}
  \label{potSM}
V^{\rm eff}(A_0) = \frac{\pi T^2}{2}\left[ \left(1 + \frac
{gA_0}{\pi T}
\right)_{mod. \ 2} - 1 \right]^2
  \end{equation}
It consists of the segments of  parabola and is periodic in
$A_0$
with the period
$2\pi T/g$. Different minima of this potential are not
distinguished by a heavy
integerly charged probe but could be distinguished by a
source with
fractional charge. Like in four dimensions, there are
topologically non-trivial
field configurations which interpolate between different
minima. These
configurations are localized (for $d=2$ there are no
transverse directions over
which they could extend) and are nothing else as high-$T$
instantons. The
minimum of the effective action in the one-instanton sector
is achieved
at the configuration \cite{bub,inst}
 \begin{equation}
  \label{inst}
A_0(x) = \left[ \begin{array}{c} \frac {\pi T}g \exp\left\{
\frac g{\sqrt{\pi}}(x - x_0) \right\}, \ \ \ x \leq x_0 \\
\frac {\pi T}g \left[ 2- \exp\left\{
\frac g{\sqrt{\pi}}(x_0 - x) \right\}\right], \ \ \ x \geq
x_0
\end{array}
\right.
 \end{equation}
The instanton (\ref{inst}) is localized at distances $x -
x_0 \sim g^{-1}$
and has the action $S_I = \pi^{3/2}T/g$. But, in spite of
that it is
time-independent, it is the essentially Euclidean
configuration and should
not be interpreted as a "soliton" with the mass $M_{sol.?} =
TS_I$ living
in the physical Minkowski space.

B. { \it Lattice Theory. }

The most known and the most often quoted arguments in favor
of the standard
conclusion of the spontaneous breaking of $Z_N$-symmetry in
hot Yang-Mills
theory come from lattice considerations. Let us discuss anew
these arguments
and show that, when the question is posed properly, the
answer {\it is}
different.

Following Susskind \cite{Sus}, consider the hamiltonian
lattice formulation
where the theory is defined on the 3-dimensional spatial
lattice and the time
is continuous. In the standard formulation, the dynamic
variables present the
unitary matrices $V({\bf r}, {\bf n})$ dwelling on the links
of the lattice
(the link is described as the vector starting from the
lattice node {\bf r}
with the direction {\bf n}).
The hamiltonian is
 \begin{equation}
 \label{hamlat}
H = \sum_{\rm links} \frac{g^2(E^a)^2}{2a} - \frac 2{ag^2}
\sum_{\rm plaq.}
{\rm Tr}\{V_1 V_2 V_3 V_4\}
 \end{equation}
where $a$ is the lattice spacing, $g$ is the coupling
constant and $E^a$ have
the meaning of canonical momenta $[E^a({\bf r}, {\bf n}),
V({\bf r}, {\bf n})]
= t^a V({\bf r}, {\bf n})$. Not all eigenstates of the
hamiltonian
(\ref{hamlat}) are, however, admissible but only those which
satisfy the Gauss
law constraint. Its lattice version is
  \begin{equation}
  \label{Gaulat}
G^a({\bf r}) = \sum_{\bf n} E^a({\bf r}, {\bf n}) = 0
  \end{equation}
It is possible to rewrite the partition function of the
theory (\ref{hamlat},
\ref{Gaulat}) in terms of the {\it dual variables}
$\Omega_{\bf r} \in SU(2)$
which are defined not at links but at the nodes of the
lattice.
$\Omega_{\bf r}$ are canonically conjugate to the Gauss law
constraints
(\ref{Gaulat}) and have the meaning of the gauge
transformation matrices
acting on the dynamic variables $V({\bf r}, {\bf n})$.
$\Omega_{\vec{r}}$
correspond to the Polyakov loop operators (\ref{P}) in the
continuum theory.
In the strong
coupling limit when the temperature is much greater than the
ultraviolet
cutoff $\Lambda_{ultr} \sim 1/a$, the problem can be solved
analytically.
The effective dual hamiltonian has 2 sharp minima at
$\Omega_{\bf r} = 1$ and $\Omega_{\bf r} = -1$ and this has
been
interpreted as the
spontaneous breaking of $Z_2$-symmetry.

Note, however, that the same arguments could be repeated in
a much simpler and
the very well known two-dimensional Ising model.
 Being formulated in terms of
the physical spin variables $\sigma$, the theory exhibits
the spontaneous
breaking of $Z_2$-symmetry at low temperatures, and at high
$T$ the symmetry
is restored. But the partition function of the Ising model
can also be
written in terms of the dual variables $\eta$ defined at the
plaquette
centers \cite{Kram}.
\be
\label{Kram}
Z \ =\ \sum_{\{\sigma_i\}} \exp\{-\beta H(\{\sigma_i\})\}\
=\
 A(\beta) \sum_{\{\eta_i\}}
\exp\{-\beta^* H^{dual}(\{\eta_i\})\}
 \ee
where $\sigma_i$ are the original spin variables sitting on
the nodes of
the two--dimensional lattice, $\eta_i$ are disorder
variables residing on
the plaquettes, $\beta^* = -1/2 \ln( \tanh \ \beta)$ and the
dual hamiltonian
$H^{dual}(\{\eta_i\})\ =\ - \sum_{ij}  (\eta_{ij}
\eta_{i+1,j}
+ \eta_{ij} \eta_{i,j+1})$  has the same functional form as
the original one.
When $\beta$ is large, $\beta^*$ is small and vice versa so
that
dual variables are ordered at
high rather than at low temperatures  and, accepting
the dual hamiltonian $H^{dual}(\{\eta_{ij}\})$ at face
value, one could
conclude that spontaneous breaking of $Z_2$ symmetry occurs
at {\it high}
temperatures. This
obvious paradox is resolved by noting that the dual
variables $\eta$ are
not measurable and have no direct physical meaning. The
"domain wall"
configurations interpolating between $\eta = 1$ and $\eta =
-1$ {\it do}
contribute in the partition function formulated in dual
terms. But one cannot
feel these configurations in any physical experiment.

Another example of a lattice theory  which is even more
similar to the lattice
pure YM theory in question is $Z_2$ three--dimensional Ising
flux model which we
will discuss in little more details later. Its
dual is the standard 3D Ising spin model. The dual spin
variables are
 ordered at high $T$ and, treating the dual spin hamiltonian
seriously, we
would have two ordered phases with domain walls which
separate them.
However, as dual spin variables are not physical and cannot
be observed,
such a physical interpretation is wrong \cite{Kiskis}.

And the same concerns the lattice pure YM theory
\cite{bub,Hanss,Kiskis}.
There {\it are}
configurations interpolating between different $\Omega{\bf
r} \in Z_2$
 and contributing to the partition function, but they do not
correspond
to any real-time object and cannot be detected as such in
any physical
experiment. That means, in particular, that metastable
states in electroweak
theory of the kind discussed in \cite{Kort} also do not
exist. The absence
of metastable states in hot $QCD$ [in the theory with
quarks, the effective
potential $V^{eff}(A_0)$ is no longer periodic under the
shift
$A_0^3 \to A_0^3 + 2\pi T/g$ but involves global minima
with $P = 1$ at
 $A_0^3 \ =\ 4\pi  T/g$
 and local metastable minima at $A_0^3 \ =\
2\pi (2n+1) T/g$ with $P = \ -1$] was shown earlier in
\cite{KogSem}.

To summarize, there is only one physical phase at high $T$.
Its
properties are relatively simple --- it is the weakly
interacting plasma of gluons.  The description in terms of
dual
variables is useful for some purposes
but one
should be very careful not to read out in it something which
is
not in Nature.

\subsection{More on phase transition.}
Bubbles are not there, but is the deconfinement phase
transition there ?
If it is not really associated with spontaneous breaking of
physical $Z_N$
symmetry, what is its physics and are the heuristic
arguments presented
in the beginning of this chapter really compelling ?
We will discuss here this question from different viewpoints
giving
the arguments pro and contra.

An additional argument {\it pro}
comes from the large $N_c$ analysis. In the plasma phase,
the energy density
$E(T)$ is proportional to $(N_c^2 - 1) T^4$ where $T^4$
appears by
dimensional reasons and the factor $N_c^2 - 1$  is (half)
the number of
colored degrees of freedom.
\footnote{The exact calculation for the coefficient of the
leading
term and the perturbative corrections is presented in
Chapter 6.}

We see
that  the energy density becomes
 infinite in the limit $N_c \to \infty$. That
means that if we start to heat the system from $T =0$, we
just cannot reach the $QGP$ state --- to this end, an
infinite
energy should be supplied !

This physical conclusion can also be reached if looking at
the problem from the low temperature end. The mass spectrum
of
the theory in the limit $N_c \to \infty$ involves infinitely
many
narrow states. The density of states grows exponentially
with mass
\footnote{One of the way to see it is to use the string
model for the hadron spectrum. A string state with large
mass is highly degenerate. The number of states with a given
mass depends on the number of ways $p(N)$ the large integer
$N \sim M^2/\sigma$ ($\sigma$ is the string tension) can be
decomposed in the sums of the form $N = \sum_i n_i$ (see
e.g. \cite{Polbook}). $p(N)$ grows exponentially with $N$.
That does not mean, of course, that in real $QCD$ with large
$N_c$ the spectrum would be also degenerate. Numerical
calculations in $QCD_2$ with adjoint matter fields show that
there is no trace of degeneracy and the spectrum displays a
stochastic behavior \cite{Kleb}. And that means in
particular that there is little hope to describe {\it
quantitatively} the $QCD$ spectrum  in the limit $N_c \to
\infty$ in the string model framework. But a qualitative
feature that the density of states grows exponentially as
the mass increases is common for the large $N_c$ $QCD$ and
for the string model.}
 \be
 \label{dens}
\rho(M) \propto e^{cM}
\ee
 The contribution of each massive state in free energy
density is
exponentially suppressed $\propto \exp\{-M/T\}$
for large $M/T$ [cf. Eq. (\ref{Fpi0}) in the next chapter].
But
  the total free energy
 \be
 \label{Zdens}
 F \sim \int \rho(M) e^{-M/T} dM
 \ee
diverges at $T \geq c^{-1}$. There is a Hagedorn
temperature $T_H = c^{-1}$ above which a system cannot be
heated \cite{Hagedorn}.

When $N_c$ is large but finite,  no limiting temperature
exists (It is seen also from the low temperature viewpoint:
at finite $N_c$ the states have finite width and starting
from some energy begin to overlap and cannot be treated as
independent degrees of freedom), and one can bring the
system to the plasma phase when supplying enough energy.
But when $N_c$ is large, the required energy is also large.
That suggests (though does not prove, of course) that at
large finite $N_c$ a first order phase transition with a
considerable latent heat takes place.

This conjecture is
supported by the lattice measurements \cite{purlat}. Indeed,
a first order phase transition is observed at $N_c = 3$
and a second order phase transition --- at $N_c = 2$. A
first
order phase transition need not be associated with an order
parameter. But a second order phase transition usually is.
The problem here that a good {\it local} order parameter
is not available in the problem. We have seen that the
Polyakov
loop expectation value as such is not a physically
observable
quantity and the only order parameter available is the
correlator
of Polyakov loops (\ref{PP}) which is nonlocal. The
situation
reminds a little bit the Berezinsky -- Kosterlitz --
Thouless
phase transition in 1--dimensional statistical systems
\cite{BKT}
where also no local order parameter is present, but some
nonlocal
correlator changes its behavior at large distances.

Another, a much more close analogy is the percolation phase
transition
which is well known in condensed matter physics (see e.g.
\cite{Shkl}) and
displays itself in the three--dimensional lattice Ising flux
model
\cite{Kis1,Kiskis}. Let us discuss it in some more details.

The partition function of the model is defined as
  \be
  \label{Isflux}
Z  =\ \sum_{\{\theta_l\}} \exp \left\{ - \frac \sigma T
\sum_l \theta_l
\right\}
  \ee
where discrete variables $\theta_l = 0,1$ are defined on the
links of the 3D
cubic lattice. When $\theta_l = 1$, the link carries a flux
with energy
$\sigma$ and, when $\theta_l = 0$, there is no flux. An
additional
constraint that the sum of the fluxes $\theta_l$ over the
links adjacent
to any node is even so that the flux lines cannot terminate.

When the temperature is small, configurations with nonzero
$\theta_l$ are
suppressed, only few links carry fluxes and they are grouped
in a set
of rarefied disconnected clusters. When the temperature is
increased, the
density of clusters is also increased. The phase transition
occurs when
a connected supercluster of flux lines extending on the
whole lattice is
formed.

The model (\ref{Isflux}) (as well as the three-dimensional
$Z_2$ gauge
lattice model which bears a {\it considerable} resemblance
to the lattice
$SU(2)$ gauge theory) is dual to the 3D Ising model. Its
partition function
just coincides up to a coefficient with the partition
function of the latter:
  \be
 \label{3Is}
Z \ =\  A(\beta)  \sum_{\{s_i\}}  \exp \left\{ \beta
\sum_{ij} (s_{ij} s_{i+1,j}
+ s_{ij} s_{i,j+1}) \right \}
  \ee
with $\tanh \beta \ =\ e^{-\sigma/T}$ and where dual spin
variables $s_i$
dwell on the nodes. Dual spins are ordered at large
temperature $T$. At
the critical temperature and below it, ordering disappears.
But, as dual
spins are not physically observed variables, $<s_i>$ cannot
serve as a physical
order parameter. The order parameter for the original theory
is the
{\it percolating flux density} which is easy to visualize
(it is the probability
that a given link belongs to the percolating supercluster),
but no nice
analytic expression in terms of original variables
$\theta_l$ can be
written for this quantity. Deconfining phase transition for
the model
(\ref{hamlat})
resembles in many respects the phase transition in this
example --- though
the order parameter which is the string tension is easy to
visualize,
 it cannot
be expressed via physical dynamic variables $V(\vec{r},
\vec{n})$.
\footnote{We will see in Chapter 5 that also the chiral
restoration phase
transition for the theory with light quarks involves a
``percolation''
in Euclidean space driven by instantons. Only in the latter
case percolation
occurs not at high, but at low temperatures.}

However, an analogy is not yet the proof.
 What is quite definite,
of course, is that the potential between heavy quarks grows
linearly at large distances at $T = 0$ and is screened in
 plasma phase. The most simple and natural assumption is,
indeed,
that a change of regime occurs at a finite critical
temperature $T_c$.
  On the other hand,
we want to emphasize that it is impossible now to conclude
from
pure theoretical premises that $T_c \ \neq \ 0$. A queer but
admissible
theoretical possibility would be that the interquark
potential is screened
at any finite temperature however small it is.

Let us discuss in more details the lattice results which
display a phase
transition at finite $T_c$. According to \cite{purlat}
  \be
 \label{Tclat}
\frac {T_c}{\sqrt{\sigma}}\ =\ \left\{
\begin{array}{cc} 0.69 \pm 0.02, \ \ \ \ \ \ SU(2) \\
0.56 \pm 0.03,\ \ \ \ \ \ \ SU(3) \end{array} \right.
  \ee
where $\sigma$ is the string tension.
The scaling (i.e. the fact that the ratio
$T_c/\sqrt{\sigma}$ does not
depend on the lattice spacing when the latter is small
enough) has been checked
which indicates that the phase transition temperature
(\ref{Tclat}) is for real
and is not just a lattice artifact. The result (\ref{Tclat})
has been confirmed
later  in \cite{Tcglue} (more recent results exceed previous
ones
by 10 --15 \% but it is not an issue here).
 What is rather surprising, however,  is
that the critical temperature turned out to be very small.
$T_c$
should be compared with the mass $M$ of the lowest $0^{++}$
glueball state
which has been measured to be $\approx 3.5 \sqrt{\sigma}$
for $SU(3)$
(see e.g. \cite{glueball}). At $T_c \approx 0.16 M$, the
system presents a
{\it very}
rarefied gas of glueballs (its density being proportional to
$\exp\{-M/T\}$).
They almost do not interact with each other and it is
difficult to understand
why the system undergoes a phase transition at this point.
\footnote{Actually, the same lattice data which indicate the
presence
of the phase transition at the low temperature (\ref{Tclat})
indicate also that
glueballs {\it do} interact at $T = T_c$. The lattice value
of the energy
density at $T = T_c$ is 5 times larger than the energy
density of free
glueball gas \cite{EKR}. And it is still completely unclear
why the interaction
of glueballs is so strong while their tree level density is
so low.}

The measurements in \cite{purlat} were performed with a
standard lattice
hamiltonian
(\ref{hamlat}). Note, however, that one can equally well
consider the lattice
theory with the hamiltonian having the same form as
(\ref{hamlat})
but involving
not the unitary but the orthogonal matrices $V^{adj}({\bf
r}, {\bf n})
 \in SO(3)$. Both lattice
theories should reproduce one and the same continuous Yang-
Mills theory
in the limit when the inverse lattice spacing is much
greater than all physical
parameters (As far as I understand, there is no unique
opinion on this issue
in the lattice community. If, however, lattice hamiltonia
involving unitary
and orthogonal matrices would indeed lead to different field
theories in the
continuum limit, it would mean that the Yang-Mills field
theory is just not
defined until a particular procedure of ultraviolet
regularization is
specified.
This assertion seems to me too radical, and I hesitate to
adopt it.).

 In the strong coupling limit $T \gg \Lambda_{ultr.}$ the
two lattice
theories are completely different. The theory with
orthogonal matrices has
the same symmetry properties as the continuum theory , and
there is no
$Z_2$-symmetry whatsoever. The effective dual hamiltonian
depending on the
gauge transformation matrices $\Omega_{\bf r}^{adj} \in
SO(3)$ also has no such
symmetry and there is nothing to be broken.

It was observed it \cite{purlat} that the phase transition
in the lattice
system with fundamental $SU(2)$ matrices
 is associated with the spontaneous symmetry breaking in the
dual
hamiltonian  $H^{eff}(\Omega_{\bf r}^{fund.})$
(We repeat that such a breaking is not a physical symmetry
breaking because
it does not lead to the appearance of domain walls
detectable in experiment.).
In our opinion, however, the additional $Z_2$-symmetry which
the
hamiltonian (\ref{hamlat}) enjoys is a nuisance rather than
an advantage.
It is a specifically lattice feature which is not there in
the continuum
theory. We strongly suggest to people who can do it to
perform a numerical
study of the deconfinement phase transition for the theory
involving
orthogonal rather than unitary matrices. In that case, no
spontaneous $Z_N$
 breaking can
occur. Probably, for finite lattice spacing, one would
observe a kind of
crossover rather than the phase transition. The crossover is
expected to
become more and more sharp  as the lattice spacing (measured
in physical
units) would become smaller and smaller.

It would be interesting also to try to observe the "walls"
(i.e. the planar
Euclidean instantons) for the orthogonal lattice theory.
They should
"interpolate" between
$\Omega_{\bf r}^{adj} = 1$ and $\Omega_{\bf r}^{adj} = 1$
along a topologically non-trivial
path. Like any other topological effect, these instantons
should become
visible only for a small enough lattice spacing (much
smaller than the
characteristic instanton size), and to detect them
 is definitely not an easy task.
But using the orthogonal matrices
 is the only way to separate from lattice artifacts.
 The only available
numerical study
\cite{Kaj} was done for the theory with unitary matrices
and  too close to  the strong coupling regime (the lattice
had just two
links in temporal direction
\footnote{We do not discuss here the measurements of
interface tension
between confined and deconfined phases in $SU(3)$  at $T_c$
which is
a perfectly
well defined physical quantity. These measurements were
performed by many
people and at larger lattices.}
) where these artifacts are
decisive. Thereby, it is not conclusive.

A comprehensive lattice study of the deconfinement phase
transition with orthogonal matrices has been performed
so far. Recently, a paper appeared \cite{Rajiv}
where the question was studied in the
mixed $SU(2)$ theory involving {\it both} the standard
Wilson action with
unitary
matrices and the term with orthogonal matrices:
  \be
\label{mixed}
S_{\rm mixed} \ = \ - \frac{\beta_F}2 \sum_\Box {\rm Tr}_F
U(\Box)
\ - \frac{\beta_A}3 \sum_\Box {\rm Tr}_A U(\Box)
  \ee
where $U(\Box)$ is the product of four unitary matrices on a
given plaquette
and the adjoint trace ${\rm Tr}_A$ is $(2 {\rm Tr}_F )^2 -
1$.
 Surprisingly, it was observed that the phase transition
which was of the
 second order at zero $\beta_A$ becomes the first order at
some admixture
of the adjoint term in the action. It was argued in
\cite{Steph}
that the authors of \cite{Rajiv} observed actually the so
called
{\it bulk} first order transition \cite{Bhanot} which is a
pure lattice
artifact and that a more accurate study indicates (not yet
quite definitely)
the presence of both first order bulk phase transition and
the deconfinement
second order phase transition in temperature. In principle,
these two
phenomena can be sorted out on lattices of larger size.

In recent \cite{sharath} the question was studied in the
$SU(2)$ theory with
pure adjoint action. Unfortunately, the numerical accuracy
was not so
high for the critical temperature in the continuum limit (as
the string
tension is zero in pure adjoint theory due to screening of
adjoint
sources, $T_c$ should be measured in the units of the lowest
glueball
mass) to be determined.

Thus the experimental situation is still not quite clear
now. In our opinion,
we would not be quite sure that the deconfinement phase
transition occurs
at a finite
temperature, and that this temperature is indeed six times
smaller than the
 glueball mass until it will be confirmed  in lattice
experiments with pure
orthogonal action.

\section{Lukewarm pion gas.}
\setcounter{equation}0

\subsection{Chiral symmetry and its breaking.}
If the theory involves besides gluons also quarks with
finite
mass, the static interquark potential $V_{Q\bar{Q}}(r)$ does
not
grow at large distances anymore even at $T=0$. Dynamic
quarks
screen the potential of static sources. One can visualize
this
screening thinking of the color gluon tube stretched between
two
static fundamental sources being torn apart in the middle
with
the formation of an extra quark-antiquark pair. Thus in QCD
with quarks, the Wilson loop average has the perimeter
rather
than the area law
\footnote{That does not mean that there is no confinement --
-
as earlier, only the colorless states are present in the
physical spectrum. But the behavior of the Wilson loop is
not a
good signature of confinement anymore.}.  The correlator of
two
Polyakov loops (\ref{PP}) tends to a constant at large
distances
universally at low and at high temperature, and this
correlator
cannot play the role of the order parameter of phase
transition.

Still, the phase transition can occur and does occur in some
versions of the theory. It is associated, however, not with
change in behavior of the correlator (\ref{PP}), but with
{\it
restoration of chiral symmetry} which is spontaneously
broken at
zero temperature. We shall discuss the dynamics of this
phase
transition in details in the next chapter. Here we will
concentrate
on the properties of the low temperature phase.

Let us first remind the well known facts on the dynamics of
$QCD$ at zero
temperature. Consider YM theory with $SU(3)$ color group and
involving $N_f$
massless Dirac fermions in the fundamental representation of
the
group. The fermion part of the lagrangian is
\be \label{Lf}
L_f
= i \sum_f \bar q_f \gamma_\mu {\cal D}_\mu q_f
\ee
where ${\cal
D}_\mu = \partial_\mu - igA_\mu^a t^a$ is the covariant
derivative. The lagrangian (\ref{Lf}) is invariant under
chiral
transformations of fermion fields:
\be \label{chi}
q_{{\small
L,R}} \rightarrow A_{{\small L,R}}\ q_{{\small L,R}}
\ee
where
$q_{{\small L,R}} = \frac 12 (1 \pm \gamma^5)q$ are  flavor
vectors with $N_f$ components and $A_{{\small L,R}}$ are two
different $U(N_f)$ matrices. Thus the symmetry of the
classical
lagrangian is $U_L(N_f) \otimes U_R(N_f)$. Not all N\"other
currents corresponding to this symmetry are conserved in the
full quantum theory. It is well known that the divergence of
the
singlet axial current $ j^5_\mu = \sum_f \bar q_f
\gamma_\mu \gamma^5 q_f$ is nonzero due to anomaly:
  \be \label{anom}
\partial_\mu j_\mu^5  \ \sim \ g^2
\epsilon^{\mu\nu\alpha\beta} G^a_{\mu\nu} G^a_{\alpha\beta}
 \ee
Thus the symmetry of quantum theory is $SU_L(N_f) \otimes
SU_R(N_f) \otimes U_V(1)$. It is the experimental fact that
(for
$N_f = 2,3$, at least) this symmetry is broken spontaneously
down to $U_V(N_f)$. The order parameter of this breaking is
the
chiral quark condensate $SU(N_f)$ matrix
  \be
\label{Sigff1}
\Sigma_{ff'}\ =\ < \bar q_{Lf} q_{Rf'}>_0
  \ee
By a proper chiral transformation (\ref{chi}) it can be
brought
into diagonal form
\footnote{Not {\it any} complex matrix $N_f \times N_f$ can
be brought
into the form (\ref{diag}) by a unitary transformation. But
the condensate
matrix is subject to a constraint that the vector $SU(N_f)$
flavor symmetry
is not spontaneously broken as it follows from the Vafa--
Witten theorem
\cite{Vafa}. All eigenvalues of  an admissible condensate
matrix (\ref{Sigff1})
are equal
and, after diagonalization, it is proportional to a unit
matrix, indeed.}
 \be \label{diag}
\Sigma_{ff'} = - \frac \Sigma  2 \ \delta_{ff'}
  \ee
In the following, the term ``quark condensate'' will be
applied to
the scalar positive quantity $\Sigma$.

Note that the phenomenon of spontaneous chiral symmetry
breaking
is specific for theories with {\it several} light quark
flavors.
In the theory with $N_f = 1$ , the non-anomalous part of the
symmetry of the lagrangian is just $U_V(1)$. It stays intact
after adding the mass term and after taking into account the
formation of the condensate $<\bar q q>$. The condensate is
still formed, but it does not correspond to spontaneous
breaking
of any symmetry.

For $N_f \ =\ 2,3$, spontaneous breaking occurs and this
{\it is}
a non-trivial feature of $QCD$. There is no way to derive
rigorously
from general premises that the chiral symmetry {\it should}
be
spontaneously broken. Indeed, recent lattice measurements
\cite{latN4,ShurRHIC}
indicate that the symmetry is probably not broken at all in
the theory
with {\it four} massless quarks. (Actually, such a theory is
easier
to analyze on the lattice than the theories with 2 or 3
light
flavors: four flavors arise quite naturally in Kogut--
Susskind
approach due to well--known doubling of massless fermion
lattice
species.) Quark condensate was measured to be very small and
compatible with zero. The calculations in instanton model
(see a detailed discussion of this model in the next
chapter) also
indicate that at $N_f = 4$ (may be, at $N_f = 5$) the quarks
condensate
disappears \cite{Schaf}.

As was just mentioned, by now we are not able to derive
theoretically
that the symmetry should be broken at $N_f = 3$ and should
not be broken
at $N_f = 4,5$, but the trend --- the larger is the number
of light
flavors, the more difficult it is to break the symmetry ---
is easy
to understand.

\vspace{.3cm}

 \centerline{\it Condensate and spectral density}

\vspace{.3cm}

The argument is based on
 the famous Banks and Casher relation
\cite{Banks} connecting quark condensate to the mean
spectral
density of Euclidean Dirac operator $\rho(\lambda)$ at
$\lambda
\sim 0$. Let us explain how it is derived. Consider the
Euclidean fermion Green's function $<q(x) \bar q(y)>$ in a
particular gauge field background. Introduce a finite
Euclidean
volume $V$ to regularize theory in  the infrared.  Then the
spectrum of massless Dirac operator is discrete and enjoys
the
chiral symmetry: for any eigenfunction $\psi_n(x)$
satisfying
the equation $\not\!\!{\cal D} \psi_n = \lambda_n \psi_n$ ,
the
function $\tilde \psi_n = \gamma^5 \psi_n$ is also an
eigenfunction with the eigenvalue $\tilde \lambda_n = -
\lambda_n$.

The idea is to use the spectral decomposition of the fermion
Green's function with a small but nonzero quark mass
\be
\label{Green} <q(x) \bar q(y)> \ = \ \sum_n \frac {\psi_n(x)
\psi_n^\dagger (y)}{-i\lambda_n + m}
  \ee
Set $x=y$ and integrate over $d^4x$. We have
\be \label{sum}
V<\bar q q> = - 2 m \sum_{\lambda_n > 0} \frac 1{\lambda_n^2
+
m^2}
\ee
 where the chiral symmetry of the spectrum has been used
and the contribution of the zero modes $\lambda_n = 0$ has
been
neglected (it is justified when the volume $V$ is large
enough
\cite{LS}).  Perform the averaging over gauge fields and
take
{\it first} the limit $V \to \infty$ and {\it then} the
limit $m
\to 0$. The sum can be traded for the integral:
\be
\label{Banks} \Sigma \ =\ - <\bar q q> = \ 2m \int_0^\infty
\frac
{\rho(\lambda)}{\lambda^2 + m^2} d\lambda = \ \pi \rho(0)
\ee
 The rightmost-hand-side of Eq.(\ref{Banks}) is only the
non-perturbative $m$-independent part of the condensate  .
There is also a perturbative ultraviolet-divergent piece
$\propto m\Lambda_{ultr}^2$ which is proportional to the
quark
mass, is related to large eigenvalues $\lambda$ and is of no
concern for us here.

Thus the non-perturbative part of the quark condensate which
is
the order parameter of the symmetry breaking is related to
small
eigenvalues of Euclidean Dirac operator.  There should be a
lot
of them --- a characteristic spacing between levels is
$\delta
\lambda \sim 1/(\Sigma V)$ which is much less than the
characteristic spacing $\delta \lambda \sim 1/L$ for free
fermions.

The average spectral density $\rho(0)$ appears after
averaging of
the microscopic spectral density
  \be
  \label{micro}
\rho_A(\lambda)\ =\ \frac 1V \sum_n \delta(\lambda -
\lambda_n[A])
  \ee
over all gauge field configurations. The weight in the
averaging
involves a fermion determinant factor
  \be
 \label{fermdet}
\left[ {\rm Det} (-i \not\!\!{\cal D} + m) \right]^{N_f} \
=\
\left[m^\nu \prod_{\lambda_n > 0} (\lambda_n^2 + m^2)
\right]^{N_f}
  \ee
where $\nu$ is the number of the zero modes of the massless
Dirac operator
(by Atiyah--Singer theorem, it coincides with the
topological charge of
the gluon field).
For small $m$ and small $\lambda$, this factor is small.
Thus the
 configurations with small eigenvalues are effectively
suppressed,
and the larger $N_f$ is, the  more prominent is suppression.
It is
 quite conceivable that at some critical $N_f$ the
suppression of small
eigenvalues becomes so strong that the averaged spectral
density
at $\lambda = 0$ disappears.

The suppression of small eigenvalues displays itself in the
dip
of spectral density at small eigenvalues which sets in
starting
from $N_f = 3$. An exact result for the spectral density in
 $QCD$ with $N_f \geq 2$ massless quarks
in the region $\lambda \ll \mu_{\rm hadr}$ ($\mu_{\rm hadr}$
is the
characteristic
hadron scale $\sim \ 0.5 \ {\rm GeV}$) reads \cite{SS}
\be
  \label{Stern}
\rho(\lambda) = \frac \Sigma \pi + \frac {\Sigma^2 (N_f^2 -
4)}{32\pi^2 N_f
F_\pi^4} |\lambda| + o(\lambda)
  \ee
The result (\ref{Stern}) was derived {\it assuming} chiral
symmetry breaking
occurred. We see, however, that the spectral density
decreases as $\lambda$
 approaches zero, and the larger is $N_f$ --- the larger is
the effect (for
$N_f = 2$
the slope is zero, and the value $N_f = 1$ is beyond the
region of applicability
of this formula).
The scenario when $\rho(0)$ hits zero at $N_f \geq 5$ is
quite probable.

\vspace{.3cm}

\centerline{\it Quenched QCD}

\vspace{.3cm}

It is instructive to make a digression  and to
discuss here the quenched case $N_f = 0$. In that
case we have a pure glue theory without dynamical quarks.
Thus
there is no chiral symmetry in the lagrangian and nothing
can be
broken, but one {\it is} allowed  to couple the quark field
as
an external probe and consider  the ``quark condensate''
(\ref{sum}),
(\ref{Banks}). Again, spectral density $\rho(0)$ is given by
averaging the microscopic spectral density (\ref{micro})
over gluon
fields where the measure {\it does}  not involve now the
determinant
factor (\ref{fermdet}) and small eigenvalues are not
suppressed whatsoever.
We think that in the quenched case the fermion condensate is
actually
infinite and that the spectral density  behaves as
\be
\label{1lam}
 \rho(\lambda) \ \sim 1/\lambda
  \ee
at small $\lambda$ \cite{SMvac}.  Three arguments in favor
of
this conjecture
can be suggested:
  \begin{itemize}
\item This happens in the Schwinger model. \footnote{The
latter is similar to $QCD$ in many respects. It involves
confinement, chiral symmetry, the quantum $U_A(1)$ anomaly,
and presents an {\it excellent} playground.}
 For standard Schwinger model
with dynamical fermions, the condensate and $\rho(0)$  have
nonzero finite
value. They
are proportional to the dimensional coupling constant $g$
with a known
coefficient. But for the {\it quenched} Schwinger model
(pure 2--dimensional
photodynamics) the condensate of massless external probes
diverges, and the
 spectral density behaves as $1/\lambda$. This
exact result can be understood
looking at characteristic field configurations in the path
integral for the
quenched vs. unquenched case   \cite{SMvac}.
  \item
{\it If} the chiral condensate would be finite in the
quenched $QCD$, the Gell-Mann -- Oakes --
Renner relation (\ref{pimass}) would imply the presence of
massless pions, i.e. a power falloff of the correlator
of quark pseudoscalar currents at large distances. In our
opinion, such a power falloff is unnatural in the quenched
case: one could expect the appearance of the Goldstone
states when a real physical symmetry of the theory acting
on the dynamic fields in the lagrangian is broken, but not
when the symmetry acts on the fields presenting external
probes. Indeed, in the Schwinger model, the correlator
$<\bar \psi \psi(x) \ \bar \psi \psi(0)>_{\rm quenched}$
grows exponentially at large distances and no trace of would
be ``pions'' is seen.
  \item An additional observation is that  path integral has
also contributions from topologically
non-trivial instanton--like
fields. The latter involve fermion zero modes $\lambda_n =
0$ which, as is seen
from (\ref{sum}), provide the
 contribution $\sim 1/m$ in the quark Green's function and
the quark condensate.
For $N_f \neq 0$, this large factor was multiplied by the
fermion determinant
factor in the measure (\ref{fermdet}) which involved the
small factor
$m^{\nu N_f}$
 and there was
no divergence. In the quenched case, the divergence $\sim
1/m$ in the condensate which suggests the divergence
(\ref{1lam}) in the spectral density survives.

For sure, this argument alone is not sufficient. It proves
that the condensate is infinite in the chiral limit $m \to
0$ when the Euclidean volume $V$ is kept fixed but cannot
formally exclude the scenario when the condensate tends to a
finite value in the thermodynamic limit $V \to \infty$ with
fixed $m$ (as was mentioned above, the contribution
of zero modes becomes irrelevant  in the thermodynamic
limit). Such a behavior would still be very strange. A
normal physical situation (which is realized in $QCD$ with
$N_f \geq 2$) is when the condensate, the order parameter of
a spontaneously broken symmetry, disappears rather than
grows to infinity in the chiral limit at finite volume.( In
QCD with $N_f = 1$ when the condensate does not play the
role of order parameter, its value is the same in the
thermodynamic and in the chiral limit).
    \end{itemize}
All that together makes us to believe that the conjecture
(\ref{1lam}) is true
\footnote{Still, it does not have a rank of {\it theorem}.}.
That means in particular that the ``quenched chiral theory''
\cite{Golt},
which was based on the
assumption inferred from old lattice measurements that the
quenched condensate
is finite and which implies that the quenched pion is
massless, may be is
relevant for finite lattices but  {\it is} not realized
in the continuum limit.

Recently, lattice people started to see that the condensate
diverges, indeed, in
the chiral limit. The lattice data suggest the behavior
$\rho(\lambda) \ \sim
\ 1/\sqrt\lambda$ for the spectral density \cite{Teper}, but
more study is
 definitely needed. Also I have to mention that the
instanton model calculations
give probably even divergent, but much more mild behavior of
$\rho^{\rm quenched}(\lambda)$ near zero \cite{Verquen} [the
prediction
(\ref{Stern}) is qualitatively confirmed]. May be, it is one
of few places
where the
instanton model gives a qualitatively wrong behavior ?
Probably, to clarify
the situation, it would make
sense to perform instanton--like simulations in the quenched
Schwinger model
where the exact result is known.

\vspace{.3cm}

\centerline{\it Chirally symmetric phase.}

\vspace{.3cm}

If the chiral symmetry is not broken spontaneously as is
suggested  for $N_f \geq 5$ by  numerical calculations, the
dynamics of the theory
has nothing in similar with $QCD$. First of all, the absence
of the condensate
implies the absence of Goldstone states. Pseudoscalars are
massive.
A $QCD$ inequality was derived which tells that the baryon
mass is always
larger that the pseudoscalar mass.
\footnote{The inequality $m_B \geq 3/4\ m_\pi$ for $N_f \geq
6$
has the rank of exact
theorem \cite{Weing}. The result $m_B \geq m_\pi$ for any
$N_f$
follows if invoking a very natural
additional assumption. A more stringent constraint $m_B \geq
N_c/2\ m_\pi$
follows
from the analysis in the framework of the constituent quark
model and/or
 string model \cite{Nuss}.}
Thus the baryons are also massive. Then
the only massless particles which {\it are} required in the
physical spectrum
to saturate
the global axial anomaly are massless colored quarks. The
theory is not
confining.
Actually, the same conclusion follows also without invoking
$QCD$
 inequalities. One cannot saturate the anomaly with massless
baryons
because the latter fail to fulfill the 't Hooft matching
conditions for $N_f \geq 3$ \cite{self} and one needs
massless
quarks anyway.

The absence of confinement in a nonabelian theory with
asymptotic freedom is
somewhat unusual, but there is  neither paradox nor
catastrophe here. Actually,
this phenomenon has been known for a long time \cite{Zaks}
(see a recent
detailed discussion of this question in \cite{ShurRHIC}).
The point is
that asymptotic freedom does not necessarily implies
infrared slavery.
Consider the 2--loop beta function of $QCD$ with $N_f$
massless quarks. It has
the form \cite{beta2}
  \be
  \label{beta2}
\beta(\alpha_s)\ = - \frac 1{2\pi} \left(11 - \frac 23
N_f\right) \alpha_s^2 \ -
\ \frac 1{4\pi^2} \left(51 - \frac {19}3 N_f\right)
\alpha_s^3 \ -\ \ldots
 \ee
Suppose $N_f = 16$. Then the first coefficient is negative
(and that corresponds
to asymptotic freedom) and very small. The second
coefficient is positive and
is not particularly small. When we evolve the coupling in
the infrared,
the evolution would be very soon affected by the second term
in (\ref{beta2}).
 We see that the equation $\beta(\alpha_s) = 0$ has the
solution
\be
\label{fpoint}
\alpha_s\ = \ \frac{2\pi}{151}
  \ee
which is rather small. At this value, the third term in the
beta function
is still much smaller than the second one [the second and
the first are
of the same
order due to accidental smallness of the first coefficient
in (\ref{beta2})].
This is the infrared fixed point. The coupling is frozen at
the small value
(\ref{fpoint}), perturbation theory works and the spectrum
involves colored
massless quark and gluon asymptotic states. No trace of
confinement.
Recently, a similar phenomenon (the so called ``conformal
window'')
was discussed in the context of supersymmetric $QCD$ with
sufficiently
large number of light flavors \cite{confwin}.

Unfortunately, no fixed point is seen in the perturbative
beta function
(\ref{beta2}) for $N_f = 4,5$. The second coefficient in the
beta function is
still negative here and the equation  $\beta(\alpha_s) = 0$
has no solution.
(One can get a solution including higher orders in the
perturbative expansion
of $\beta(\alpha_s)$ in a particular renormalization scheme,
but, besides it
is scheme--dependent,
 the stable point would correspond to large values of
$\alpha_s$
where perturbative
expansion makes no sense)  Thus the absence of confinement
at $N_f = 5$
is still
surprising. More studies (lattice, instanton, etc.) are
required.

 \vspace{.3cm}

\centerline{\it Effective chiral theory.}

\vspace{.3cm}

For $N_f = 3$, the spontaneous
breaking occurs and leads to appearance of the octet of
pseudoscalar
Goldstone   states in the spectrum. Of course,
the quarks are not exactly massless in  real $QCD$, and the
mass term is not
invariant with respect to the symmetry (\ref{chi}).
 As a result, in real world we have the octet
of light (but not massless) pseudo-goldstone pseudoscalar
states ($\pi, K, \eta$). But the small mass of
pseudogoldstones
and the large splitting between the massive states of
opposite
parity ($\rho/A_1$, etc.)  indicate beyond reasonable doubts
that the exact chiral symmetry (\ref{chi}) would be broken
spontaneously in the massless case. As the masses of the
strange
and, especially, of $u$- and $d$- quarks are small
\cite{GLmass}
, the mass term in the lagrangian can be treated as
perturbation.
 E.g. the pion mass satisfies the relation
\footnote{ We assumed here that the quark mass matrix is
diagonal. Otherwise, the ground state of the hamiltonian
would correspond to a nondiagonal quark condensate matrix
(\ref{Sigff1}).
See the detailed discussion of the ``disoriented quark
condensate''
in the next chapter.}
\be
\label{pimass}
F_\pi^2 M_\pi^2 = (m_u + m_d) \Sigma
\ee
($F_\pi = 93$  Mev is the pion decay constant) and turns to
zero
in the chiral limit $m_{u,d} \rightarrow 0$.

It is noteworthy that the symmetry breaking pattern
\be
\label{br3}
SU_L(N_f) \otimes SU_R(N_f) \rightarrow SU_V(N_f)
\ee
depends crucially on the assumption that the gauge group
involves at least 3 colors. For $SU(2)$ color group where
quarks
and antiquarks belong to the same representation (the
fundamental representation of the $SU(2)$ group is
pseudoreal:
${\bf 2} \equiv {\bf \bar 2}$), the symmetry group of the
lagrangian (\ref{Lf}) is much higher. It is $U(2N_f)$ and
involves also mixing between quarks and antiquarks. $U_A(1)$
-
part of this symmetry is anomalous and the formation of
chiral
condensate breaks spontaneously the remaining $SU(2N_f)$
down
to a simplectic group
\cite{simpl}:
  \be
  \label{br2}
SU(2N_f) \rightarrow Sp(2N_f)
  \ee
As a result, $2N_f^2 - N_f -1$ Goldstone bosons living on
the
coset space appear. For $N_f =2$, we have not 3 as usual,
but 5
``pions''. Two extra ``pions'' are diquarks.
This fact is important to understand for people who
would wish to study numerically on lattices  spontaneous
chiral symmetry breaking  with $SU(2)$ gauge group.

The presence of light pseudogoldstones in the spectrum is of
paramount importance for the physics of low temperature
phase. When
we heat the system a little bit, light pseudoscalar states (
pions
in the first
place) are excited, while $\rho$--meson, nucleon and other
massive degrees
of freedom are still frozen.
We can study {\it analytically} the properties of the system
at low temperatures when the medium presents a rarefied
weakly
interacting gas of pions with low energies.  Their
properties
are described by the effective nonlinear chiral lagrangian
\be
\label{Lchi}
{\cal L} = \frac 14 F_\pi^2 {\rm Tr} \{\partial_\mu
U \partial_\mu U^\dagger\}\ + \ \Sigma {\rm Re\ Tr}\{{\cal
M} U^\dagger\}
\ +\ \ldots
\ee
 where $U\ =\ \exp\{2it^a \phi^a/F_\pi\}$ is the
$SU(N_f)$ matrix ($\phi^a$ are the pseudogoldsone fields),
${\cal M}$ is
the quark mass matrix
 and the dots stand for higher derivative terms
and the terms of higher order in quark masses.
 When the characteristic
energy and the quark masses are small, the effects due to
these
terms are suppressed and a perturbation theory (the {\it
chiral
perturbation theory}
\cite {GLold,CPT}) can be developed.

The particular nonlinear form of the lagrangian (\ref{Lchi})
is dictated by its symmetry properties:
it is invariant under
multiplication of $U$ by an arbitrary unitary matrix on the
right or on the left  in the massless case . This symmetry
exactly
corresponds to
the symmetry (\ref{chi}) of the original lagrangian. One can
also
realize the symmetry (\ref{chi}) linearly including an
additional
$\sigma$ field. Actually, it is equivalent for many purposes
to the nonlinear
$\sigma$ -- model approach, but with some  particular
assumptions
on the coefficients of higher--derivative terms in
(\ref{Lchi})
 (they appear if integrating $\sigma$--field out).
Expanding (\ref{Lchi}) up to the quartic terms in $\phi^a$
and using
 (\ref{pimass}), we obtain for $N_f = 2$
  \be
  \label{chiexpan}
{\cal L} \ =\ \frac 12 (\partial_\mu \phi^a)^2 \
- \ \frac{M_\pi^2}{2}(\phi^a
\phi^a)\  + \ \frac 1{6F_\pi^2}
\left[ (\phi^a \partial_\mu \phi^a)^2 \ -\ (\phi^a \phi^a)
(\partial_\mu \phi^b)^2 \right] \nonumber \\
 + \ \frac{M_\pi^2}{24 F_\pi^2}(\phi^a
\phi^a)^2\ +\ \ldots
   \ee
When $M_\pi$ is zero, the vertex of the four--pion
interaction
 involves the second power of momentum.
That means, in particular, that the cross section of
scattering
of {\it massless} pions occurs at $P$--wave and behaves as
 $\sim E^2/F_\pi^4$ at small energies.

\subsection{Thermodynamics}
After these preliminary remarks about the physics of zero--
temperature
$QCD$, we are ready to discuss what happens if we switch on
the temperature.
In this chapter, we assume the temperature to be small so
that only the lightest
degrees of freedom, the pions, are excited. Their density is
small and their
interaction is weak.
The basic thermodynamic characteristic of  a finite $T$
system is its free energy. We have in the lowest order

\be
\label{Frpi}
F_0 = -T \ln Z = -T \ln \left[ \prod_\vec{p} \left(
\sum_{n=0}^\infty e^{-\beta n E_{\vec{p}}} \right)^{(N_f^2-
1)} \right]
\ee
where $E_{\vec{p}} = \sqrt{\vec{p}^2 + M_\pi^2}$ and
 $(N_f^2-1)$ is the number of degrees of freedom of
the pseudogoldstone field with the mass $M_\pi$
(we derive free of charge the result for
any $N_f$, but, physically, one should put $N_f = 2$ or, if
$N_f = 3$,
take into account the different masses of $\pi$, $K$ and
$\eta$ mesons).
The sum $\sum_{n} e^{-\beta n E_{\vec{p}}}$
is the free energy of a single boson field oscillator with
a given momentum. Trading the sum for the
integral, we obtain for the volume density of the free
energy
 \be
\label{Fpi0}
 \frac{F_0}{V} = (N_f^2 -1) T \int \frac {d^3p}{(2\pi)^3}
\ln \left[ 1 - e^{-\beta \sqrt{\vec{p}^2 + M_\pi^2}} \right]
\ee
A very important corollary of this simple formula is that
the order
parameter of the spontaneously broken chiral symmetry, the
quark
condensate, decreases with temperature. The temperature--
dependent
condensate $\Sigma(T)$ is defined as a logarithmic
derivative of
free energy with respect to quark mass:
  \be
\label{SigTdef}
\Sigma(T)\ = \ - \frac 1{N_f} \frac 1V \frac {\partial F(T)}{\partial m}
  \ee
(we endowed all quark flavors with a common small mass $m$ to be set to zero
after differentiation).
The expansion
of (\ref{Fpi0}) in pion mass in the region $M_\pi \ll T$ reads
  \be
\label{Fexpan}
\frac {F_0}V \ =\ (N_f^2 - 1) \left[ -\frac {\pi^2 T^4}{90}
+ \frac {T^2 M_\pi^2}{24} \ + \ O(M_\pi^3 T) \right]
  \ee
Substituting it in (\ref{SigTdef}) and taking into account
the Gell-Mann
-- Oakes -- Renner relation (\ref{pimass}), we obtain
\cite{Gaillard,GLT}
  \be
\label{SigT1}
\Sigma(T)\ =\ \Sigma(0)\left[ 1 - \frac{N_f^2 - 1}{12 N_f}
\frac {T^2}{F_\pi^2} - O(T^4/F_\pi^4) \right]
\ee
Actually, a similar phenomenon has been known for a long
time in condensed
matter physics. At finite temperature, spontaneous
magnetization of ferromagnet
starts to fluctuate and its average value decreases with
temperature. Technically, these fluctuations are described
by  massless magnon degrees of freedom which are excited in
the heat
bath and contribute in the free energy (magnons are the
Goldstone
particles which appear due to spontaneous breaking of
rotational symmetry
in ferromagnet by the same token as pions appear due to
spontaneous
breaking of chiral symmetry in $QCD$). Only the shift in
spontaneous
magnetization
is proportional to $T^{3/2}$ ( see e.g. \cite{staty2},
p.299)
 rather than to $T^2$ as in our case.

Also higher terms of the expansion of $\Sigma(T)$ in
$T^2/F_\pi^2$ were
 found. To this end, one has to calculate the two loop and
three
loop corrections to the free energy density which take into
account
pion interactions. This was done in \cite{GLT,Ger}, but we
choose to
postpone the presentation and the discussion of these
results to the
next chapter.

\subsection{Pion collective excitations.}
At $T = 0$ the particles display themselves as asymptotic
states. As was
mentioned
before, asymptotic states do not exist in heat bath. Their
role is taken over
by {\it collective excitations}. This notion can be easily
visualized when
recalling the familiar college physics problem of
propagation of electromagnetic
waves in the medium. The frequency and the wave vector of
such a wave do not
satisfy the vacuum relation $\omega = c|\vec{k}|$ anymore. A
refraction index may
appear.
\footnote{In the system involving free electrons (plasmas) a
mass gap can
develop.
That means that plasma supports a standing electromagnetic
wave with
$\vec{k} = 0$ and
nonzero frequency (it is called the {\it plasma frequency}).
Quark--gluon
plasma has a similar property, and we are going to discuss
it at length
in Chapter 6.}
The amplitude of the classical wave decreases with time due
to dissipative
effects. That means that the wave frequency acquires a
negative imaginary
part (a positive imaginary part would correspond to
instability).

Technically, collective excitations display themselves as
poles in the retarded
thermal Green's functions. The position of the pole
$\omega_{\rm pole}(\vec{k})$
is called the {\it dispersive law} of the corresponding
collective
excitation.

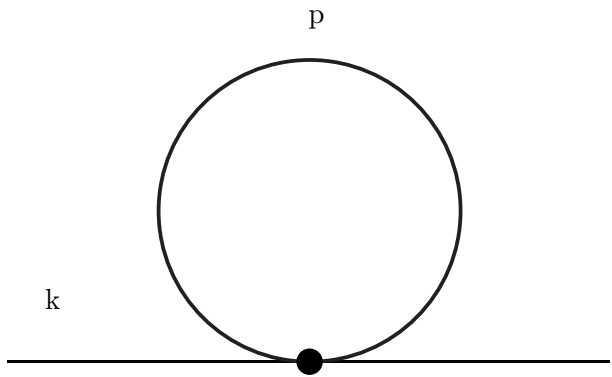
\begin{figure}

\begin{center}
\begin{picture}(80,50)
\SetScale{2.845}
\thicklines

\put(0,0){\line(1,0){80}}

\CArc(40,20)(20,0,360)
\GCirc(40,0){1.5}{0}

\thinlines
\put(40,45){p}
\put(5,7){k}
\end{picture}
\end{center}
\caption{Pion polarization operator in one loop.}
\label{pi1loop}
\end{figure}

Let us find the dispersive law for pion collective
excitations in the heat
bath. To this end, we have to calculate the retarded
polarization operator
including thermal effects and solve the dispersive equation
  \be
 \label{de}
D_R^{-1}(\omega, \vec{k}) \ =\ \omega^2 \ -\ \vec{k}^2 \ - \
\Pi_R(\omega, \vec{k}) \ =\ 0
   \ee
The simplest graph is shown in Fig. \ref{pi1loop} where the
4-pion interaction
vertex
can be inferred from (\ref{chiexpan}).
The corresponding calculation was first done in \cite{GLT}.
As a result, it was
found that the pion mass acquires a temperature  correction
  \be
 \label{MpiT}
M_\pi^2(T) \ =\ M_\pi^2(0) \left[1 + \frac {T^2}{24 F_\pi^2}
\ + \ldots \right]
  \ee
It is not difficult to generalize this formula on the case
when we have
$N_f$ light quarks with the same mass (it is not so, of
course, in the
real world). The dependence is qualitatively the same, only
the coefficient of the $T^2$ term is changed from $1/24$ to
$1/(12 N_f)$.

We see that the pion mass slightly increases with
temperature, but, if the
quarks
masses and $M_\pi(0)$ would be zero, pion mass would remain
zero also at finite
temperature. This is easy to understand: massless pions are
Goldstone particles
appearing due to spontaneous breaking of chiral symmetry.
But if the temperature
is small, chiral symmetry is still spontaneously broken and
the massless
Goldstone
particle should still be there as it is.

\newpage

\newpage

It is instructive to see how this result appears in the
direct calculation of
the
1--loop graph in Fig. \ref{pi1loop}. The quickest way is to
use the
 Keldysh technique. Actually,  we do not need here the full-
-scale Keldysh
technique.
The situation is the same as in the model example discussed
at the end of
Chapter 2. $\Pi_R = \Pi_{11} + \Pi_{12}$ , but 12 --
component is zero here
and it
 suffices to consider only the temperature--dependent part
in the $T$--ordered
pion propagator in the loop. For massless pions, the latter
is $-2\pi i
n_B(|\vec{p}|) \delta(p^2)$. The 4-point vertex involves two
momenta. If one
or two
 of these momenta is the loop momentum, the corresponding
integrals
  $$I_{1\mu} \ \sim \ \int \frac {d^4p}{(2\pi)^3} \ \frac
{p_\mu \delta(p^2)}
{e^{\beta |\vec{p}|} - 1},\ \ \ \ \ \
I_2 \ \sim \ \int \frac {d^4p}{(2\pi)^3} \ \frac {p^2
\delta(p^2)}
 {e^{\beta |\vec{p}|} - 1}
  $$
are just zero. The only contribution to the polarization
operator comes from the
term where both momenta sit on external lines. This
contribution has the form
\be
\label{intPol}
  \Pi_R(\omega, |\vec{k}|) \ =\  \frac {N_fk^2}{3F_\pi^2}
\int \frac {d^4p}{(2\pi)^3}
\ \frac {\delta(p^2)}
{e^{\beta |\vec{p}|} - 1}  \ = \ k^2 \frac
{N_fT^2}{36F_\pi^2}
  \ee
Substituting it in the dispersive equation (\ref{de}), we
find that
$\omega = |\vec{k}|$
is still a solution. Only the residue of the pole is
renormalized.
(Incidentally, this
residue, in contrast to the pole position, does not have a
universal meaning.
Its
value depends on the particular parameterization
$U = \exp\{2\pi i \phi^a t^a/F_\pi \}$
used. In Weinberg parameterization \cite{Wein} it would be
different.)

Another temperature effect is the renormalization of
$F_\pi$. The invariant
way to define $F_\pi$ at finite temperature is via the
residue of the
polarization operator of axial currents. At large distances,
the latter
behaves as
 \be
\label{axcor}
<A_\mu(\vec{x}) A_\nu(0)>_T \ = \ \propto F_\pi^2(T) \exp\{-
M_\pi(T)
 |\vec{x}|\}
 \ee

\newpage

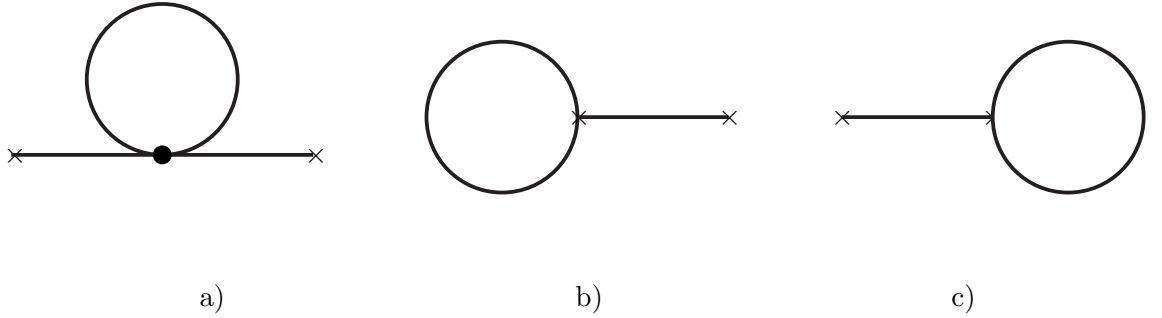
\begin{figure}

\begin{center}
\begin{picture}(150,50)
\SetScale{2.845}

\put(25,0){a)}
\put(75,0){b)}
\put(125,0){c)}

\put(-1,19){$\times$}
\put(39,19){$\times$}

\BCirc(20,30){10}
\Line(0,20)(40,20)
\GCirc(20,20){1}{0}

\BCirc(65,25){10}
\put(74,24){$\times$}
\Line(75,25)(95,25)
\put(94,24){$\times$}

\put(109,24){$\times$}
\Line(110,25)(130,25)
\put(129,24){$\times$}
\BCirc(140,25){10}

\end{picture}
\end{center}
\caption{Axial correlator in one loop.}
\label{axcorgr}
\end{figure}

\newpage

 The graphs which contribute to the axial correlator on the
one--loop level
are depicted in Fig. \ref{axcorgr}. The first one was
already discussed
for the problem of
pion mass renormalization. Here we need to know, however,
also the renormalization
of the residue of the pion propagator which contributes in
the
renormalization of the residue of the axial correlator
$F^2_\pi(T)$.

The latter depends also on two other graphs in Fig.\ref{axcorgr}
 [which do not affect the pion propagator]. The corresponding vertex
$<0|A_\mu|3\pi>$ can be found by ``covariantizing'' the
derivatives
 in (\ref{Lchi}) in a proper way \cite{GLold}:
$\partial_\mu U \to \partial_\mu U - i (A_\mu + V_\mu)U + iU
(V_\mu -A_\mu)$
where
$A_\mu \equiv A_\mu^a t^a$, $V_\mu \equiv V_\mu^a t^a$, and
$A_\mu^a$,
$V_\mu^a$
are external sources coupled to axial and vector currents.
The result
coming from
the sum of all three graphs is parameterization--independent
and has
the form \cite{GLT,Hansen}
  \be
  \label{FpiT}
F_\pi(T) \ = \ F_\pi(0)\left[ 1 - \frac {N_f}{24} \frac
{T^2}{F_\pi^2}
\ + \ \ldots \right]
  \ee
Note that the temperature--dependent condensate given by
Eq.(\ref{SigT1}),
the pion mass given by Eq.(\ref{MpiT}), and $F_\pi(T)$ given
by Eq.(\ref{FpiT})
 still
satisfy the Gell--Mann -- Oakes -- Renner relation
(\ref{pimass}).

\newpage

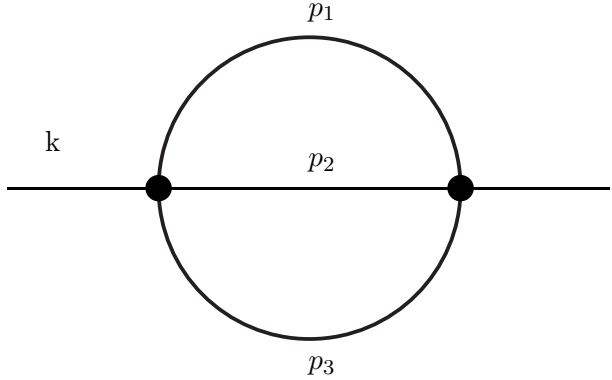
\begin{figure}

\begin{center}
\begin{picture}(80,50)
\SetScale{2.845}
\thicklines

\put(0,25){\line(1,0){80}}

\CArc(40,25)(20,0,360)
\GCirc(20,25){1.5}{0}
\GCirc(60,25){1.5}{0}
\thinlines
\put(40,48){$p_1$}
\put(40,1){$p_3$}
\put(40,28){$p_2$}
\put(5,30){k}
\end{picture}
\end{center}
\caption{Pion polarization operator in two loops.}
\label{pi2loop}
\end{figure}

\newpage

The next in complexity is the two--loop graph depicted in
Fig. \ref{pi2loop}.
The essential novel feature is that two--loop polarization
operator has not
only real, but also an imaginary part. Correspondingly, the
solution of the
dispersive equation (\ref{de}) becomes complex, the
imaginary part
of $\omega_{\rm pole} (\vec{k})$ describing the {\it
damping}
 of pion waves.
\footnote{In this review, we consistently define damping
$\zeta(\vec{k}) =
-{\rm Im}\
\omega_{\rm pole}(\vec{k})$ as the coefficient determining
the
exponential attenuation
of the {\it amplitude} of the classical wave with time.
People often
discuss also the twice as large quantity $\gamma(\vec{k})$
which is the
coefficient in the exponential time decay of energy or
probability.}
 This damping was first found in \cite{Goity}.

Let us  derive it in a slightly different way than it was
done in
\cite{Goity},
 using Keldysh technique.
The analytic expression corresponding to the graph in
Fig.\ref{pi2loop}
is
  \be
\label{Pi2loop}
\Pi_R(k)\ =\ \Pi_{11}(k) +  \Pi_{12}(k)
\ =\ -\frac 1{3!} \ \int  \frac{d^4p_1}{(2\pi)^4}\
\frac{d^4p_2}{(2\pi)^4}\
\frac{d^4p_3}{(2\pi)^4}
\ (2\pi)^4 \ \delta^{(4)}(k - p_1 - p_2 - p_3) \nonumber \\
|T(k,p_1,p_2,p_3)|^2 \
\left[D_{11}(p_1) D_{11}(p_2) D_{11}(p_3) \ -\ D_{12}(p_1)
D_{12}(p_2)
D_{12}(p_3)
\right]
  \ee
where $T(k,p_1,p_2,p_3)$ is the 4--pion vertex and the
negative sign of the
second
term is due to the fact that the corresponding vertex comes
from the
anti--T--ordered
 exponential [cf. (\ref{Gam12})]. $|T|^2$ summed over all
flavor
polarizations in the loop has the form
  \be
 \label{T2}
|T|^2 \ =\ \frac {2(s^2 + t^2 + u^2) - 9M_\pi^4}{F_\pi^4}
  \ee
where $s = (k-p_1)^2, t = (k - p_2)^2, u = (k - p_3)^2$ and
we suppressed the
trivial flavor factor
$\delta^{aa'}$. Substitute now the expansion (\ref{invers})
for the matrix
components
$D_{11}$, $D_{12}$ in Eq.(\ref{Pi2loop}).

 Note that the terms involving products
$D_R(p_1) D_A(p_2)$ etc. give zero after integration.
Really, fix $p_3$ and
consider
the integral
 \be
 \label{exRA}
\sim\ \int d^4p_1\ D_R(p_1) D_A(k-p_3 - p_1)
  \ee
Recalling that $D_A = D_R^*$ and that $D_R$ which has the
form (\ref{DR0})
is analytic in the upper $\omega$ half--plane, it is easy to
see that
the integrand in (\ref{exRA}) is analytic in the upper
$p_{10}$ half--plane.
Closing the contour there, we get zero.

Having that in mind, one obtains after some transformations
  \be
\label{Pi2RPP}
{\rm Im} \ \Pi_R(k)
\ =\ -\frac 1{24} \ \int \frac{d^4p_1}{(2\pi)^4}\
\frac{d^4p_2}{(2\pi)^4}\
 \frac{d^4p_3}{(2\pi)^4} \ (2\pi)^4 \
\delta^{(4)}(k - p_1 - p_2 - p_3)\
|T|^2 \   \nonumber \\
\left[D_R(p_1) D_R(p_2) D_R (p_3) \ +\ D_R(p_1) D_P(p_2) D_P
(p_3) \ +\
D_P(p_1) D_R(p_2) D_P (p_3)  \right.\nonumber \\
\left. +\ D_P(p_1) D_P(p_2) D_R (p_3) \right]
 \ee
 Now we have to take the imaginary part of (\ref{Pi2RPP})
and subtract its
zero--temperature value (the latter does not affect the
dispersive relation
and does not concern us here). Substituting the expression
(\ref{PbRA})
for $D_P$
and noting that \ ${\rm Im}\ D_R(p)\ =\ -\pi \delta(p^2 -
m^2) \epsilon(p_0)$
[$ \epsilon(p_0)$ is the sign factor], we obtain
  \be
\label{Piint}
{\rm Im}\ \Pi_R(k)
\ =\ -\frac{\pi^3}6 \ \int \frac{d^4p_1}{(2\pi)^4}\
\frac{d^4p_2}{(2\pi)^4}\
 \frac{d^4p_3}{(2\pi)^4}
\ (2\pi)^4 \
\delta^{(4)}(k - p_1 - p_2 - p_3)\
|T|^2 \   \nonumber \\
\delta(p_1^2 - M_\pi^2) \ \delta(p_2^2 - M_\pi^2)\
\delta(p_3^2 - M_\pi^2)
\left[ \coth \frac {\omega_1}{2T} \coth \frac
{\omega_2}{2T}\ -
\epsilon(\omega_1)
\epsilon(\omega_2) \right. \nonumber \\
\left. + \ \coth \frac {\omega_1}{2T} \coth \frac
{\omega_3}{2T}\ -
\epsilon(\omega_1)\epsilon(\omega_3) \ +\
\coth \frac {\omega_2}{2T} \coth \frac {\omega_3}{2T}\ -
\epsilon(\omega_2)
\epsilon(\omega_3)  \right]
\epsilon(\omega_1)\epsilon(\omega_2)\epsilon(
\omega_3)
 \ee
where $\omega_i \equiv p_{i0}$. The integral acquires
contributions both from
positive
 and negative $\omega_i$. Note the essential difference with
usual $T = 0$
 Cutkosky
rules where only positive energies in the direct channel are
allowed.
Physically,
negative energies correspond to the particles not in finite
but in {\it initial}
state --- there are a lot of them in heat bath. In our case,
the
contribution from the kinematic region where all
frequencies are positive is just zero. Indeed,
 we want eventually to substitute the polarization
operator (\ref{Piint}) in the dispersive equation (\ref{de})
to find a {\it
small}
imaginary pole frequency shift: $\omega_{\rm
pole}^T(\vec{k}) = \
\omega_{\rm pole}^{T=0}(\vec{k})
\ -\ i\zeta(\vec{k})$. We are allowed then to calculate the
integral assuming
$\omega\ =\ \omega_0(\vec{k})
\ = \ \sqrt{\vec{k}^2 + M_\pi^2}$. In that case kinematics
does not allow to
have all frequencies
positive:  a massive particle cannot decay in 3 particles of
the same mass.
A massless particle can, in principle, do it, but all final
particles
would go parallel to each other and the corresponding phase
space
is zero. Nonzero
contribution comes from the kinematic region when one of the
frequencies is
negative
and two others --- positive. Choose for definiteness
 $\omega_1 < 0$ and $\omega_{2,3} > 0$.
Physically, it corresponds to {\it scattering} of the
ingoing pion with the
energy $\omega$ on the pion from heat bath with the energy  $-\omega_1$
producing two other pions with energies $\omega_2$ and
$\omega_3$ in the
final state.
\footnote{You may ask: how can we talk of scattering now
while
we stated earlier that scattering matrix is an ill--defined
notion
at finite temperature. That is true, and $S$--matrix does
not exist as a
transition
amplitude between {\it asymptotic} states. But now we are
dealing with a
``scattering
on the way''. We need not follow the fate of scattered pions
up to $t = \infty$.
The wave loses energy and never reaches infinity exactly
because of the damping
effects we are currently studying.}
Taking into account three possibilities (either $\omega_1$
or $\omega_2$ or
$\omega_3$ is negative) and using the nice relation
$$
\coth \kappa_1 \coth \kappa_2\ +\ \coth \kappa_1 \coth
\kappa_3
\ -\ \coth \kappa_2 \coth \kappa_3\ -\ 1 \ =\ \frac {\sinh
\kappa}{\sinh
\kappa_1
\sinh \kappa_2 \sinh \kappa_3}, $$
where $\kappa = \kappa_2 + \kappa_3 - \kappa_1$, we
eventually
obtain
 \be
\label{pizeta}
\zeta(\vec{k}) \ =\ -\frac {{\rm Im} \Pi_R[\omega,
\vec{k}]}{2\omega} \ =
\ \frac {\sinh( \omega/{2T})}{4\omega}
\int d\nu_1 d\nu_2 d\nu_3 \ (2\pi)^4 \ \delta^{4}(k+p_1 -
p_2 - p_3) \ |T|^2,
\nonumber \\
d\nu_i \ =\ \frac 1{(2\pi)^3} \frac {d^3p}{4E_i \sinh
({E_i}/{2T}) };
\ \ \omega = \sqrt{M_\pi^2 + \vec{k}^2},\  E_i =
\sqrt{M_\pi^2 + \vec{p}_i^2}
 \ee
(all energies are positive now).

One readily checks that, if the gas is dilute which is true
in the region
$T \ll M_\pi$, this formula reduces to the relativistic
expression for the
collision rate
  \be
\label{pivir}
\zeta(\vec{k}) \ = \ \frac 12 \int \frac {d
\vec{p}_1}{(2\pi)^3} \exp(-E_1/T)
\sigma^{\rm tot} v^{\rm rel}
(1 - \vec{v} \vec{v}_1 )
\ee
where $\vec{v}$, $\vec{v}_1$ are the velocities of the two
collision
partners and where $v^{\rm rel}$ is the velocity of one of
them in the rest
frame of the other.
$\sigma^{\rm tot}$ is the total cross section summed over
the flavors of
the initial pion from the heat bath.

\newpage

\begin{figure}

\begin{center}

\begin{picture}(80,50)
\SetScale{2.845}
\thicklines

\put(40,38.5){$|$}
\put(0,0){\line(1,0){80}}

\CArc(40,20)(20,0,360)
\GCirc(40,0){3}{0.5}

\thinlines
\put(40,45){p}
\put(5,7){k}
\end{picture}
\end{center}
\caption{Pion polarization operator in the first order in
density.
Vertical dash stands for the temperature insertion.}
\label{virialpi}
\end{figure}
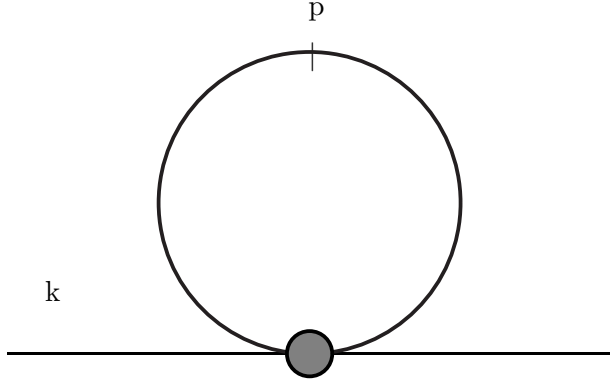

\newpage

This result may be obtained by the {\it virial
expansion} over the pion density which corresponds to taking
into account
the temperature correction $-2\pi i n_B(E_i) \delta (p_i^2 -
M_\pi^2)$
in only one of the virtual pion propagators
in the loop in Fig. \ref{pi2loop}. The sum of the graphs in
Figs.
\ref{pi1loop}, \ref{pi2loop} and many other graphs with only
one
temperature insertion in a virtual pion line can be drawn as
in Fig.
\ref{virialpi} where the grey blob stands for a {\it zero-
temperature}
amplitude of forward $\pi\pi$ scattering $T^{\pi\pi}(s)$
summed over
the isotopic species of one of the pions. We have
  \be
\label{virpi}
 \Pi_R(\omega, \vec{k}) \ =\ -\int \frac {d\vec{p}}{(2\pi)^3
2 E}
\ n_B(E)\ T^{\pi\pi}(s) \ \ + \ O(n_B^2)
  \ee
Imaginary part of $T^{\pi\pi}(s)$
gives the total cross section and we arrive at the result
(\ref{pivir}) (when the density $n_B(E)$ is small, it
coincides with the
Boltzmann distribution $\exp\{-E/T\}$).

Substituting $|T|^2$ in the form (\ref{T2}) in the general
expression
(\ref{pizeta}), an angular integral can be done and we
obtain \cite{Goity}
  \be
\label{grob}
  \zeta(k) \ =\ \frac {\sinh (\omega/2T)}{8k\omega} \frac
\pi{(4\pi F_\pi)^4}  \ \times \nonumber \\
\int_{M_\pi}^\infty dE_1 \int_{M_\pi}^{\omega + E_1
-M_\pi} dE_2 \ \frac{\theta(k_+ - k_-) f_1 + 4 \theta(q_+ -
q_-) f_2}
{\sinh (E_1/2T) \sinh (E_2/2T) \sinh [(\omega + E_1 -
E_2)/2T ]}\ ,
\nonumber \\
f_1 \ =\ (k_+ - k_-) [2(\omega + E_1)^4 - 9 M_\pi^4 ]
- \frac 43 (k_+^3 - k_-^3) (\omega + E_1)^2 + \frac 25
(k_+^5 - k_-^5)\ ,
\nonumber \\
f_2 \ =\ (q_+ - q_-) (\omega - E_2)^4 - \frac 23 (q_+^3 -
q_-^3)(\omega -
E_2)^2 + \frac 15(q_+^5 - q_-^5) \ ,
  \ee
where
 \be
\label{kqpm}
k_+ \ =\ {\rm min} (k + p_1, \ p_2 + p_3), \ \ \ k_-\ =\
{\rm max} (|k -
p_1|, \ |p_2 - p_3|),  \nonumber \\
q_+ \ =\ {\rm min} (k + p_2, \ p_1 + p_3), \ \ \ q_-\ =\
{\rm max} (|k -
p_2|, \ |p_1 - p_3|),
 \ee
$k \equiv |\vec{k}|$ and $p_i \equiv |\vec{p}_i|$.
This complicated expression can be greatly simplified in the
theoretically
interesting region $T \gg \omega \gg M_\pi$. Assuming also
$E_i \sim T \gg
\omega$ and expanding over $\omega/E_i$, $\omega/T$, we
obtain
  \be
\label{zetchir}
\zeta^{\rm chiral}(k)\  =\ \frac{k^2 }{192\pi^3 TF_\pi^4}
\int_0^\infty dE_1 \int_0^{E_1} dE_2 \frac{E_1^2 + 2 E_2^2}
{\sinh (E_1/2T) \sinh (E_2/2T) \sinh [( E_1 - E_2)/2T] }
  \ee
The integral here involves a logarithmic divergence at $E_2
\sim 0$ or at
$E_3 \sim  E_1 - E_2 \sim 0$. This divergence was absent in
the original
expression
(\ref{grob}) and is actually an artifact of our assumption
 $E_{2,3} \gg \omega$. It is effectively cut off at $E_{2,3}
\sim \omega \sim k$.
Finally, we obtain with a logarithmic accuracy
  \be
 \label{zetlog}
\zeta^{\rm chiral}(k) \ =\ \frac{k^2 T^3}{24\pi^3 F_\pi^4}
\ln \frac Tk
\int_0^\infty \frac{x^2 dx}{\sinh^2 (x/2)} \ =\ \frac {k^2
T^3}{18\pi F_\pi^4}
\ln \frac Tk
  \ee
This is not, however, the end of the story. At very large
characteristic
times (much larger than the relaxation time), the system is
in {\it
hydrodynamic regime} and the propagation of a classical pion
wave is
described by an effective wave equation with dissipative
term
 \be
\label{hydro}
\ddot{\pi} \ -\ \Delta \pi \ -\ C \Delta \dot{\pi} \ + \
{\rm higher\
 derivatives \ terms}\ =\ 0
 \ee
From this, one gets
the behavior $\zeta = Ck^2$ at small $k$ and there should be
no
logarithmic singularity. The matter is that, in the region $
k < \zeta(k)$,
one {\it is} not allowed to restrict oneself with the graph
in Fig.
\ref{pi2loop}, and higher order corrections are important.
Physically,
one cannot assume in this region that the imaginary parts of
loop
 propagators in
Fig. \ref{pi2loop} are $\delta$--functions, but should take
into account
the ``smearing'' of $\delta$--functions due to the nonzero
imaginary part
$\zeta(p_i)$ of the poles of the propagators. For very small
$k$ it is
the  damping
 itself rather than $k$ which provides an infrared
cutoff in the integral (\ref{zetchir}). The result
(\ref{zetlog}) has
a limited  range of applicability
\be
\label{ranzet}
\frac {T^5}{24 F_\pi^4} \ \ll k \ \ll \ T \ \ll \ \mu_{\rm
hadr}
 \ee
where $T^5/(24 F_\pi^2)$ is the approximate value
\cite{Goity} of the damping
 $\zeta(k)$ averaged
over the thermal distribution. At still lower $k$, the
damping is estimated
with logarithmic accuracy as
 \be
\label{zetkll}
\zeta(k) \ = \ \frac {2 k^2 T^3}{9\pi F_\pi^4}
\ln \frac {2.2F_\pi}T
  \ee
Note that a similar damping--induced
cutoff plays an important role in some kinetic problems
for quark--gluon plasma. We will discuss this issue in
Chapter 6.

\begin{figure}
  \begin{center}
        \epsfxsize=400pt
        \epsfysize=0pt
        \vspace{-5mm}
       \epsfbox[0 0 370 300]{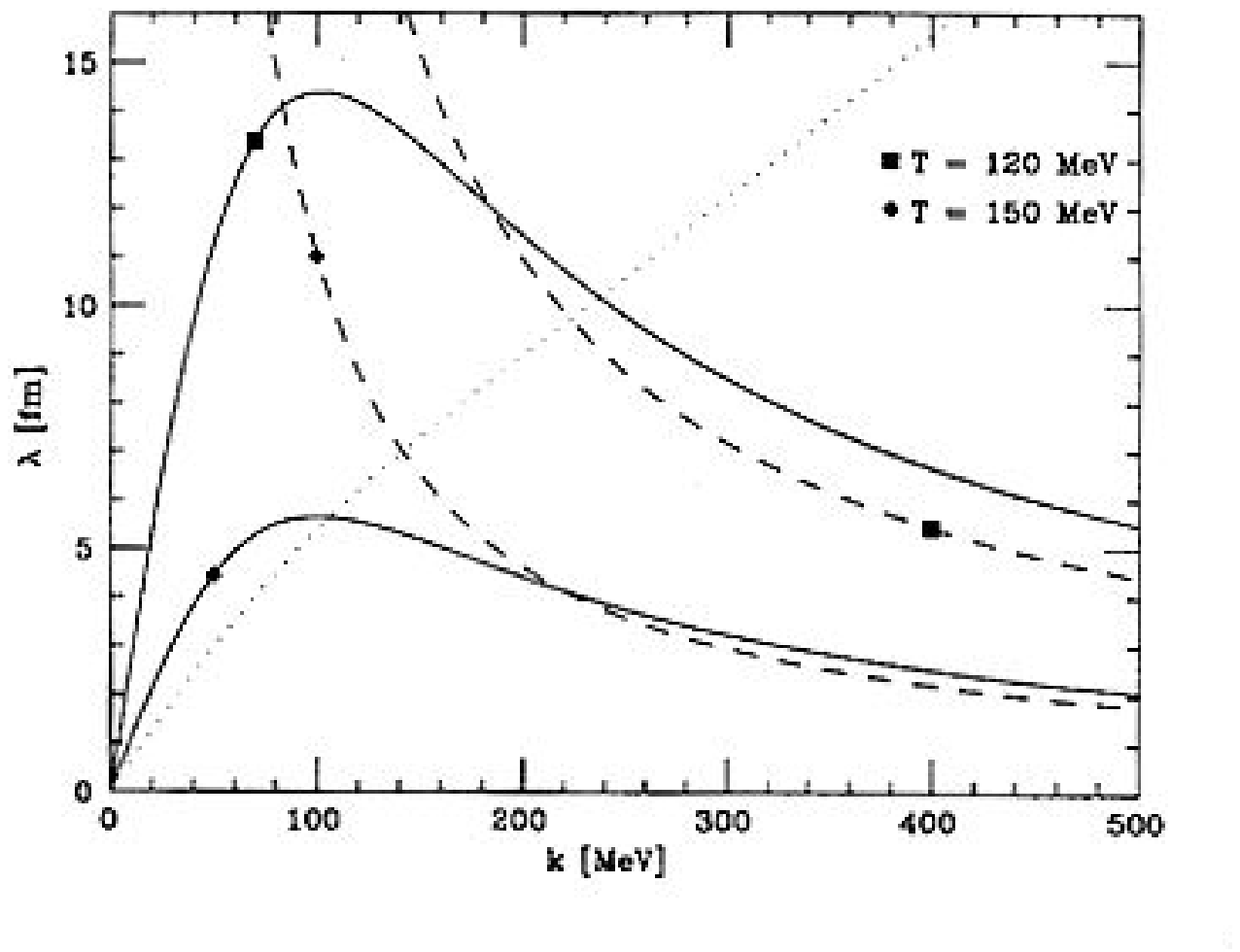}
        \vspace{5mm}
    \end{center}
\caption{ Absorption length as a function of pion
momentum. The dashed lines describe the behavior in chiral
limit, the solid
lines correspond to the experimental value of $M_\pi$ and
the dotted line
gives the qualitative behavior in the $\lambda \phi^4$
theory.}
\label{Goity}
\end{figure}

In the general case when the temperature is not assumed to
be much larger
than $\omega$ and the latter is not assumed to be much
larger than the pion
mass
(and, as far as  the real $QCD$ system is concerned, the
region
$T \gg M_\pi$ is well above
the phase transition temperature, and the soft pion approach
is not
justified there),
the integral in (\ref{grob}) can be done numerically. The
results
of Ref. \cite{Goity} for the
{\it absorption length}
\footnote{The term ``mean free path'' used in \cite{Goity}
is not quite
exact. The latter is usually applied not to the problem of
absorption of
the wave with a definite value of $k$, but rather to the
problem of
relaxation of a wave packet with a broad momentum
distribution. The latter
depends not on the total cross section as the absorption
length, but on
the so called {\it transport} cross section $\sigma^{tr} =
\int d\sigma
(1 - \cos \theta)$.
An additional
factor $1 - \cos \theta$ suppresses the contribution of
scattering at small angles. In the problem under
consideration, the
 differential cross section (\ref{T2}) has
a broad angular distribution, and the mean free path is of
the same order
as the absorption length. In other systems, in particular in
quark -- gluon
plasma where the Rutherford differential cross section is
strongly
peaked around
$\theta = 0$, the difference of these two quantities is very
essential.
 We will
discuss it in details is Chapter 6.} $\lambda = \frac
{d\omega/dk}{2\zeta}$
, i.e. the distance over which the energy density of the
classical pion wave is attenuated by the factor $1/e$ ,
 are displayed in Fig. \ref{Goity}.

\newpage

\newpage

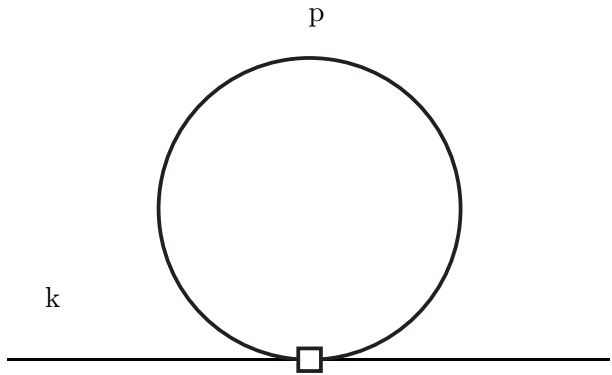
\begin{figure}

\begin{center}
\begin{picture}(80,50)
\SetScale{2.845}
\thicklines

\put(0,0){\line(1,0){80}}

\CArc(40,20)(20,0,360)
\BBox(38.5,-1.5)(41.5,1.5)

\thinlines
\put(40,45){p}
\put(5,7){k}
\end{picture}

\end{center}
\caption{Contribution of four-derivative vertices in the
 pion polarization operator.}
\label{RepiL4}
\end{figure}

Also the real part of the pion dispersion relation is
modified on the
two--loop level. To study this, one has to calculate the
real part of the
polarization operator (\ref{Pi2RPP}) from the graph in Fig.
\ref{pi2loop}
and also take into account the graph in Fig. \ref{RepiL4}
where the
box stands for the vertices involving four momenta in the
derivative
expansion of the effective chiral lagrangian (\ref{Lchi}).

\newpage

\newpage

An {\it explicit}
calculation along these lines has not yet been done so far.
A. Schenk
\cite{Schenk} calculated the thermal shift of ${\rm Re}\
\omega(\vec{k})$
in a ``mixed approach'' : he used the virial expansion
formula
(\ref{virpi}) and substituted there the forward scattering
amplitude
$T^{\pi\pi}(s)$ (and a similar 3 $\to$ 3 forward scattering
amplitude which
 enters
the term of the second order in pion density) as it follows
from the chiral
perturbation theory. Numerical plots for ${\rm Re}\
\omega(\vec{k})$
at different temperatures assuming nonzero pion mass were
drawn.
Unfortunately, the results were presented in a way which
makes an
analytic analysis difficult.

In the old paper by Itoyama and Mueller \cite{ItMul} and in
the recent
\cite{Pisrefr}, the same problem was studied in the linear
sigma--model
framework.
A beautiful qualitative effect was discovered: the
dispersion law of slow
massless pions is modified to
  \be
\label{refr}
{\rm Re} \ \omega(\vec{k}) \ =\ c(T) |\vec{k}|, \nonumber \\
c(T) \ =\ 1 \ -\ \frac {4\pi^2}{45} \frac {T^4}{F_\pi^2
M_\sigma^2}
  \ee
where $M_\sigma$ is the mass of $\sigma$--meson which was 
assumed to be
small compared to $2\pi F_\pi$. A ``refraction index''
 appears. It is not clear for us now whether this  result is
reproduced
also in the nonlinear sigma--model [If it is, the
correction should be of order $T^4/F_\pi^4$ with a possible
dependence  on the coefficients of higher--derivative
terms in
(\ref{Lchi})]. But I do not see a reason why it should not.

In \cite{Pisrefr} also the two--loop contribution to the
residue
$F_\pi(T)$ in (\ref{axcor}) was calculated and the validity
of the Gell-Mann
-- Oakes -- Renner relation (\ref{pimass}) to the order
$\sim T^4$
was checked.

\subsection{Nucleons.}

 Not only  pions but  also other hadron states change their
properties when
temperature
is switched on. Nucleons, vector mesons etc. are not easily
excited when the
temperature is  small, but  still it makes sense to study
the problem
of propagation of a massive state in pion gas and find out
how the presence
of the thermal heat bath affects its dispersive properties.
Such a study was
first carried out  for nucleons in \cite{nucl}.

 We restrict ourselves with the case when the nucleon is at
rest
and will be interested with the temperature dependence of
the pole position
of the nucleon Green's function $\Omega_N^T(\vec{0}) \equiv
 M_N^T - i\zeta_N^T$. The real part of the pole will be
called
the \ temperature--dependent \ nucleon \ mass
\footnote{Note that at finite temperature, alternative
definitions of mass
can be considered which in general are not equivalent to the
one adopted
here (see e.g. \cite{Gr,polar}). In particular, the
correlation length at
large space--like distances is not given by the inverse of
$M_N^T$. The
two quantities differ even for free fermions where the pole
position is
temperature independent while the correlation length is
given by
$(M_N^2 + \pi^2T^2)^{-1/2}$. [according to (\ref{p0ferm}),
the Euclidean
fermion frequencies are quantized
to an odd multiple of $\pi T$]. It is important to
understand when theoretical
predictions are compared  with the lattice data
\cite{Elcor}.}
and the imaginary part $\zeta_N(T)$   --- the nucleon
damping rate.

What we need to know is the nucleon polarization operator at
finite
temperature. Virial expansion in the pion density is very
handy here.
In the first order in density, the polarization operator is
given by
the graph of the same form as in Fig. \ref{virialpi} and,
quite
analogously to Eq. (\ref{virpi}), we obtain
  \be
\label{nucvir}
M_N^T - i\zeta_N^T  = M_N^0 + \frac {{\rm Tr}[(\gamma^0 +1)
\Sigma_R^N(M_N^0, \vec{0})] }{2}  =
M^0_N -  \sum_{\pi^\pm, \pi^0} \int \frac{d^3p}{(2\pi)^3 2E}
n_B(E)
\frac {T_{\pi N}(E)}{2M_N^0},
  \ee
$E = \sqrt{M_\pi^2 + \vec{p}^2}$. $\pi N$ scattering is very
well studied
in experiment.
We may use  the available phenomenological
information \cite{Npi1,Npi2} on the $\pi N$ scattering
amplitude and
need not come to grips with evaluating it in the
 chiral perturbation theory framework. The only
assumption in (\ref{nucvir}) is that the heat bath include
only pions and
that their density is small enough for the virial expansion
to be
justified. In reality, these conditions are fulfilled up to
$T \sim 130 \ {\rm MeV}$. The limitations are the same as
for the mesonic
sector.

\begin{figure}
\begin{center}
        \epsfxsize=300pt
        \epsfysize=0pt
        \hspace{15mm}
       \epsfbox[0 0 570 500]{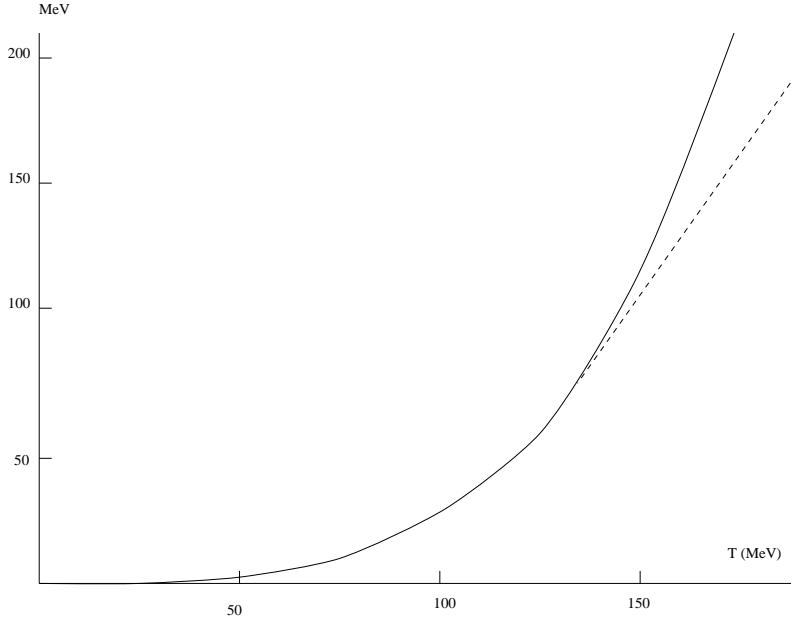}
        \hspace{-15mm}
    \end{center}
\caption{Damping rate $\gamma^T_N = 2\zeta^T_N$
of nucleon collective excitations as a function of
temperature. The dashed
line indicates the contribution due to the $\Delta$--
resonance.}
  \label{nucdamp}
 \end{figure}

Consider first the damping rate.
According to the optical theorem, the imaginary part of the
forward
scattering amplitude is related to the total cross section,
 \be
\label{opt}
{\rm Im} T_{\pi N} (E) \ =\ 2M_N^0 \sqrt{E^2 - M_\pi^2}
\ \sigma^{\pi N} (E)
  \ee
We have
\be
\label{zetaN}
\zeta_N^T \ = \ \frac 1{4\pi^2} \int_{M_\pi}^\infty dE\
\frac{E^2 - M_\pi^2}{e^{\beta E} - 1} \left[
\sigma_{\pi^+ p}(E) + \sigma_{\pi^0 p}(E) + \sigma_{\pi^-
p}(E) \right]
  \ee
To evaluate the integral, we insert the experimental cross
section, which
in the energy region of interest is dominated by the
$\Delta$--resonance.
In the narrow resonance limit, this gives
  \be
  \label{zetaNres}
\zeta_N^T \ =\ \frac 1{8\pi} \Gamma_r \sigma_r
\frac{E_r^2 - M_\pi^2}{e^{\beta E_r} - 1}
  \ee
where $\Gamma_r = 115 \ {\rm MeV}$ is the width of the
resonance and
$E_r \approx 330 {\rm MeV}$ is the corresponding pion lab
energy. At the
peak of the resonance, the cross section saturates the
unitarity limit,
$\sigma_r^{\pi^+p} = 8\pi/(q_r)^2$, where $q_r$ is the c.m.
momentum at
resonance. Using the isospin ratios $\sigma_r^{\pi^+ p} :
\sigma_r^{\pi^0 p} :
\sigma_r^{\pi^- p} = 1: \frac 23 : \frac 13$, this implies
$\sigma_r
\approx 380 \ {\rm mb}$. Eq. (\ref{zetaNres}) is plotted in
Fig.
\ref{nucdamp} (dashed line) together with a numerical
evaluation of Eq.
(\ref{zetaN}), based on the Karlsruhe--Helsinki data
\cite{Npi1}
(full line).
Damping rapidly grows with temperature.

\newpage

\newpage

It is instructive to compare this result
with a situation
where the physics is well understood: atoms exposed to a
heat bath of
photons. The virial expansion in the photon density can be
carried out
and a formula similar to Eq.(\ref{zetaN}) but involving the
total cross
section $\sigma_{\gamma A}$ for the collision of an atom
with a photon from
the heat bath can be written. This cross section grows as
$\omega^4$ at
small frequencies. As a result, the atom damping grows as
$T^7$. This
estimate is correct while the temperature is still
much less than the characteristic distance between atomic
levels
$\sim m_e \alpha^2$. If it is not the case, excitation and
ionization
processes set it. In other words, when the temperature of
the heat bath
is high enough, the atom does not live in the ground state
most of the
time, but is constantly excited by thermal photons and gets
eventually
ionized.

Also in our case, when the temperature is high enough, the
nucleon
does not stay in the ground state, but is constantly
converted
into $\Delta$ and higher excited baryon states, absorbing
and emitting
pions. Our  calculation of the nucleon damping
makes contact with reality up to the temperatures of order
$T \approx
150 \ {\rm MeV}$. At this temperature, the ratio of
probabilities for finding
the nucleon in excited state and in the ground state,
respectively, is
given by $4 \exp[-(M_\Delta - M_N)/T] \approx 0.6$. Also, at
temperatures
of order $150 \ {\rm MeV}$ and higher, the medium contains
an increasing
number of massive excitations and is not adequately
represented by a dilute
gas of pions.

Consider now the shift of the real part of the pole $M_N^T$
compared to its
zero temperature value. We use again the virial expansion
formula
(\ref{nucvir}). The real part of the forward $\pi N$
scattering amplitude
involves the Adler zero: in the chiral limit,
it behaves as $\propto E^2$ when the energy
of an ingoing pion is small. As a result, the thermal shift
of the nucleon
mass is proportional to $T^4$ at small temperatures.

Note that the term $\propto T^2$ is absent here in contrast
to what we had
for the quark condensate (\ref{SigT1}) and $F_\pi$
(\ref{FpiT}). It
particular, early attempts \cite{Deyold} to use directly the
Ioffe formula
for the
nucleon mass \cite{Iofnuc} derived from $QCD$ sum rules
  \be
 \label{nucIof}
M_N \ \approx \ 1.2 (4\pi^2 \Sigma)^{1/3}
  \ee
and substitute there the temperature dependent condensate
(\ref{SigT1})
are not justified.
\footnote{There are two  reasons why, contrary
to original hopes  \cite{Tsum}, $QCD$ sum rules method
cannot be
straightforwardly generalized to the finite temperature
case. First, sum rules
are contaminated by new
parameters: thermal expectation values of Lorentz - non-
invariant
operators. The second and even more important reason
is that
the standard ``resonance + continuum'' model for the
spectral density
of a current correlator does not hold at finite temperature
anymore.
At $T \neq 0$, the spectral density acquires contributions
due to
{\it scattering} of ingoing current on the pions from the
heat bath
\cite{Elsum}.
One can show that this additional contribution to the
spectral density
plus the temperature renormalization (\ref{lambda}) of the
nucleon residue
exactly match the thermal shift of the quark condensate
(\ref{SigT1})
in the theoretical side of the corresponding sum rule
\cite{Koike}
(the same is true for vector mesons \cite{El1}).
See \cite{front} for a related discussion.}
Speaking of the pion mass, its {\it relative} shift
$(M_\pi^T - M_\pi^0)/M_\pi^0$ is of order $T^2/F_\pi^2$,
but, as $M_\pi$
is small, the absolute value of the shift is also small, and
there is no
shift at all in the chiral limit.

\begin{figure}
\begin{center}
        \epsfxsize=250pt
        \epsfysize=0pt
        \vspace{-5mm}
       \epsfbox[0 0 310 300]{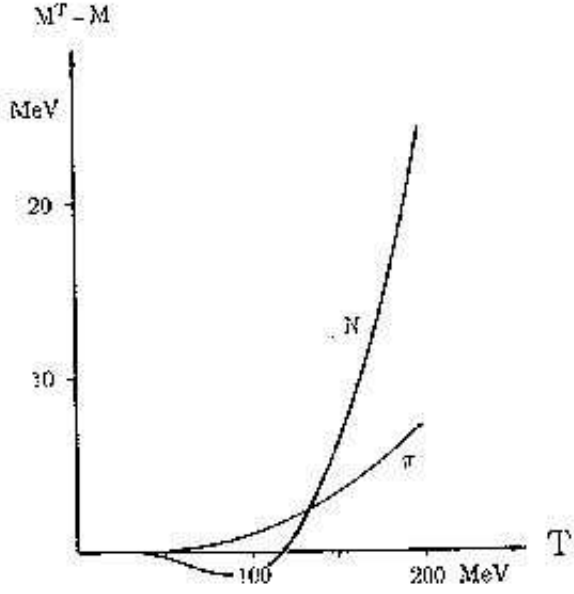}
        \vspace{5mm}
    \end{center}
\caption{Shift in the effective nucleon and pion masses
caused by
temperature.}
  \label{ReNpi}
 \end{figure}

Like the damping rate, the shift of the nucleon mass can be
determined
from the formula
(\ref{nucvir}) without invoking chiral theory, but using the
available
experimental information on the $\pi N$ forward scattering
amplitude
\cite{Npi1}. The result is shown in Fig. \ref{ReNpi}. For
comparison,
we also plot the chiral perturbation theory result for the
temperature
dependence of the pion mass (\ref{MpiT}).

\newpage

\newpage

The main feature here is that the shift in the real part of
the nucleon
pole position is small compared to the shift in the
imaginary part in the
region of applicability of the whole approach.
This finding  of \cite{nucl} was confirmed in recent
\cite{IzmN}.

Pion exhibits itself as a pole in the axial current
correlator
(\ref{axcor}). Likewise, a nucleon exhibits itself as a pole
in a
correlator of currents with nucleon quantum numbers. If we
restrict
ourselves with the currents not involving derivatives of the
quark fields,
two choices are possible \cite{Iofnuc},
  \be
\label{nuccur}
 \eta_1(x) \ =\ \epsilon^{abc} \left\{ \left[(u_R^a)^T C
d_R^b \right]
u_L^c \ -\ \left[(u_L^a)^T C d_L^b \right]
u_R^c \right\}, \nonumber \\
 \eta_2(x) \ =\ \epsilon^{abc} \left\{ \left[(u_R^a)^T C
d_R^b \right]
u_R^c \ -\ \left[(u_L^a)^T C d_L^b \right]
u_L^c \right\}
  \ee
where $C$ is the charge conjugation matrix, $\gamma_\mu^T =
-C\gamma_\mu
C^{-1}$.
 
\newpage

   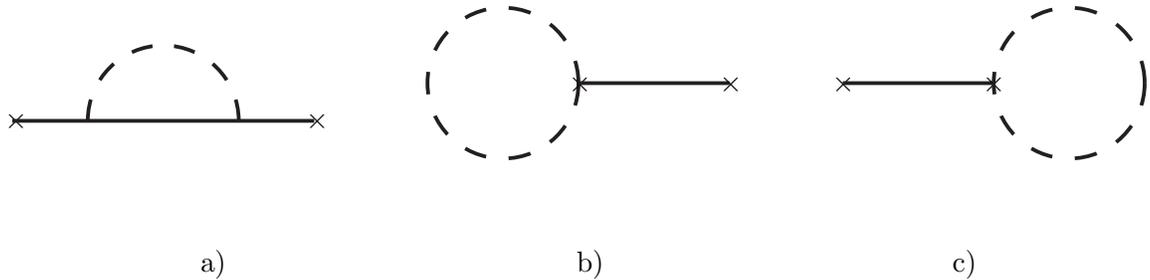
\begin{figure}

\begin{center}
\begin{picture}(150,50)
\SetScale{2.845}

\put(25,0){a)}
\put(75,0){b)}
\put(125,0){c)}

\put(-1,19){$\times$}
\put(39,19){$\times$}

\DashCArc(20,20)(10,0,180){3}
\Line(0,20)(40,20)

\DashCArc(65,25)(10,0,360){3}
\put(74,24){$\times$}
\Line(75,25)(95,25)
\put(94,24){$\times$}

\put(109,24){$\times$}
\Line(110,25)(130,25)
\put(129,24){$\times$}
\DashCArc(140,25)(10,0,360){3}

\end{picture}
\end{center}
\caption{Renormalization of the nucleon residue. Dashed
lines stand for
pions.}
\label{nucres}
\end{figure}

\newpage

Finite temperature brings about the renormalization of the
residues of
these currents in the nucleon states $\lambda$ by the same
token as
it brings about the renormalization of the residue of the
axial current
$F_\pi$. (while the current $\eta_1(x)$ plays a preferred
role in the
context of
$QCD$ sum rules \cite{Iofnuc}, the  temperature effects on
the residue
of these two  currents are the same). The corresponding
graphs,
which exactly correspond to the graphs in Fig.
\ref{axcorgr}
in the pion case, are shown in
Fig. \ref{nucres} . Using the Goldberger--Treiman relation
and
the  current algebra relation for the soft pion
vertex $<0|\eta|\pi\pi N>$,
 \be
\label{Npipi}
<0|\eta|N\pi^a \pi^b> \ =\ - \frac{\delta^{ab}}{4 F_\pi^2}
<0|\eta|N>,
  \ee
and calculating the graphs in Fig. \ref{nucres} in the same
way as we did
it earlier for pions, we arrive at the low temperature
theorem
\cite{nucl}
  \be
  \label{lambda}
\lambda^T \ =\ \lambda^0 \left\{ 1 \ -\ \frac{g_A^2 + 1}{32}
\frac {T^2}{F_
\pi^2 } \ + \ \ldots \right \}
  \ee
Like $F_\pi$, the residue drops with temperature.

One can also derive similar formulae for other hadrons. For
example,
$F_K^T = F_K(1 - T^2/32 F_\pi^2 + \ldots )$,  etc.

\subsection{Vector mesons. Experiment.}

Let us ask now whether the theoretical results presented in
the two last
sections can be directly confronted with experimental data
in heavy ion
collisions. Unfortunately, the answer is negative.

We hasten to comment that the pole positions of the
collective excitations
with pion and nucleon quantum numbers are quite physical
quantities and can
be measured in a {\it gedanken} experiment.

Suppose we study the spectrum of invariant masses of two
$\gamma$ emitted
from hot hadronic matter. The spectrum has a sharp peak
associated with the
$\pi^0 \to 2\gamma$ decay. For free pions, the width of the
peak is very
small $\Gamma_\pi^0 \approx 8\ eV$. But {\it thermal} pions
have a finite
width $\Gamma_\pi^T = 2 \zeta_\pi^T$ where $\zeta_\pi^T$ is
the thermal
average of the expression (\ref{grob}) for $\zeta_\pi(k)$.
Also
the maximum of the
resonance is shifted towards larger mass values according to
Eq.(\ref{MpiT}).
Likewise, for nucleons. Proton is believed to decay
eventually, and most
of the decay modes (like
$e^+ \pi^0$) involve hadrons which are stuck within the
hadronic medium
and do not go out undisturbed, but there is also a mode $e^+
\gamma$.
For the argument
sake, let us assume that $e^+$ and $\gamma$ do not interact
with the heat
bath, but only with detector. Measuring the distribution in
invariant masses
of $e^+\gamma$
and focusing on the ``nucleon resonance'', one could, in
principle, extract
information on the position of the nucleon pole.

Of course, these gedanken experiments are quite unrealistic.
Not even
speaking of the ``nucleon resonance experiment'' which is a
pure science
fiction, also the studying of $2\gamma$ spectrum in  heavy
ion collisions
would provide little information about pion properties in
heat bath. The
matter is that $\pi^0$ lifetime is at least six orders of
magnitude larger
than the lifetime of the fireball produced in the collision
of heavy nuclei
. Pions would decay on flight in vacuum and their mass and
width would be
the same as in any other experiment.

As far as experiment is concerned, the situation is much
better for vector
mesons, especially for $\rho$--meson. Its vacuum lifetime is
pretty small,
 almost all  $\rho$-s produced in the collision  decay {\it
inside} the
fireball, and measuring the spectra of $e^+e^-$ or
$\mu^+\mu^-$ pairs {\it
can} provide an information about the properties of $\rho$
in hot hadron
medium. Things are not so good with $\omega$ and $\phi$
mesons. Lifetime
of hadronic fireball is estimated to be 25 fm/c or less
\cite{Hung}. It is
of the same order as the $\omega$ lifetime and almost 2
times less than $\phi$
lifetime. A considerable fraction of $\omega$-s and most of
$\phi$-s would
decay outside the fireball.

Unfortunately, Life and Nature usually ingeniously resist
our attempts to
eliminate Trouble and put things in Order. A manifestation
of this general
Law in this particular case is that the problem to describe
$\rho$--meson
in hot matter involves much more theoretical uncertainties
than in the
pion or nucleon cases. The most straightforward way to study
the position of
$\rho$ pole is to perform virial expansion in the pion
density and to write
the formula like (\ref{virpi}), (\ref{nucvir}) involving the
forward $\pi
\rho$ scattering amplitude $T_{\pi\rho}$. The Trouble here
is that the
experimental information on $T_{\pi\rho}$ is not available.
We have to
calculate it in some theoretical model, and that brings
about
uncertainties.

Something can be said, however. First of all,  the amplitude
of pion
forward scattering on {\it any} hadron $h$ involves
the Adler zero due to Goldstone nature of
pion. In the chiral limit, this amplitude summed over all
pion species
behaves as

\be
\label{Tchilim}
T_{\pi^+ h} + T_{\pi^0 h} + T_{\pi^- h}
(E) \ \sim \ \frac {E^2}{\mu_{\rm hadr}^2}\left(A + iB
\frac {|E|}{\mu_{\rm hadr}}\right)
 \ee
where $\mu_{\rm hadr}$ is some characteristic hadron scale,
and we assumed that the hadron $h$
has a nonzero isospin.
\footnote{If the isospin of the target is zero, the
imaginary part of the
amplitude involves an
additional suppression. The situation is quite parallel
\cite{nucl} to the
problem of soft photon scattering. If the target is charged
(Compton
scattering), the total cross section is constant in the
limit
$\omega \to 0$, and if the target is
neutral (like the atom), the
cross section behaves as $\sim \omega^4$.}
It applies also to $\rho$. And that means that, like in the
nucleon case,
the real part of the $\rho$ mass shift does not have a
contribution $\sim T^2$ in the chiral limit.

The same conclusion can be reached considering the
correlators of  vector and
axial currents in pion heat bath. Some complication there is
that the
correlators themselves {\it are} shifted in the order $\sim
T^2$, but an
accurate analysis displays that this shift occurs not due to
a shift of
$\rho$ and $A_1$ poles, but due to admixture of the graph
with $A_1$ pole
in the vector correlator and the admixture of the graph with
$\rho$ pole
in the axial correlator \cite{Dey}.

However, $\rho$--meson mass {\it is} shifted in the order
$\sim T^4$.
It cannot be excluded right now that its temperature
dependence is more
profound than for the nucleon mass (Fig. \ref{ReNpi} shows
that the latter
is almost constant in the region $T \ ^<_\sim\  150\ MeV$
where the whole
approach based on the virial expansion is justified), and it
can, in principle,
be  essentially modified at  temperatures $T ^>_\sim 100--
150$ MeV. Whether it
is modified, indeed, and, if it is, does it eventually grow
with temperature
\cite{Pisrho} or decreases \cite{El2,GBrown} --- is the
question under
discussion now.

Experimental data on dilepton mass spectra in heavy ion
collisions
\cite{CERES,HELIOS} favor more the possibility that the
effective mass
of $\rho$ decreases with temperature. Experimentalists
observe an excess
of dilepton pairs in the small invariant mass region in
nucleus--nucleus
collisions compared to the spectrum in proton--nucleus
collisions.
It seems that this excess cannot
be explained in a conservative framework, with the processes
of direct dilepton production in the hadronic fireball
like
$\pi\pi \to \rho l^+l^-$ or $\pi\rho \to \pi l^+l^-$
\cite{Baierdil,Hung1}.
 The assumption
that $M_\rho$ drops with temperature {\it can} explain it
\cite{GBrown,Hung1}.

\section{Chiral symmetry restoration.}
\setcounter{equation}0

\subsection{General considerations. Order of  phase
transition and critical
behavior.}

In $QCD$ with 2,3, and possibly 4 massless quark flavors
(see the discussion
in Sect. 4.1), chiral symmetry is spontaneously broken at
zero
temperature. In the preceding chapter, we studied the
dynamics of
the system at comparatively low temperatures when chiral
symmetry
is still broken, and the spectrum of the system involves
massless
Goldstone states. But a spontaneously broken symmetry is
usually
restored under sufficient heating, like a spontaneously
broken rotational
symmetry in ferromagnet is restored above the Curie point.

There are few exceptions of this general rule. First of all,
it does not
apply to supersymmetry. Supersymmetry is always broken
spontaneously
at nonzero temperature irrespectively of whether it is
broken or not
at $T = 0$. The most immediate way to see it is to notice
that different
boundary conditions in  Euclidean time in the Matsubara
formalism, periodic
for bosons and antiperiodic for fermions, are not invariant
under
supersymmetry transformations which mix bosonic and
fermionic fields.
This is also seen in the real time hamiltonian formalism.
The standard
density matrix $\exp\{-\beta H\}$ is transformed under
supersymmetry
transformation into $\exp\{-\beta (H - Q \bar \epsilon -
\epsilon \bar Q)\}$
where the parameters of SUSY transformation $\bar \epsilon$,
$ \epsilon$
have the meaning of chemical potentials corresponding to
conserved
supercharges $Q$, $\bar Q$ \cite{sound}. Supersymmetry is
spontaneously
broken by the same reason why the Lorentz symmetry is
spontaneously
broken at finite $T$. In the latter case, the density matrix
is non-trivially
transformed under
the Lorentz boost and acquires the form $\exp\{-\beta (H -
\vec{v}\vec{P})
/\sqrt{1-\vec{v}^2}\}$ where
$\vec{P}$ are conserved momenta and the velocity $\vec{v}$
plays the role of
corresponding  chemical potentials.

Supersymmetry is a special case, but there are also some
systems where
spontaneous breaking of a usual (bosonic) global symmetry
persists until
arbitrary high temperatures. Moreover, heating the system
can even
{\it induce} the spontaneous breaking which was absent at
zero temperature
\cite{Moh}. Such systems are, however, rather exotic, and
there are
some special dynamic reasons for this peculiar phenomenon.
There are
no such reasons in $QCD$, it behaves as most of the physical
systems  do, and that means that
 a critical
temperature $T_c$ exists above which the chiral symmetry is
restored
and the fermion condensate $<\bar q
q>_T$ is zero. This is the temperature of phase transition
and
$<\bar q q>_T$ is the order parameter associated with the
transition.

Note that the statement of the existence of the phase
transition point
does not apply to the theory with only one massless flavor.
As was
already noted in Sect. 4.1, quark condensate does not signal
there a
spontaneous symmetry breaking [$U_A(1)$ symmetry is anyway
broken
explicitly by anomaly], is not an order parameter,
and need not vanish at high temperature. So, it
does not. At high temperatures when the effective coupling
is
small, it can be evaluated semi-classically in the instanton
approach
\cite{GPJ,KY}, and one can show that it falls down as a
power of temperature and never reaches zero.

Let us return to the case of several
 massless flavors when spontaneously broken chiral
symmetry {\it is} restored at some temperature.
The first question to be asked is what is the order of this
phase
transition, i.e. whether the order parameter  $<\bar q q>_T$
itself,
or only its derivative $\partial<\bar q q>_T/\partial T$ are
discontinuous
at the phase transition point.

This question can be studied theoretically. There are rather
 suggestive arguments which
indicate that the phase transition is of the second order
for
2 massless flavors. When $N_f \geq 3$, the phase transition
is
probably of the first order.
(If  it is there, of course. We mentioned earlier that, for
$N_f \geq 4,5$,
chiral symmetry is  probably not broken at all, and
there is nothing to be restored.) The arguments are the
following
\cite{PW}:

The starting point is the observation that, in theories
involving {\it scalar} fields,  phase transition of the
first
order often occurs when the potential involves a cubic in
fields
term. One can recall in the first place a cubic Van-der-
Vaals
curve $P(\rho, T)$ which describe the first order water
$\leftrightarrow$ vapor phase transition. The simplest field
theory example is the theory of real scalar field with the
potential
\be \label{Vphi}
V(\phi) = \lambda(\phi^2 - v^2)^2 -
\mu \phi^3
\ee
Assume for simplicity $\mu \ll \lambda v$. At
$T=0$, the potential has one global minimum at $\phi \approx
v +
3\mu/8\lambda$ and a local minimum at  $\phi \approx -v +
3\mu/8\lambda$. At nonzero temperature, the term $\sim
\lambda
T^2 \phi^2$ is added to the effective potential. At high
temperature $T \gg v$, the only minimum occurs at $\phi =
0$. One can
be easily convinced that, as the temperature increases, the
left
local minimum moves on the right and reaches the point $\phi
= 0$
(after which it does not move anymore)
 at some temperature $T^*$ when the global minimum at
positive $\phi$ still exists. When the temperature is
further increased,
the right minimum moves up and becomes degenerate with the
minimum
at $\phi = 0$ at some temperature $T_c$. The former
disappears
altogether at a still larger
larger temperature $T^{**}$. In a certain temperature range,
two
minima of the potential coexist, one
being a metastable state with respect to the other.  This is
exactly the physical situation of the first order phase
transition.

Let us go back to $QCD$. A direct application of this
reasoning
is not possible because the $QCD$ lagrangian does not
involve
scalar fields. The effective chiral lagrangian (\ref{Lchi})
is
also of no immediate use because higher-derivative terms
which
stand for dots cannot be neglected in the region close to
critical temperature. Suppose, however, that in the region
$T
\sim T_c$ some other effective lagrangian in Ginzburg-Landau
spirit can be written which depends on the composite
colorless
fields
\be \label{Fia}
\Phi_{ff'} = \bar q_{Rf} q_{Lf'}
  \ee
A general form of the effective potential which is invariant
under $SU_L(N_f) \otimes SU_R(N_f)$ is
\be
 \label{V3}
V[\Phi] \sim  g_1 {\rm Tr}\{ \Phi \Phi^\dagger\}
+ g_2 ({\rm Tr}\{ \Phi \Phi^\dagger\})^2 \nonumber \\
+ g_3{\rm Tr}\{ \Phi \Phi^\dagger \Phi \Phi^\dagger\}
+ g_4 (\det \Phi + \det \Phi^\dagger ) + \ldots
 \ee
(the coefficients may be smooth functions of $T$).  Now look
at
the determinant term. For $N_f = 2$, it is quadratic in
fields
while, for $N_f = 3$, it is cubic in fields and the
effective
potential acquires the structure similar to Eq.(\ref{Vphi})
which is characteristic for the systems with first order
phase
transition. A more refined analysis \cite{PW} shows that the
first order
phase transition is allowed also for $N_f \geq 4$, but not
for
$N_f =2$ where the phase transition is of the second order.

Most of the existing lattice data \cite{Columb,Bernard,Iwas}
indicate that,
at $N_f = 3$, the phase transition is of the first order,
indeed,
while, at $N_f = 2$, it is of the second order. The latter
is not, however,
a generally accepted statement yet, and the discussion on
this issue is
still alive
(we will discuss lattice data in some more details a bit
later).

\vspace{.3cm}

\centerline{\it Critical behavior}

\vspace{.3cm}

A second order phase transition is characterized by
critical behavior, i.
e. a power--like scaling behavior of the specific heat, the
 expectation value of the
 order parameter, and  the correlators. Different {\it
critical
exponents} can be defined.

Let us list them briefly here (for more details, see e.g.
Chapter XXIV
in \cite{staty1}). Suppose we have a system with an order
parameter
$<\phi>$ which turns to zero at the second order phase
transition point.
Introduce also a weak external field $h$ which is
canonically conjugate
to $\phi$. For standard ferromagnet, $<\phi>$ is
magnetization and $h$ is
the magnetic field. For $QCD$, $<\phi>$ is the quark
condensate and $h$ is
the small common quark mass.

The exponent $\alpha$ is related to the scaling behavior of
the specific
heat $c_P(T)$ slightly above or slightly below a critical
temperature:
  \be
\label{aldef}
c_P(T) \ =\ A_\alpha^\pm |T - T_c|^{-\alpha}
  \ee
The coefficients $A_\alpha^\pm$ are called {\it critical
amplitudes} and
the index + or - refers to whether we are dealing with the
system in the
disordered phase $T > T_c$ or in the ordered phase $T <
T_c$.

The scaling behavior of $c_P$ at $T = T_c$ but at nonzero
external field
$h$ is determined by the critical exponent $\epsilon$:
 \be
\label{epsdef}
c_P(h) \ =\ A_\epsilon h^{-\epsilon}
  \ee
When we approach $T_c$ from below, the order parameter tends
to zero as a
power
  \be
 \label{betdef}
<\phi>_T \ =\ A_\beta (T_c - T)^\beta
  \ee
At $T = T_c$ and $h \neq 0$, the order parameter also
exhibits a
power--like behavior
  \be
 \label{deldef}<\phi>_h \ =\ A_\delta h^{1/\delta}
  \ee
When $T \neq T_c$, the shift in $<\phi>$ provided by the
external field $h$
is proportional to $h$. The proportionality coefficient is
called the
generalized susceptibility $\chi$ which scales as
  \be
 \label{gamdef}
\chi \ =\ A_\gamma^\pm |T - T_c|^{-\gamma}
  \ee
At $T = T_c$ and $h = 0$, the mass gap in the physical
spectrum is
absent and the correlation length is infinite. The
correlator of the order
parameters falls down at large distances as a power
  \be
 \label{zetdef}
<\phi(\vec{x}) \phi(0)>_{T=T_c, h=0} \ = \ A_\zeta
|\vec{x}|^{-(d-2 + \zeta
)}
  \ee
where $d$ is the spatial dimension.
At $T \neq T_c$ and/or $h \neq 0$, the mass gap appears. It
scales as
  \be
\label{mudef1}
M_{\rm gap} \ =\ A_\mu h^\mu,\ \ \ \ \ \ \ \ \ \ \ T=T_c,\ h
\neq 0
  \ee
and
     \be
\label{nudef}
M_{\rm gap} \ =\ A_\nu^\pm |T - T_c|^\nu,\ \ \ \ \ \ \ \ \ \
\
T \neq T_c,\ h = 0
  \ee
Eight critical exponents $\alpha, \beta, \gamma, \delta,
\epsilon, \zeta,
\mu, \nu, $ are related by six scaling relations (so that
only two of them are
independent parameters):
 \be
\label{scalrel}
\alpha + 2\beta + \gamma \ =\ 2,\ \ \ \ \ \ \ \ \beta
\delta\  =\ \beta +
\gamma,   \nonumber \\
\epsilon(\beta + \gamma) \
=\ \alpha, \ \ \ \ \ \ \ \ \ \ \mu(\beta + \gamma) \ =\ \nu,
\nonumber \\
\nu(2 - \zeta)\ =\ \gamma, \ \ \ \ \ \ \ \ \ \ \nu d \ =\ 2-
\alpha
  \ee

A very important property of the critical exponents is their
{\it
universality}. That means that the values of $\alpha ,
\beta$ etc. do not
depend on the details of microscopic hamiltonian, but only
on gross
symmetry features of the theory. It follows basically from
the fact that
critical behavior is determined by the dynamics of the
theory at
distances which are much larger than the microscopic scale
(much larger
than $1/\mu_{\rm hadr}$ in our case). At the critical point,
the effective
theory describing large--distance dynamics is a conformal
theory not
involving any dimensional parameter. At the vicinity of
critical point,
it is a {\it perturbed} conformal theory corresponding to
adding to the
effective lagrangian an energy operator
or an order parameter operator with small
coefficients. The effective theory thus obtained involves
now a mass scale
, but as long as it is small compared to a characteristic
energy scale in
the microscopic hamiltonian ($\mu_{\rm hadr}$), this
effective theory is
still universal.

And that means that, in order to determine the critical
exponents
 in $QCD$ with 2
light quarks at the vicinity of the second order phase
transition point, we
need not to deal explicitly with $QCD$ which is a
complicated theory, but
are allowed to consider some other theory belonging to the
same
universality class, i.e. with the same symmetry breaking
pattern. The
exponents will be same. The universality class of $QCD$ with
$N_f = 2$
corresponds to the symmetry breaking pattern $SU_L(2)
\otimes SU_R(2)
\to SU(2)$ which is the same as $O(4) \to O(3)$. A rather
simple
example of  the theory in this universality class is the so
called Heisenberg
$O(4)$ magnet (Imagine a 3--dimensional cubic lattice
involving 4--dimensional
spin variable $S_\mu$ at each node. The hamiltonian is the
interaction
$S_\mu S'_\mu$ for all nearest neighbors). At low
temperatures, spins are
ordered $<S_\mu>_{T=0} \neq 0$. When temperature is
increased, the
system undergoes a second order phase transition and goes
over into
the disordered phase.
The values of critical exponents in
this model are known \cite{Baker}
(they were obtained in $\epsilon$ expansion technique). The
universality
argument tells us that their values in $QCD$ are exactly the
same
\cite{Wilcrit,Rajag}. Specifically,
 \be
\label{critexp}
\alpha\  =\ -.19 \pm .06,\ \ \beta\  = \ .38 \pm .01, \ \
\gamma\ =\
 1.44 \pm .04, \ \
\delta \ =\ 4.82 \pm .05, \nonumber \\
 \epsilon \ =\ -.10 \pm .03,\ \  \zeta\ =\ .03 \pm .01,\ \
\mu \ =\ .40 \pm
.02,\ \  \nu \ =\ .73
\pm .02
 \ee
A distinct feature of the model is the negative value of
$\alpha$ so that
the specific heat does not diverge at the critical
temperature, but rather
goes to zero with a cusp.

The predictions (\ref{critexp}) can be confronted with
experimental lattice
data. At the moment, an experimental situation is uncertain.
 The values of the exponents determined  in \cite{Kar1}
agree with the
theory rather well. In particular, their result $\delta^{-1}
= .24 \pm .03$
coincides with the theoretical prediction within the error
bars.
This finding was confirmed in recent \cite{Iwasexp} by
another group.
However,
a preferred value for $\delta^{-1}$ given in a later work
of the Bielefeld group done on larger lattices
\cite{Kar2}  is close to zero which actually indicates that
the phase
transition observed is of the first order and which sharply
contradicts
theoretical expectations. Also nobody
has seen the predicted cuspy zero for the specific heat so
far. One has to
mention also the recent work \cite{Kogut} where critical
exponents
were measured in  2+1 -- dimensional Gross--Neveu model.
(Like $QCD$, it
involves light fermions and enjoys a chiral symmetry. Like
in $QCD$,
chiral symmetry is broken spontaneously at zero temperature
and is restored
at high temperature)
The values of the exponents measured in \cite{Kogut} do
 not conform with the predictions based
on universality and the $\epsilon$--expansion (which is
rather surprising
as these predictions work
remarkably well in the condensed matter systems where {\it
laboratory}
experimental
information is in abundance), but are  instead rather
close to the values predicted in the Landau mean field
model.

Being not a lattice expert, I cannot judge, of course, who
is right and who
is wrong. Four--dimensional calculations {\it are}
complicated. The most
modern, specially dedicated for this purpose computers
should work for
hundreds of hours to produce statistically reliable results.

However, there are also systematic errors in any experiment.
Taking them under
control depends on the skill of an experimentalist, and a
layman has no say
in that. The situation is much better when the experiment
can be {\it gauged},
i.e. performed in the region where some solid theoretical
results
exist. If something known is reproduced in experiment with
good
accuracy, one can trust the experiment also when Unknown is
measured.
Fortunately, such a possibility (which has not been yet
fully
exploited) {\it does} exist for lattice numerical
calculations.
I mean here certain 2--dimensional models involving fermions
where  {\it exact} theoretical
predictions can be made and where, on the other hand,
lattice calculations
are much more easy and require much less computer time.

That refers in
particular to Schwinger model (i.e. 2--dimensional $QED$)
with several (
$N_f > 1$) light
fermion flavors. Much like as $QCD$, this model enjoys the
chiral symmetry
$SU_L(N_f) \otimes SU_R(N_f)$. In two dimensions, a
continuous symmetry
cannot be broken spontaneously due to the Mermin -- Wagner -- Coleman theorem
\cite{Coleman}.
So it does not and the quark condensate is zero in the
massless theory.
It can be shown, however, that at small temperatures and
small fermion
masses, the system displays a critical behavior like a
system with second
order phase transition at temperatures slightly {\it above}
critical. One
can say that the second order phase transition occurs at
zero temperature
(so that there is no ordered phase, indeed) \cite{JacSM}.
The exact values
of critical exponents can be determined. They are
\cite{SMind}
 \be
 \label{expSM}
\alpha = -1,\ \ \gamma = \frac 2{N_f},\ \ \delta =  \frac
{N_f+1}{N_f-1},\ \
\mu = \frac {N_f}{N_f+1},\ \ \nu = 1,\ \ \zeta = 2 - \frac
2{N_f}
  \ee
When $N_f = 2$,
also, critical amplitudes entering (\ref{aldef}) -
(\ref{zetdef}) can be
exactly calculated \cite{critamp}. In particular,
  \be
\label{ampSM}
<\bar \psi_1 \psi_1>_{T=0} \ =\ <\bar \psi_2 \psi_2>_{T=0} \
=\ -.388\ldots
m^{1/3} g^{2/3}, \nonumber \\
M_{\rm gap} \ =\ 2.008\ldots m^{2/3} g^{1/3}
  \ee
There is only one comparatively old
lattice paper where Schwinger model with 2 flavors was
studied \cite{Grady}. The measured values of the critical
exponents agree
well with theoretical predictions. However, the experimental
values of the
critical amplitudes exceed their theoretical values by 35\%
for the
condensate and by 25\% for the mass gap. It would be very
interesting, indeed,
 to repeat this calculation on modern computers and with
different modern
lattice algorithms (Wilson vs. Kogut--Susskind fermions,
perfect vs.
improved vs. simple--minded actions etc.).
Confirmation of the theoretical predictions (\ref{expSM},\
\ref{ampSM})
with good accuracy would provide an efficient test of
lattice methods.

\subsection{Insights from soft pion physics.
Large $N_c$.}

 We have already
seen an indication for the chiral symmetry restoration
in the previous chapter: according
to Eq.(\ref{SigT1}), the order parameter
$\Sigma(T) = -<\bar q q>_T$ {\it decreases} as the
temperature is
increased. That was a one--loop calculation which had only a
limited
range of validity $T < \ 80-100 \ MeV$ where the temperature
correction
is still small. One can, however, improve
the accuracy and find out the temperature dependence of the
condensate
on the 2--loop and 3--loop level in chiral perturbation
theory.
This calculation done in
\cite{GLT,Ger} allows one to monitor  melting down of the
quark condensate
at larger temperatures and make
a reasonable estimate for the  temperature  of phase
transition
where the condensate disappears.

\newpage

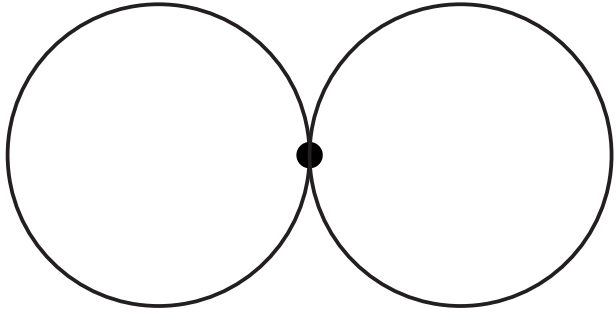
\begin{figure}

\begin{center}
\begin{picture}(80,50)
\SetScale{2.845}

\CArc(20,25)(20,0,360)
\GCirc(40,25){1.5}{0}
\CArc(60,25)(20,0,360)

\end{picture}
\end{center}
\caption{Two loop contribution in the pion free energy.}
\label{F2loop}
\end{figure}

\newpage

\begin{figure}

\begin{center}
\begin{picture}(120,50)
\SetScale{2.845}

\put(30,0){a)}
\put(90,0){b)}

\BCirc(10,25){10}
\BCirc(30,25){10}
\BCirc(50,25){10}

\GCirc(20,25){1}{0}
\GCirc(40,25){1}{0}

\BCirc(110,25){10}
\BCirc(90,25){10}
\BBox(98.5,23.5)(101.5,26.5)

\end{picture}
\end{center}
\caption{Some graphs contributing in the free energy on the
three--loop
level.}
\label{F3loop}
\end{figure}
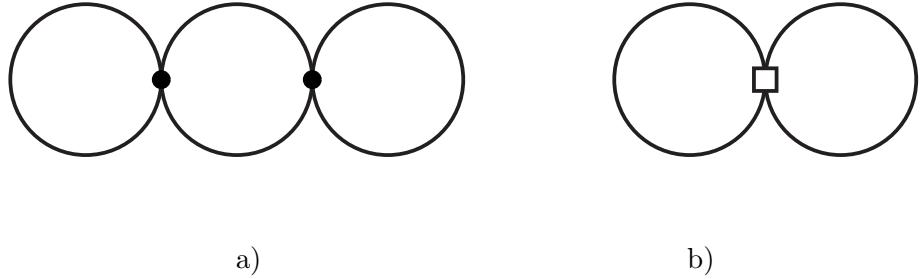

\newpage

\begin{figure}
 \begin{center}
        \epsfxsize=400pt
        \epsfysize=0pt
        \vspace{-5mm}
       \epsfbox[132 335 566 610]{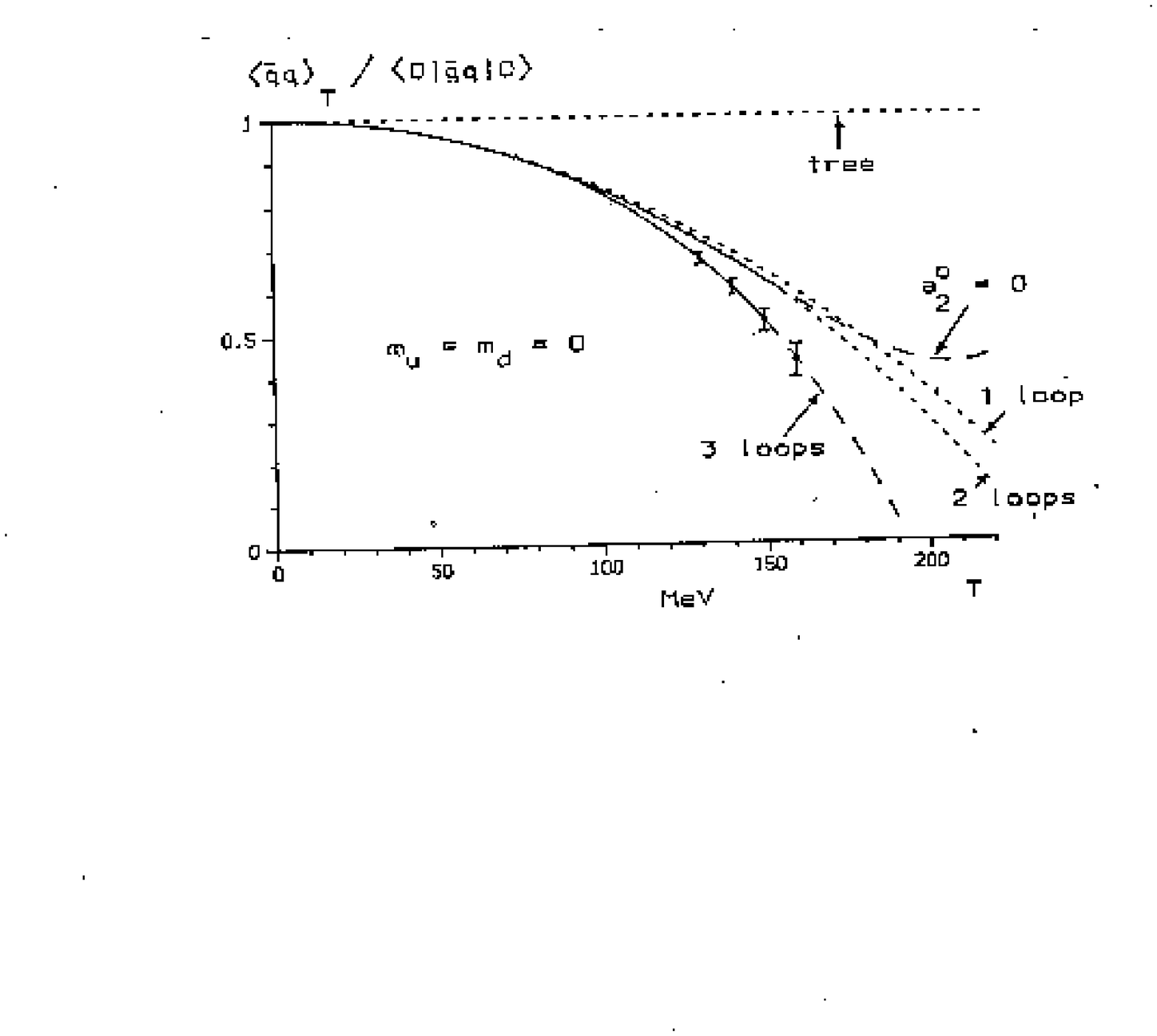}
        \vspace{5mm}
    \end{center}
\caption{Temperature dependence of the condensate in three
loops.}
\label{Gerberl}
\end{figure}

 Let us describe the results of
\cite{Ger} where the
temperature dependence of $<\bar q q>_T$ has been determined
on
the 3-loop level.  To this end, one has to calculate the
free energy density
on the three loop level at small nonzero quark masses and
differentiate
 the result over the quark mass. There is only one relevant
two--loop
graph depicted in Fig. \ref{F2loop}.
By a  simple power counting [We have $T^2 M_\pi^2 \sim
\Sigma m (T^2/F_\pi^2)$
from the pion mass dependent term
in the free energy and an additional factor $\sim
T^2/F_\pi^2$ from the
vertex (\ref{chiexpan}) involving two derivatives, i.e. the
square of a
 characteristic
momentum $p_{\rm char} \sim T$ of pions in the thermal loop
],
it gives the correction
$\sim T^4/F_\pi^4$ in the condensate.
 There are many 3--loop graphs. We will depict
two of them. The graph in Fig. \ref{F3loop}a involves only
the standard
4--pion vertex  (\ref{chiexpan}) and depends only on $F_\pi$
(it depends also
on the ultraviolet cutoff, but disregard it for a moment).
The graph in
 Fig. \ref{F3loop}b contains something new. It looks as a
two--loop graph,
but contributes actually in the order $\sim T^6$ because the
square vertex
in this
graph stands for the terms in the effective lagrangian
(\ref{Lchi})
involving four derivatives. There are several chiral
invariant structures
of this kind. One of them has the form
  \be
\label{L2}
 {\cal L}^{(4)} \ =\ \alpha \left( {\rm Tr} \left\{
\partial_\mu U^\dagger
 \partial_\mu U \right\} \right)^2
  \ee
The dimensionless coefficient $\alpha$ (another name for it
is
$L_1$ \cite{GLold})  is a new independent parameter
of the lagrangian. In this order, there are 4 more such new
dimensionless
coefficients  $L_{2,3,4,5}$ which are
relevant to our problem. All of them affect the temperature
renormalization of
the condensate in the order $\sim T^6$.
These new parameters are largely fixed,
however, from experimental data on pion (and also $K$,
$\eta$ ) interactions
 at intermediate
range of energies which are also affected by $L_i$
\cite{CPT}.
\footnote{Actually, $L_i$ appear not only from the higher--
derivative
vertices in the tree lagrangian, but also from counterterms
which are required
to renormalize ultraviolet divergent pieces in the loop
graphs of the kind
depicted in Fig.\ref{F3loop}a. The effective lagrangian
involves also many
terms with six derivatives
etc. On each new level of the expansion in momenta and in
pion masses, many new
independent constants appear, and their total number is
infinite (which is not
surprising --- after all, the lagrangian (\ref{Lchi}) is not
renormalizable
and involves an infinite number of counterterms and
subtraction constants.).
The point is, however, that if we restrict ourselves by not
too high energies
(and we have to do it anyway: at higher energies, neglecting
other than pions
degrees of freedom is not justified), the contribution of
these still higher
derivatives terms is suppressed. Chiral perturbation theory
is not the expansion
over some dimensionless lagrangian coupling as usual, but
rather an expansion
 over the parameters
$\sim M_\pi^2/(2\pi F_\pi)^2$, $\sim p^2/(2\pi F_\pi)^2$
where $p$ is a
characteristic momentum scale of the process under
consideration.}
The final
 result for the condensate in massless theory in soft pion
approximation
(i.e. when the effects due to $K$ and $\eta$ are disregarded
together with
effects coming from other resonances)
 has a rather simple form

\be \label{qqT}
<\bar q q>_T = <\bar q q>_0
\left[ 1 - \frac {T^2}{8F_\pi^2} - \frac {T^4}{384F_\pi^4} -
\frac {T^6}{288F_\pi^6} \ln
\frac \Lambda T + \ldots \right]
  \ee
Here all effects from the higher-derivative term are
described by
the constant $\Lambda$. Experimental data on pseudogoldstone
interactions
give the value
$\Lambda \sim 500 \pm 100$  Mev. The dependence (\ref{qqT})
together with the curves where only the 1-loop correction
$\propto T^2$ or also the 2-loop correction $\propto T^4$
are taken into account
(please, do not put attention to the ``technical'' curve
marked $a^0_2 = 0$)
 is drawn in Fig. \ref{Gerberl} taken from Ref. \cite{Ger}.

\newpage

\newpage

The expansion in the parameter $\sim T^2/8F_\pi^2$ makes
sense
when this parameter is small, i.e. when $T \ ^<_\sim\  100-
150$ Mev.
Strictly speaking, one cannot extrapolate the dependence
(\ref{qqT}) for larger temperatures, especially having in
mind
that, at $T > 150$ Mev,  the heat bath includes a
considerable
fraction of other than pion hadron states. But as we know
anyhow
that the phase transition with restoration of chiral
symmetry
should occur, the estimate of the phase transition
temperature
(i.e. the temperature when $<\bar q q>_T$ hits zero, and we
assume
here that $N_f = 2$ so that the transition is of the second
order)
based on such an extrapolation is not altogether stupid.
This estimate is

\be \label{est}
T_c \approx 190 \ {\rm Mev}
\ee
or, roughly, one inverse fermi.
A more accurate
treatment which takes into account nonzero $m_{u,d}$ and
also
the presence of other mesons in the heat bath gives
practically
the same estimate for the phase transition temperature
(assuming, of course, that the phase transition is there
which,
 as we will see in the next
section, is probably not the case): nonzero quark masses
smoothen the
temperature dependence of the condensate and push $T_c$ up
while
 excitation of other degrees of freedom sharpen the
temperature dependence
and pulls it down. These
 two effects  practically cancel each other.

\vspace{.3cm}

\centerline{\it Large $N_c$.}

\vspace{.3cm}

It is interesting to discuss what happens in the large $N_c$
limit
\cite{HoftNc}. As usual, we assume $N_c \to \infty,\ \ g^2
\to 0$ such
that the product $g^2N_c$ stays  constant. For simplicity,
we
also assume zero quark masses.

In this limit, the spectrum of $QCD$
presents a set of infinitely narrow meson resonances.
In the first place, we have $N_f^2$ massless Goldstone
states. An additional
Goldstone particle is $\eta'$. When $N_c$ is finite, $\eta'$
is
massive because $U_A(1)$ axial symmetry of the $QCD$
lagrangian
is broken explicitly by anomaly. However, in the limit $N_c
\to \infty$,
the effects due to anomaly are suppressed, $U_A(1)$ becomes
a good
symmetry of the full quantum theory, and its spontaneous
breaking
by quark condensate brings about a new massless Goldstone
state
\cite{Venez1}. When $N_c$ is large but not infinite, the
mass of
$\eta'$--meson has the order $M_{\eta'} \propto
1/\sqrt{N_c}$
\cite{Venez2}.
Besides that, there are also massive meson excitations. The
mass
of the first such excitation $\equiv \mu_{\rm hadr}$ does
not
depend  on $N_c$. The density of higher excited states grows
exponentially with mass.

Low energy properties of $N_f^2$ Goldstone states are
described
 by a modified chiral lagrangian which has exactly the same
form
as (\ref{Lchi}), only $U$ is now a full $U(N_f)$ rather than
$SU(N_f)$
matrix (there is also a determinant term \cite{Venez2}, but
it is
 suppressed in the large $N_c$ limit). The temperature
corrections
to the quark condensate can be calculated by the same token
as earlier,
only the coefficients are modified a little bit. In
particular, the first
temperature correction to the condensate (\ref{SigT1})
involves
now the factor $N_f^2$ rather than $N_f^2 - 1$. A glance on
the large
$N_c$ analog of the expression (\ref{qqT}) displays an
apparent paradox.
Indeed, $F_\pi$ scales as $\sqrt{N_c}$ in the large $N_c$
limit. And
that seems to mean that the temperature corrections to the
condensate
 become essential at $T \propto \sqrt{N_c} \mu_{\rm hadr}$.
Thereby, the
condensate should turn to zero and chiral symmetry be
restored at
$T_c \sim \sqrt{N_c} \mu_{\rm hadr}$.

This estimate for the phase transition temperature
contradicts, however,
physical intuition. The relevant physical scale in large
$N_c$ theory
is not $\sqrt{N_c} \mu_{\rm hadr}$, but just $\mu_{\rm
hadr}$. In particular,
in the infinite $N_c$ limit, the temperature $T_c \sim
\sqrt{N_c}
\mu_{\rm hadr}$ is much larger than the Hagedorn maximal
temperature $T_H \sim
\mu_{\rm hadr}$ and just cannot be reached (see the
discussion in the
beginning of Sect. 3.3).

If $N_f \geq 3$ (when the phase transition is of the first
order), there
is a way out. We may suppose that the condensate stays
practically unchanged
until $T_c$ and then abruptly jumps to zero. But in the
theory with $N_f =2$,
the phase transition is of the second order, the condensate
should
approach zero when we approach $T_c$ from below, and we do
not see yet
how is it possible in the soft pion framework.

The resolution of this paradox will be given in a moment,
but first note
that, after all, the fact that at,
say, $T = T_c/2$ the condensate is practically not changed
in the large
$N_c$ limit compared to its zero
temperature value is quite natural and reasonable.
Eventually, the condensate
melts down due to nonlinear pion interactions. The latter
can be
estimated as being due to a resonance exchange, say, $\pi\pi
\to \rho
\to \pi\pi$.
Three--meson vertices scale as $1/\sqrt{N_c}$ in the large
$N_c$ limit (this
suppression brings about the suppression $\propto 1/N_c$ for
resonance widths).
Thus four--pion vertex scales as $1/N_c$ and, indeed, it is
seen also
from Eq.(\ref{chiexpan}). On the other hand, when the
temperature is increased
up to $\sim \mu_{\rm hadr}$, the density of thermal pions $n
\sim T^3$
becomes so large that the characteristic distance between
pions is of the
same order as their size (the latter has the order $\sim
\mu_{\rm hadr}^{-1}$
and does not scale with $N_c$). Pions overlap, and something
{\it is} bound
to happen at this point.

And it does. We have seen that the correction
$\propto T^6$ in (\ref{qqT}) involves, besides $F_\pi$, also
a new
constant $\Lambda$ which
describes the contribution of the graphs of the kind drawn
in
 Fig. \ref{F3loop}b involving higher--derivative terms in
the chiral
lagrangian. The corresponding dimensionless constants $L_i$,
i = 1,\ldots, 5
scale as $N_c$ \cite{GLold}. (Indeed, at $p_{\rm char} \sim
\mu_{\rm hadr}$,
all terms in the pion lagrangian should be of the same
order, and hence all
 coefficients in ${\cal L}^{(4)}$,  ${\cal L}^{(6)}$, etc.
scale in the same way
as $F_\pi \propto N_c$.
\footnote{If the lagrangian is written in terms of only
$N^2_f - 1$ Goldstone
fields, a certain coefficient in  ${\cal L}^{(4)}$ called
$L_7$ scales
as $N_c^2$ \cite{GLold}. But {\it i)} $L_7$ does not
contribute in $\Sigma(T)$;
{\it ii)} If it would, that would only help us to trace back
the origin of
large temperature correction; {\it iii)} Anyway, we are
thinking in terms
of the effective lagrangian involving $N_f^2$ Goldstone
mesons.}
). The contribution in $\Sigma(T)/\Sigma(0)$ due to the
graph in
Fig. \ref{F3loop}b is estimated to be $\sim N_c
T^6/F_\pi^6$, i.e., for large
$N_c$,  it dominates over the contribution coming from the
graph in
Fig. \ref{F3loop}a depending only on $F_\pi$. Staying on the
3--loop level,
we would conclude that the correction is of order 1 when $T
\sim N_c^{1/3}
 \mu_{\rm hadr}$ which is better than $T \sim N_c^{1/2}
 \mu_{\rm hadr}$, but is not satisfactory yet.

However, one should, of course, include in the estimate also
the contributions
due to still higher order terms  ${\cal L}^{(6)}$,  ${\cal
L}^{(8)}$, etc.
Drawing the same graph as in Fig. \ref{F3loop}, but with a
4--pion vertex
coming from  ${\cal L}^{(6)}$ (the corresponding
coefficients scale
as $N_c/\mu_{\rm hadr}^2$), we arrive at the estimate
  \be
 \label{est4loop}
\frac {\Sigma(0) - \Sigma^{\rm 4-loop}(T)} {\Sigma(0)} \
\sim \
\frac 1{N_c \mu_{\rm hadr}^3} \ (\mu_{\rm hadr}T^2) \
\frac1{F_\pi^4} \
\frac{N_c}
{\mu_{\rm hadr}^2} \ T^6 \ \sim \ \frac {T^8}{N_c^2 \mu_{\rm
hadr}^8}
  \ee
Here the first factor is $[\Sigma(0)]^{-1}$, $\mu_{\rm
hadr}T^2$ comes
from the expansion
of the free energy in pion mass $F_M \propto M_\pi^2 T^2$
and its subsequent
differentiation over quark mass m, $1/F_\pi^4$
comes from the
expansion of $U$ up to the fourth order in $\phi^a$,
$N_c/\mu_{\rm hadr}^2$
is the estimate of a coefficient of six--derivative term,
and $T^6$
is the sixth power of a characteristic momentum.

Summing up the leading in $N_c$ terms, we obtain
  \be
\label{estN}
\Sigma(T) \ \sim \ \Sigma(0) \left[ 1 - \frac{x^2}{N_c^2}
F(x) \right]
  \ee
where $F(x)$ is some function of a dimensionless parameter
$x = T^2/\mu_{\rm hadr}^2$. We do not now how the function
$F(x)$ behaves
at $x\ ^>_\sim\ 1$, but a quite natural assumption is that
the Taylor series for
this function has a  finite radius of convergence so that
$F(x)$ becomes
infinite at some $x_0$. After all, the condensate is the
derivative
of free energy density over  quark mass, and free energy
definitely becomes infinite at Hagedorn temperature [see
Eq.(\ref{Zdens})].
Let us assume further that $F(x)$ has the simple pole at $x
= x_0$.
Now if we keep temperature fixed and send $N_c$ to infinity,
the
correction to the condensate vanishes. If, on the other
hand,
 we keep $N_c$ fixed, the quark condensate (\ref{estN})
would vanish and
the phase transition would occur at some point $x_{\rm
crit}$
somewhat below $x_0$,
$x_0 - x_{\rm crit} \sim 1/N_c^2$. The characteristic width
of the region where
$\Sigma(T)$ is essentially changed is also estimated then as
  \be
 \label{WidthT}
\frac {\Delta T}{T_c} \ \sim \ \frac 1{N_c^2}
  \ee
If one assumes that $F(x \sim x_0) \ \sim \ 1/(x-x_0)^2$,
the width of the
transition region would be somewhat larger $\propto 1/N_c$.
If the singularity
is weaker than $1/(x - x_0)$, the width of the transition
region would be
smaller (but is always suppressed at large $N_c$).

Thus the paradox is resolved. We see that, at $N_f = 2$ and
large $N_c$,
the condensate is
practically not changed, indeed, up to very vicinity of the
phase transition
point where it drops sharply and eventually hits zero. At
$N_f \geq 3$, there
is no ``need'' for the condensate to approach zero smoothly
when the temperature
approaches $T_c$. It can still be essentially changed in the
transition
region below $T_c$, approach some half--way value of order
$\Sigma(0)/2$
and then abruptly jump to zero at the phase transition
point. On the other
hand, it can remain constant until the very phase transition
point and then
jump to zero ``the whole way''. Finally, it can also
approach zero when
$T \to T_c$ and the phase transition can still be the second
order.
(Ginzburg--Landau arguments presented above {\it allow} the
first order
phase transition at $N_f \geq 3$ but do not {\it dictate} it
to be the case.)
Currently, no further conclusions on this point can be
drawn.

\subsection{ The real world.}

Up to now, we discussed only $QCD$ with
massless quarks. But the quarks have nonzero masses: $m_u
\approx$ 4 Mev, $m_d \approx$ 7 Mev, and $m_s \approx$  150
Mev \cite{GLmass}. The
question arises whether the nonzero masses affect the
conclusion on the existence or non-existence and the
properties
of the phase transition.

\begin{figure}
\begin{center}
        \epsfxsize=300pt
        \epsfysize=0pt
       \epsfbox[0 0 320 290]{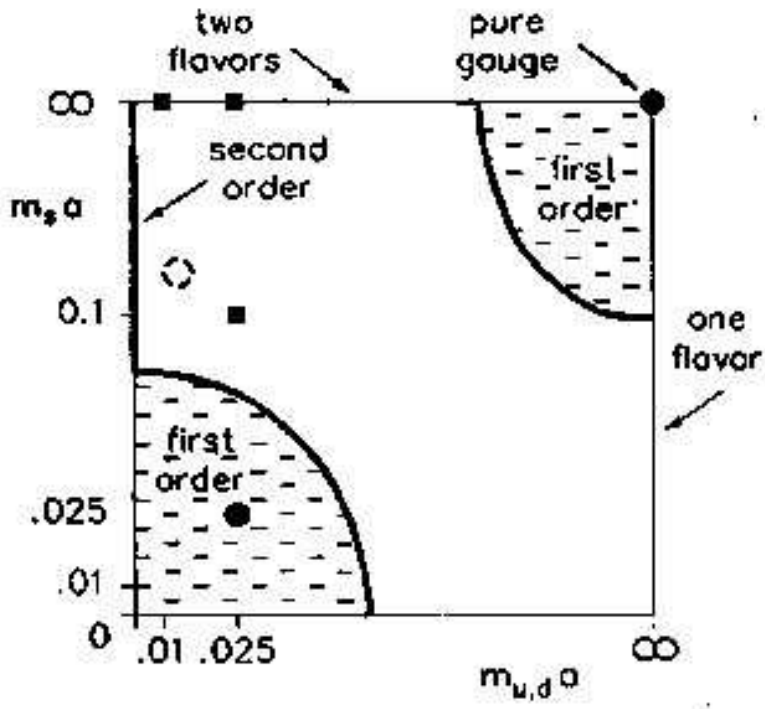}
    \end{center}
\caption{Phase diagram of $QCD$ according to
Ref.\protect\cite{Columb}.
Mass values (measured in the units of the inverse lattice spacing 
$a^{-1}$ for 
which the transition is and is not seen are denoted respectively
by solid circles and squares. Dashed circle corresponds to
the physical
values of masses.}
\label{Columbl}
\end{figure}

Two different
experimental (i.e. lattice) works where
this question was studied are available now, and the results
of these two
studies drastically disagree with each other. The results of
Columbia
collaboration \cite{Columb} display high sensitivity of the
phase
transition dynamics to the values of light quark masses. In
Fig.
\ref{Columbl},
a phase diagram of
$QCD$ with different values of quark masses $m_s$ and $m_u =
m_d$ as drawn in Ref.\cite{Columb} is plotted.

\newpage

\newpage

Let us discuss different regions on this plot. When the
quark
masses are large, quarks effectively decouple and we have
pure
YM theory with $SU(3)$ gauge group where the phase
transition is
of the first order. When all the quark masses are zero, the
phase transition is also of the first order. When masses are
shifted from zero a little bit, we still have a first order
phase transition because a finite discontinuity in energy
and
other thermodynamic quantities cannot disappear at once when
external
parameters (the quark masses) are smoothly changed.

But when all the masses are nonzero and neither are too
small
nor too large, phase transition is absent. Notice the bold
vertical line on the left. When $m_u = m_d = 0$ and $m_s$ is
not
too small, we have effectively the theory with two massless
quarks and the phase transition is of the second order. The
experimental values of quark masses (the dashed circle in
Fig.\ref{Columbl})
 lie close to this line of second order phase transitions
but
in the region where no phase transition occurs. It is the
experimental fact as measured in Ref. \cite{Columb}.

This statement conforms nicely with a semi-phenomenological
theoretical argument of Ref. \cite{KK} which displays that
even
{\it if} the first order phase transition occurs in QCD, it
is
rather weak. The argument is based on a generalized
Clausius-Clapeyron relation. In college physics, it is the
relation connecting the discontinuity in free energy at the
first-order phase transition point with the sensitivity of
the
critical temperature to pressure. The Clausius-Clapeyron
relation in $QCD$ reads
\be
\label{KK}
  {\rm disc} <\bar q q>_{T_c}\  = \ \frac 1{T_c} \ \frac
{\partial T_c}
{\partial m_q} {\rm disc} \ \epsilon
  \ee
where ${\rm disc}\ \epsilon$ is the latent heat.
The derivative $\frac {\partial T_c}{\partial m_q}$ can the
estimated from theoretical and experimental information of
how
other essential properties of $QCD$ depend on $m_q$ and from
the
calculation of $T$ - dependence of condensate at low
temperature
in the framework of chiral perturbation theory (see
Fig.\ref{Gerberl} and the
discussion thereof).  The
dependence on quark masses is not too weak: $\partial
T_c/\partial m_q \approx
0.9 - 1.0$. From that, assuming
 that the discontinuity in quark
condensate is as large as $<\bar q q>_0$,
we arrive at an estimate
 \be
 \label{Lheat}
{\rm disc}\ \epsilon \ < 0.4 \ {\rm GeV/fm}^3
 \ee
This upper limit for ${\rm disc}\ \epsilon$ should be
compared with the
characteristic energy density of the hadron medium at $T
\sim T_c \sim 190$ MeV.
 The
latter cannot be calculated analytically and we have to rely
on numerical
estimates. The lattice study in \cite{Blum} (for the theory
with 2 massless
flavors) gives a rather large value $\epsilon(T \sim 200\
{\rm MeV}) \
\approx \ 3\ {\rm GeV}/{\rm fm}^3$.
May be, this value is even too large, it is of the same
order as the
Stefan--Boltzmann energy (\ref{Fg0}, \ref{Fq0}, \ref{EFP})
of free quarks and gluons  in $QGP$ phase. This is
surprising as, at such low
temperatures, quarks and gluons are certainly not free and
the interaction
effects are important --- see a detailed discussion in the
next chapter.
Anyway, one can safely conclude that the limit (\ref{Lheat})
for the latent heat
 is several times smaller than $\epsilon(T_c)$.
In reality, the discontinuity in $<\bar q q>$ at the phase
transition point
(if it is there)
is, of course, much smaller than the value of the condensate
at zero
temperature. It is seen from the graph in Fig.\ref{Gerberl}.
The condensate
drops significantly still in the region where $K$ and $\eta$
degrees of
freedom are not yet effectively excited. These degrees of
freedom
(and also
$\rho$ and $\omega$ degrees of freedom) become relevant only
at
$T \ ^>_\sim\  150$ MeV when the tendency of the condensate
to drop down is
already well established. Moreover,
as was mentioned earlier in the paragraph
after Eq.(\ref{est}), taking into account these degrees of
freedom
{\it sharpens} the condensate dependence rather than
smoothens it.
This is also seen from Eq.(\ref{SigT1}): The more is the
number of
massless quark flavors $N_f$, the sharper is the temperature
dependence of
the condensate.

Thus latent heat of the first order phase transition in the
theory with 3
massless quarks
must be rather small [significantly smaller than the
estimate (\ref{Lheat})]
 which means that the phase transition is likely to
disappear
under a relatively small perturbation due to nonzero $m_s$.
The numerical
findings of Ref.\cite{Columb} are thereby rather appealing
from theoretical
viewpoint.

\begin{figure}
\grpicture{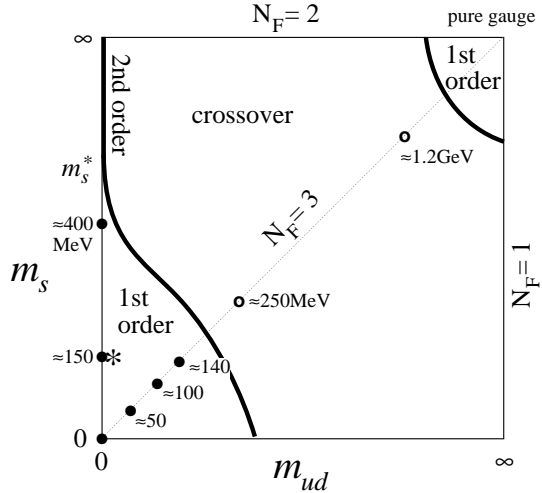}
\caption{Phase diagram of $QCD$ according to
Ref.\protect\cite{Iwas}.
First order phase transition is seen at solid circles and is
not seen at
blank circles.
The star marks out the physical values of quark masses.}
\label{Iwasl}
\end{figure}

However, as far as experiment is concerned, the question is
far from
being  completely resolved by now. A recent lattice study
\cite{Iwas} done
with Wilson rather than Kogut--Susskind fermions displayed
quite a
different picture shown in Fig.\ref{Iwasl}.

\newpage

\newpage

The bend of the second order phase transition line
separating the region
with first order phase transition and the region with no
phase transition
reflects a recent finding \cite{Rajag1} that the shape of
this line
near the tricritical point $m_s = m_s^*,  m_q \equiv m_u =
m_d = 0$,
where $m_s^*$ is the smallest value of the strange quark
mass where we
still have a second order phase transition in a theory with
2 massless
quarks
\footnote{This point is usually called {\it tri}critical
because we can
continue the phase diagram in the region of negative $m_q$.
Then the point
$(m_s^*,0)$ signifies the meeting of three regions: the
region with first
order phase transition, the region with no phase transition
for positive
$m_q$, and the region with no phase transition for negative
$m_q$. Generally, the ratio $m_q/m_s$ can be a complex
number, and it is
better to think of the point  $(m_s^*,0)$ as of the point
where the {\it
surface} of second order phase transitions in $(|m_s|, \
m_q/m_s)$ 3-space
degenerates into the line. } is
  \be
  \label{tricr}
m_s^* - m_s \ \sim \ m_q^{5/2}
  \ee
What is very surprising is the large value of $m_s^* \approx
450\  {\rm MeV}$ as measured in Ref.\cite{Iwas}. If they are
right, the point
corresponding to our
physical world lies well inside the first order phase
transition region in
which case the first order phase transition should be of
strong variety
which contradicts Ref.\cite{Columb} and also the theoretical
arguments above.

 Again, we do not know what of these two measurements is
more correct.
Possibly, intrinsic lattice artifacts are more dangerous
when an algorithm
with Wilson fermions is used \cite{Rajag1,deTar}, but I do
not have my own
opinion on this point. To resolve the controversy, it would
make sense, first,
 as was already mentioned, to
test lattice methods on a two--dimensional playground and,
second, to
perform a {\it quantitative} comparison of the temperature
dependence of
the quark condensate as measured on lattices with the
theoretical curve
in Fig.\ref{Gerberl}. Such a comparison has not been done so
far.

To summarize, at the current level of understanding, the
picture
where the hadron gas goes over
to quark-gluon plasma and other way round without any phase
transition looks more probable. We have instead a sharp
crossover in
a narrow
temperature range which is similar in properties to
second-order phase transition (the ``phase crossover'' if
you
will).

  \subsection{Instantons and percolation.}

In the discussion of the properties of the system at the
vicinity of phase
transition in this chapter, we relied so far on the fact
that chiral symmetry {\it is} broken at zero temperature. It
is
an experimental fact in real $QCD$, but it is important to
understand from pure theoretical premises {\it why} it is
broken
and what is the mechanism of its restoration at higher
temperatures.

A completely satisfactory answer  to this question has not
yet
been obtained. The problem is that $QCD$ at zero temperature
is
a theory with strong coupling and it is very difficult (may
be
impossible) to study the structure of $QCD$ vacuum state
analytically from the first principles.  However, a rather
appealing
qualitative physical
picture exists \cite{Shurbook,SSrev} which is based on the
model of
instanton-antiinstanton liquid and on the analogy with the
so
called percolation phase transition in doped superconductors
\cite{Shkl}.

The starting point is the Banks and Casher relation
(\ref{Banks}).
As we have seen, nonzero quark condensate means nonzero
spectral density
of the Euclidean Dirac operator at $\lambda = 0$ and that
implies the
presence of rather small characteristic eigenvalues
\be
\label{lam1V}
\lambda \sim 1/(\Sigma
V) \ll \lambda_{\rm free} \sim 1/L
\ee
 in the spectrum.
The question is what is the physical reason for these small
eigenvalues to appear.

As far as we know, the first pioneer
paper where a mechanism for generating small eigenvalues was
proposed is Ref.\cite{Flor} where small eigenvalues appeared
as
zero modes of monopole-like gauge field configurations. The
disadvantage of this model is that the monopole
configurations
are static whereas it is natural to expect that
characteristic
gauge fields contributing to the Euclidean path integral at
$T=0$ are more or less symmetric in all four directions with
no
particular axis being singled out. The model of
instanton-antiinstanton liquid formulated in
\cite{liquid,Diak}
 and
developed later in \cite{Shur1} is much better in this
respect. We refer
the reader to a recent comprehensive review \cite{SSrev} for
details and
just elucidate here main physical point of the reasoning.

\begin{figure}
\begin{center}
\begin{picture}(90,55)
\SetScale{2.845}
\CArc(10,10)(7.5,0,360)
\put(9.5,8){{\Large I}}
  \BCirc(30,20){7.5}
  \put(28.5,18){{\Large A}}
\BCirc(29,36){7.5}
\put(28.5,34){{\Large I}}
  \BCirc(12,30){7.5}
  \put(10.5,28){{\Large A}}
\BCirc(48,34){7.5}
\put(47.5,32){{\Large I}}
\BCirc(45,7){7.5}
\put(44.5,5){{\Large I}}
  \BCirc(60,10){7.5}
  \put(58.5,8){{\Large A}}
 \BCirc(63,30){7.5}
  \put(61.5,28){{\Large A}}
\BCirc(82,20){7.5}
\put(81.5,18){{\Large I}}
\BCirc(58,48){7.5}
\put(57.5,46){{\Large I}}
  \BCirc(44,50){7.5}
  \put(42.5,48){{\Large A}}
\BCirc(12,45){7.5}
\put(11.5,43){{\Large I}}
\end{picture}
\end{center}
\caption{Instanton-antiinstanton liquid.}
\label{liq}
\end{figure}
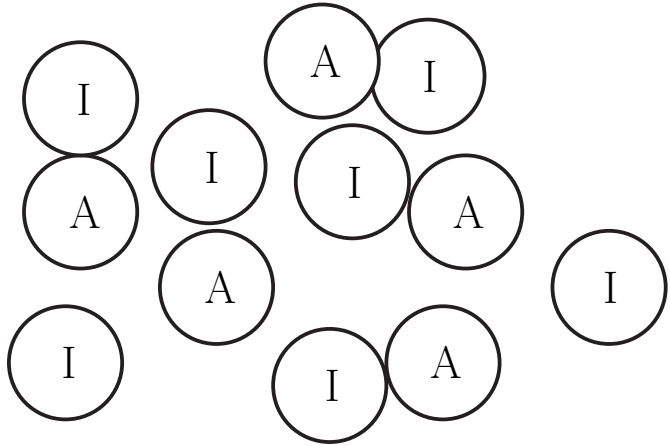

The basic assumption of the model is that a characteristic
gauge
field contributing in  $QCD$ path integral is a medium of
instantons and instantons as shown in Fig.\ref{liq}.  It is
not a ``gas''
of Callan, Dashen, and Gross \cite{CDG} because the
interaction
between instantons and antiinstantons bringing about a
short-range correlations between instanton positions and
orientations cannot be neglected. A ``liquid'' is a more
proper
term \cite{liquid}.

\newpage

\newpage

Each individual instanton and antiinstanton involves a
fermion
zero mode \cite{Hooft}. Assuming the constant density of
quasi-particles $\propto \mu_{hadr}^4$, the total number of
zero
modes in the Euclidean volume $V$ is $N\ = \ N_I + N_A \
\sim \ V\mu_{hadr}^4$.
However, these are not {\it exact} zero modes. They are
shifted
from zero due to interaction between instantons and
antiinstantons. In other words, individual instanton and
antiinstanton zero modes $\psi^{(0)}_{I,A}(x)$ are not
eigenfunctions of
the full Dirac operator $\not\!\!{\cal D} \ =\ \gamma_\mu
(\partial_\mu -
iA_\mu^{\rm IA\ liquid})$, where
  \be
 \label{liquid}
A_\mu^{\rm IA\ liquid}(x) \ =\ \sum_{i=1}^{N_I} A_\mu^I(x-
x_i) \ +
\ \sum_{j=1}^{N_A} A_\mu^A(x-x_j) \
  \ee
If choosing $\psi_I^{(0)}(x - x_i)$ and  $\psi_A^{(0)}(x -
x_j)$ as a basis
in the Hilbert space, the Dirac operator presents a matrix
with
off--diagonal elements

$$[\not\!\!{\cal D}]_{ji} \ \sim \ \int d^4x \
\psi_A^{\dagger (0)}(x - x_j)
\not\!\!{\cal D} \psi_I^{(0)}(x - x_i) $$
 \be
 \label{overlap}
 [\not\!\!{\cal D}]_{ij} \ \sim \ \int d^4x \
\psi_I^{\dagger (0)}(x - x_i)
\not\!\!{\cal D} \psi_A^{(0)}(x - x_j)
  \ee
Diagonalizing this matrix gives $N$ eigenvalues which are
not zero anymore,
but
are spread over the characteristic range
$\Delta \lambda \sim \mu_{hadr}$. Assuming a uniform spread,
the volume density
of {\it quasi-zero modes} is $\rho(0)
\sim N/(V\Delta \lambda) \sim \mu_{hadr}^3$. Due to
Eq.(\ref{Banks}), a nonzero quark condensate appears.
\footnote{The assumption of quasi-uniform spreading of
eigenvalues is not so
innocent. It probably holds only in the theory with several
light dynamical
quarks $N_f < 4,5$, but not in the quenched theory ($N_f =
0$) where it is
natural to
expect a singular behavior of the spectral density near
zero: $\rho(\lambda)
\sim 1/\lambda$ so that the ``fermion condensate'' (i.e. the
vacuum
expectation value $<\bar q q>_0$ where quark fields are
treated as external
sources) is infinite \cite{SMvac}. In the theory with $N_f >
4$, we expect
on the contrary a {\it suppression} of the spectral density
at
$\lambda \sim 0$ (see Sect.4.1 for detailed  discussion ).}

This picture is rather similar to what happens in a doped
semiconductor with high enough doping. When a characteristic
distance between individual atoms of the admixture is not
large,
the wave functions of outer electrons of these atoms
overlap,
and the electrons can jump from site to site. If the set of
atoms of admixture with a noticeable overlap of wave
functions
forms a connected network in the space, the electrons can
travel
through this network at large distances and the specimen is
a
{\it conductor}.  Note that it is not a standard metal
mechanism
of conductivity when the medium is a crystal, has the long-
range
order, and the electron wave functions are periodic Bloch
waves.
Here the distribution of the dope whose electrons are
responsible for conductivity is stochastic and wave
functions
are complicated.  The essential is that they are {\it
delocalized}.

Thus one can say that the vacuum of $QCD$ is the
``conductor'' in
a certain sense. For sure, there is no conductivity of
anything
in usual Minkowski space-time. Only the Euclidean vacuum
functional has ``conducting'' properties. In principle, one
can
introduce formally the fifth time and write an analog of
Kubo
formula for conductivity in $QCD$, but the physical meaning
of
this ``conductivity'' is not clear.  It is sufficient to say
that,
in a characteristic Euclidean gauge field background, the
eigenfunctions of Dirac operator corresponding to small
eigenvalues are delocalized. It would be interesting to
check the latter
statement explicitly for the lattice vacuum configurations
or in the
instanton model framework.  Instanton study of this question
is now in
progress \cite{Jacdel}.

What happens if we heat the system ? The effective coupling
constant $g^2(T)$ decreases, the action of individual
instantons
$S = 8\pi^2/g^2(T)$ increases, and the density of
quasi-particles $\propto \exp\{-S\}$ decreases.
Correspondingly, the
characteristic
value of off-diagonal matrix elements (\ref{overlap}) in
Dirac operator
decreases. Note also that the
 overlap integrals (\ref{overlap}) decrease as the
temperature increases
even if the distance between instanton and antiinstanton is
not changed.
The matter is that the individual fermion zero modes of
thermal instantons
\cite{Grossman}
 fall down exponentially $\sim \exp\{-\pi T r\}$ at large
spatial
distances, not just as a power $\sim 1/r^3$ as is the case
at $T = 0$. Both
effects lead to suppression of overlap matrix elements at
finite $T$.
(Concrete numerical calculations \cite{Schaf,SSrev} show
that the second
effect is more important in the temperature region $T \sim
200$ MeV. Speaking
of the instanton density, it falls down only by a factor 2
in this region.
 That means, in particular, that the value of the gluon
condensate
$<(\alpha_s/\pi) G^2>_T$ at the critical temperature is only
twice as small as
at zero temperature, and the gluon condensate is far from
being melted down.
This is another manifestation of the fact that the gluon
condensate is
{\it not}
an order parameter of the chiral restoration phase
transition. Neither
 it is an order parameter of the deconfinement phase
transition in pure
Yang--Mills theory.)

Let us look first at our doped semiconductor when we
decrease
the density of admixture. Below some critical density, the
set
of atoms with essential overlapping of wave functions does
not
form a connected network in 3-dimensional space anymore.
The
electrons can no longer travel far through this network,
wave functions become
localized, and the
conductivity drastically falls down. This is called the
percolation
 phase transition (see
\cite{Shkl} for detailed discussion).

Likewise, there is a critical temperature in $QCD$ above
which
instantons and antiinstantons do not form anymore a
connected
cluster with an essential overlap of individual fermion zero
modes (what overlap is ``essential'' and what is not  is a
numerical question.  For condensed matter systems,
  computer estimates for the critical
admixture density were performed long time ago.
Corresponding
calculations in thermal $QCD$ have been done only
recently \cite{Schaf}). At high temperatures, few
remaining quasi-particles tend to form ``instanton-
antiinstanton
molecules''. These molecules tend to be ``polarized'' in
imaginary time
direction (the overlap integrals and, correspondingly, the
``binding energy''
for a molecule polarized along imaginary
time axis is larger as  the zero modes involve an
exponential
suppression factor only in spatial, but not in Euclidean
time direction).

\begin{figure}
\begin{flushleft}
\begin{picture}(90,85)
\SetScale{2.845}
\Line(10,10)(155,10)
\Line(10,50)(155,50)

\BCirc(30,20){7.5}
\put(28.5,18){{\Large I}}
  \BCirc(34,40){7.5}
  \put(32.5,38){{\Large A}}

\BCirc(70,17.5){7.5}
\put(68.5,15.5){{\Large I}}
  \BCirc(67,42){7.5}
  \put(65.5,40){{\Large A}}

\BCirc(130,30){7.5}
\put(128.5,28){{\Large I}}
\CArc(132,50)(7.5,180,360)
  \put(137.5,43){{\Large A}}
\CArc(132,10)(7.5,0,180)
\put(135.5,16){{\Large A}}

\DashLine(10,10)(10,25){2}
\DashLine(10,35)(10,50){2}
\put(10,30){$\beta$}

\end {picture}
\end{flushleft}
\caption{ Gas of instanton-antiinstanton molecules (high
$T$).}
\label{molec}
\end{figure}
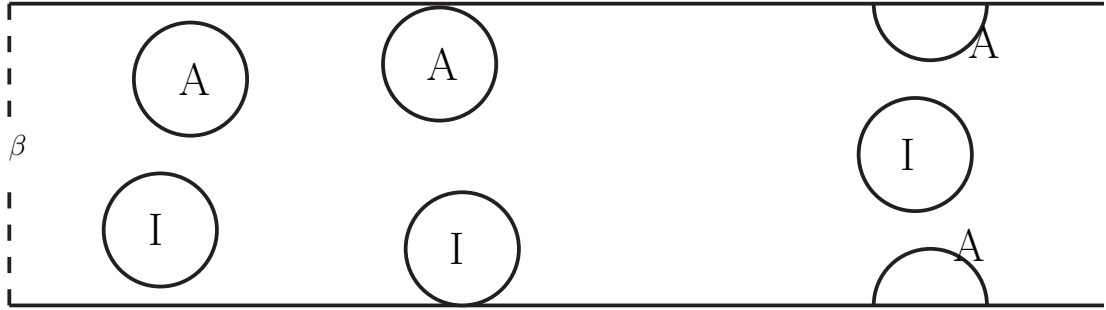

A characteristic Euclidean path integral configuration at
high temperature
(here ``high'' means just several hundred MeV)
is drawn schematically in Fig.\ref{molec}.
 In this picture, the individual zero modes are not {\it
spread out} uniformly in the range $\Delta \lambda \sim
\mu_{hadr}$ after diagonalization, as was the case at zero
temperature
where instantons
and antiinstantons formed an infinite cluster, but are just
{\it
shifted} by the value $\sim \mu_{hadr}$ due to interaction
in
individual molecules. Small eigenvalues in the spectrum of
Dirac
operator are absent and the fermion condensate is zero.

\newpage

\newpage

It is worthwhile to emphasize once more that this scenario
of percolation
phase transition leading to the molecular high-temperature
phase is expected
to hold only at $N_f \geq 2$. For $N_f =1$ with arbitrary
small but nonzero
fermion mass, molecules are effectively ionized even at very
high
temperatures and the ``medium'' presents a very dilute
instanton-antiinstanton gas --- the instanton density
involves a product of two
small factors: $\exp\{-8\pi^2/g^2(T)\}$ and the fermion mass
$m$.
Differentiating ${\rm log} Z$ over $m$ and sending $m$ to
zero, one obtains a
small but nonzero quark
condensate \cite{GPJ,KY}. Cf. the analogous situation in the
Schwinger model
\cite{inst}.
\footnote{Whether molecules {\it are} ionized or not depends
on the
particular quantity we are interested in. An accurate
statement is that,
for $N_f =1$, the
gas component gives a non-vanishing contribution to the
quark condensate in
the chiral limit. Thermodynamic quantities like the energy
density would
not depend on the gas component at all in the limit $m \to
0$ and would be
determined by ``molecular'' topologically trivial
configurations.
For $N_f = 2$, the gas
component does not contribute to the condensate in the
chiral limit, but
contributes to the so called scalar susceptibility which is
related to the
double derivative of the free energy over mass. The
susceptibility is small
at high temperatures  but never vanishes. In the theory with
3 massless
flavors, also the
susceptibility vanishes in the high temperature phase, but
the expectation
value $<\bar u u \ \bar d d\ \bar s s>_T$ is still nonzero.
Gas component is
responsible for the explicit $U_A(1)$ breaking via anomaly.
If we are specifically
interested in $U_A(1)$ breaking effects, we {\it always}
have to take it
into account. Disregarding it can lead to wrong results
\cite{Cohen}.}

\subsection{Disoriented chiral condensate.}

When we talked in previous sections about ``experimental''
tests
of theoretical predictions, we meant  numerical lattice
experiment.
However, speaking of the particular problem of the phase
transition in $QCD$ associated with chiral symmetry
restoration,
an intriguing possibility exists that a direct experimental
evidence for such a transition can be obtained at the
high-energy heavy ion collider RHIC which is now under
construction.

 After a head-on collision of two energetic heavy
nuclei, a  high temperature hadron ``soup'' is created.
We do not call this soup the quark-gluon plasma because,
even at
RHIC energies, the temperature $T_{\rm RHIC} \sim 0.5$ GeV
\cite{RHICT}
 would not be high enough to
provide a sufficient smallness of the effective coupling
$g^2(T)$ and to make the perturbation theory over this
parameter
meaningful. (As will be discussed in details in Chapter 6, a
proper
normalization point for the coupling constant is probably
$2\pi T$
rather than $T$. Still perturbative corrections to the
physical quantities
like the free energy density appear to be rather large at $T
\sim .5$ GeV,
 and the perturbative
series does not display a sign of convergence.)
What is important, however, is that, at RHIC
energies, the temperature of the soup would be well above
the
estimate (\ref{est}) for the phase transition temperature.
The
high-temperature state created in heavy nuclei collision
would
exist for a very short time, after which it expands, cools
down and decays eventually into  mesons.

\begin{figure}
\begin{center}
        \epsfxsize=300pt
        \epsfysize=0pt
       \epsfbox[0 320 600 600]{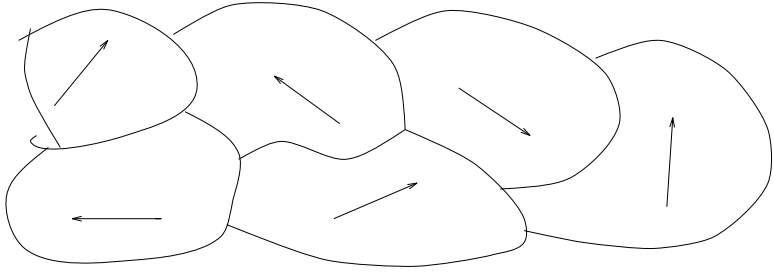}
    \end{center}
\caption{Domains of disoriented chiral condensate in cooling
hadron soup.}
\label{disor}
\end{figure}

Let us look in more details at the cooling stage. At high
temperature, the fermion condensate is zero. Below phase
transition, it is formed and breaks spontaneously chiral
symmetry. This breaking means that the vacuum state is not
invariant under the chiral transformations (\ref{chi}) and a
direction in isotopic space is distinguished. What
particular
direction --- is a matter of chance. This direction is
specified
by the condensate matrix
\be \label{condmat}
\Sigma_{f f'} = <\bar q_{Lf} q_{Rf'}>
  \ee
For simplicity, we have assumed up to now that the
condensate matrix is
diagonal $\Sigma_{ff'} = -\frac \Sigma 2 \delta_{ff'}$.
But any unitary matrix can be substituted for $\delta_{ff'}$
(of course, it can be brought back in the form
$\delta_{ff'}$ by
a chiral transformation ).  In different regions of space,
cooling occurs independently and flavor directions of
condensate are
not correlated. As a result, domains with different
directions
of condensate shown in Fig.\ref{disor} are formed (cf.
cooling down of a
ferromagnetic below the Curie point).

\newpage

\newpage

In our world, we do not observe any domains, however. The
direction of the condensate in all spatial points is
identical.
This is a consequence of the fact that $u$- and $d$- quarks
have
nonzero masses which break chiral symmetry explicitly, the
vacuum energy involves a term
\be \label{EMSig} E_{vac} \sim
{\rm Tr} \{{\cal M}^\dagger \Sigma \} + {\rm c.c.}
\ee and the
only true vacuum state is (\ref{diag}) (in the basis where
the
quark mass matrix ${\cal M}$ is diagonal).

However, the masses of $u$- and $d$- quarks are rather small
and
one can expect that the domains with ``wrong'' direction of
the
condensate are sufficiently developed during the cooling
stage
before they eventually decay into true vacuum (\ref{diag})
with
emission of pions.
\footnote{Fig.\ref{disor} implies the existence of several
domains
and describes better the physical situation immediately
after
the ``phase crossover'' in early Universe. Probably, the
size of
the hot fireball produced in collision of two nuclei is too
small and the cooling occurs too fast for several domains to
be
developed. The popular ``Baked Alaska'' scenario \cite{Bj}
implies
the formation of only one domain with (generally) wrong
flavor
orientation. }

This is a crucial assumption. A theoretic estimate of the
characteristic size of domains they reach before decaying is
very difficult and there is no unique opinion on this
question
in the literature. But if this assumption is true, we can
expect
to observe a very beautiful effect \cite{Bj,Rajag} (for a
recent
 review, see \cite{Rajag1}).  From the true
vacuum viewpoint, a domain with disoriented $\Sigma_{ff'}$
is a
classical object --- kind of a ``soliton'' (quotation marks
are put
here because it is not stable) presenting a {\it coherent}
superposition of many pions. The mass of this quasi-soliton
is
much larger than the pion mass. The existence of such multi-
pion
coherent states was discussed long ago in pioneer papers
\cite{Ans} but not in relation with thermal phase
transition.

Eventually, these objects decay into pions. Some of the
latter
are neutral and some are charged. As all isotopic
orientations
of the condensate in the domains are equally probable, the
{\it
average} fractions of $\pi^0$, $\pi^+$, and $\pi^-$ are
equal:
$<f_{\pi^0}> = <f_{\pi^\pm}> =
\frac 13$ as is also the case for incoherent production of
pions in, say, $pp$ collisions where no thermalized high-$T$
hadron soup is created.

But the {\it distribution} $P(f)$ over the fraction of, say,
neutral pions is quite different in the case of incoherent
and
coherent production. In incoherent case, $P(f)$ is a very
narrow
Poissonic distribution with the central value $<f_{\pi^0}> =
1/3$. The events with $f_{\pi^0} = 0$ or with $f_{\pi^0} =
1$
are highly unprobable: $P(0) \sim P(1) \sim
\exp\{ - C N \}$ where $N \gg 1$ is the total number
of pions produced.

For coherent production, the picture is quite different.
$\Sigma_{ff'}$ is proportional to a $SU(2)$ matrix.
Factorizing
over $U(1)$, one can define a unit vector in isotopic space
$\in
S^2$. The fraction of $\pi^0$ produced would be just $f =
\cos
^2 \theta$ where $\theta$ is a polar angle on $S^2$.  The
probability to have a particular polar angle $\theta$
normalized
in the interval $0 \leq \theta
\leq \pi/2$ [ the angles $\theta > \pi/2$ do not bring about
anything new as $f(\pi - \theta) = f(\theta)$ ] is
$P(\theta) =
\sin \theta$.  After an elementary transformation, we get a
normalized probability in terms of $f$:
\be \label{Pf}
P(f) df =
\frac {df}{2 \sqrt{f}}
\ee As earlier, $<f> = 1/3$, but the
distribution in $f$ is now wide and the values $f = 0$ and
$f =
1$ are quite probable.

Thus a hope exists that, in experiments with heavy ion
collisions at RHIC, wild fluctuations in the fractions of
neutral and charged pions would be observed. That would be a
direct experimental indication that a quasi-phase-transition
occurs where domains of disoriented chiral condensate of
noticeable size are developed in a cooling stage. One can
recall
in this respect mysterious Centauro events with anomalously
large fraction of neutral or of charged particles observed
in
cosmic ray experiments \cite{Cent}.
Cosmic ray experiment is not a heavy ion collision
experiment
and there are no
strong theoretical reasons to expect the formation of hot
hadron
 matter there (see, however, \cite{Bj}), but
who knows, may be that still {\it
was} the first experimental observation of the $QCD$ phase
transition ?

\unitlength=0.7mm
\section{Quark--gluon plasma.}
\setcounter{equation}0

The last  chapter of the review is devoted to the properties
of
high--temperature
phase, the quark--gluon plasma. It  is a very rich and
interesting physical
system which attracted lately a considerable attention of
theorists. The number
of the papers in SLAC archive involving the words ``quark--
gluon plasma''
in the title exceeds seven hundred (not yet exceeds right
now when I write it,
but will certainly pass this mark by the time you read it).
Thereby,
this chapter is longer than others and still many issues
were left undiscussed.
In particular, we will touch only superficially a
fascinating, but technical
subject of effective lagrangian of soft modes and of
nonabelian Vlasov
equations.
 Instead, we will discuss at some length different physical
phenomena
which are specific for QGP with an
emphasize on the problem of {\it physical observability} of
various
characteristics of QGP discussed in the literature.

\subsection{Static Properties of $QGP$: a bird eye's view.}
We start with discussing static (thermodynamic)
characteristics of QGP
and find it useful to do it in two ``rounds''. In this
section, we describe
gross physical features of the system and go down to details
in the next
one.

\vspace{.3cm}

\centerline{\it Thermodynamics}

\vspace{.3cm}

The basic thermodynamic characteristic of  a finite $T$
system is its free energy.  The lowest order results
are derived exactly in the same way as for free pion gas
(see Sect. 4.2).
We have for a pure gluon system
 \be
\label{Fg0}
 \frac{F^g}{V} = 2(N_c^2 -1) T \int \frac {d^3p}{(2\pi)^3}
\ln \left[ 1 - e^{-\beta |\vec{p}|} \right] =
- \frac{\pi^2 T^4}{45} (N_c^2 -1)
\ee
where $2(N_c^2-1)$ is the number of degrees of freedom of
the gluon field (the factor 2 comes due to two
polarizations).
 This is nothing else as the Stefan--Boltzmann formula
multiplied by the color factor $N_c^2-1$. The quark
contribution is obtained quite similarly. Taking the Pauli
principle
into account, we have instead of Eq.(\ref{Frpi})
 \be
\label{Fqdef}
F^q = -T \ln Z = -T \ln \left[ \prod_\vec{p} \left( 1 + e^{-
\beta n |\vec{p}|} \right)^{4N_c N_f} \right]
\ee
where $4N_c N_f$ is the number of degrees of freedom.
Trading
 the sum for the integral, we obtain
 \be
\label{Fq0}
 \frac{F^q}{V} \ = \ - 4N_c N_fT \int \frac {d^3p}{(2\pi)^3}
\ln \left[ 1 + e^{-\beta |\vec{p}|} \right] \ =\
 - \frac{7\pi^2 T^4}{180} N_c N_f
\ee
Notice that in our world with $N_c = N_f = 3$, the lowest
order
quark contribution  to the free
energy is roughly 2 times larger than the gluon one.

All other thermodynamic quantities of interest can be
derived
from the free energy by standard thermodynamic relations.
For example, the pressure just coincides with the free
energy with the sign reversed. The energy density is
  \be
 \label{Edef}
E = F - T \frac {\partial F}{\partial T}
 \ee
For massless particles in the lowest order the relation
 \be
 \label{EFP}
E = 3P = -3F
 \ee
holds.

\vspace{.3cm}

\centerline{\it Debye screening.}

\vspace{.3cm}

Consider the gluon polarization operator in $QGP$ with
account
of thermal loop corrections. It is transverse, $k_\mu
\Pi_{\mu\nu}(k) = 0$. At zero temperature, transversality
and Lorentz-invariance dictate the form $\Pi_{\mu\nu}(k) =
\Pi(k^2) (g_{\mu\nu} - k_\mu k_\nu /k^2)$. At finite $T$,
Lorentz-invariance is lost and the polarization operator
presents a combination of two different (transverse and
longitudinal) tensor structures. Generally, one can write
\be
\label{tens}
\Pi_{00} = \Pi_l(\omega, |\vec{k}|) \nonumber \\
\Pi_{i0} = \frac{k_i\omega}{\vec{k}^2} \Pi_l(\omega,
|\vec{k}|) \nonumber \\
\Pi_{ij} = - \Pi_t(\omega, |\vec{k}|) (\delta_{ij} - k_ik_j
/\vec{k}^2) + \frac{\omega^2}{\vec{k}^2} \frac {k_i
k_j}{\vec{k}^2} \Pi_l(\omega, |\vec{k}|)
\ee
Consider first the longitudinal part of the polarization
operator in the kinematic region where $\omega$ is set to
zero in the first place after which $k \equiv |\vec{k}|$ is
also sent
to zero. By the reasons which will be shortly seen, we
denote this quantity $m_D^2$:
  \be
  \label{mDdef}
m_D^2 = \lim_{k \to 0} \Pi_l(0, k)
 \ee
To understand the physical meaning of this quantity,
consider the correlator
  \be
\label{screen}
 <A_0^a(\vec{x}) A_0^b(0)> \sim \delta^{ab} \int \frac
{d\vec{k}}{\vec{k}^2 + m_D^2} e^{i\vec{k} \vec{x}} \propto
e^{-m_D |\vec{x}|}
 \ee
Thus $m_D$ coincides with the inverse screening length of
chromoelectric potential $A_0$.

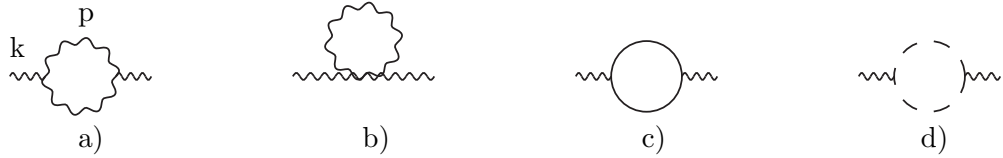
\begin{figure}
\begin{center}
\SetScale{1.333}
\begin{picture}(200,50)(0,0)
\Photon(0,20)(10,20){1}{3}
\Photon(30,20)(40,20){1}{3}
\PhotonArc(20,20)(10,0,360){1}{10}
\put(13,0){a)}
\put(0,17){k}
\put(13,24){p}

\Photon(80,20)(120,20){1}{10}
\PhotonArc(100,30)(10,0,360){1}{10}
\put(67,0){b)}

\Photon(160,20)(170,20){1}{3}
\Photon(190,20)(200,20){1}{3}
\CArc(180,20)(10,0,360)

\Photon(240,20)(250,20){1}{3}
\Photon(270,20)(280,20){1}{3}
\DashCArc(260,20)(10,0,360){5}

\put(120,0){c)}
\put(173,0){d)}
\end{picture}

\vspace{0.5cm}
\caption{Gluon polarization operator in one loop.}
\label{Gl1loop}
\end{center}
\end{figure}

There is a clear analog with the usual plasma. A static
electric charge immersed in the plasma is screened by the
cloud of ions and electrons so that the potential falls down
exponentially $\propto \exp\{-r/r_D\}$ where the {\it Debye
radius} $r_D$ in an ordinary nonrelativistic
plasma is given  by the expression (see e.g. \cite{LP})
 \be
\label{rD}
r_D^{-2} = \frac {4\pi n e^2}T + \frac {4\pi n (Ze)^2}T
 \ee
where $n$ is the electron and ion density, $Z$ is the ion
charge (so that the second term describes the ion
contribution), and,  to
avoid unnecessary complications,
  we assumed that the electron and ion
components of the plasma have the same temperature $T$
Calculating the one-loop thermal contribution to the gluon
polarization operator (see Fig.\ref{Gl1loop}), one can
easily obtain an
analogous
formula for $QGP$:
 \be
 \label{mD}
m_D^2 = r_D^{-2} = \frac {g^2 T^2}3 \left(N_c  + \frac
{N_f}2 \right)
 \ee
where the first term describes the screening due to thermal
gluons and the second term --- the screening due to thermal
quarks.
The result (\ref{mD}) was  first obtained
 by Shuryak \cite{ShurD}. It has the same
structure as (\ref{rD}) (Note that the density of particles
in ultrarelativistic plasma is expressed via the
temperature, $n \propto T^3$).
It is worth mentioning that the Debye screening is
essentially a classical effect and not only quarks but also
gluons induce  screening rather than antiscreening. That
should be confronted with the famous antiscreening of the
charge in Yang-Mills theory at zero temperature due to
quantum effects.

\newpage

\newpage

In $QED$, the correlator $<A_0(\vec{x}) A_0(0)>_T$ is a
gauge
invariant object and also the notion of charge screening is
unambiguous and well defined --- there {\it are} classical
electric charges and one can measure the potential of such a
charge immersed in plasma by standard classical devices. Not
so in QCD. We do not have at our disposal classical color
charges due to confinement and the experiment of measuring
the chromoelectric field of a test color charge cannot be
carried out even in principle. Also the gluon polarization
operator which enters the definition (\ref{screen}) of  a
Debye screening mass is generally speaking gauge-dependent.
Thus a question of whether the Debye screening mass is a
physical notion in a nonabelian theory is fully legitimate.

 The answer to this question is positive. But with
reservations.

Note first of all that though the gluon polarization
operator is generally gauge-dependent, the Debye screening
mass defined in (\ref{mDdef}) does not depend on the gauge
{\it in the leading order} (we shall discuss also nonleading
corrections in the next section). That suggests that
the result (\ref{mD}) has an invariant physical meaning.

Indeed, one can consider a gauge-invariant correlator
  \be
  \label{CP}
C(\vec{x}) = <P(\vec{x}) P^*(0)>_T
  \ee
where
  \be
  \label{Pdef}
P(\vec{x}) = \frac 1{N_c} {\rm Tr} \exp \left\{
i \int_0^\beta A_0(\vec{x}, \tau) d\tau \right\}
  \ee
is the Polyakov line [both the Polyakov line  and the
correlator
(\ref{CP})  were  discussed at length in Chapter 3].
If $r = |\vec{x}|$ is not too large (the exact meaning of
this will be specified later), the  connected part of the
correlator
$C_c(r) = C(r) - C(\infty)$ (as was explained in Chapter 3,
$C(\infty)$ is
a nonzero constant) has an
exponential behavior
  \be
  \label{CmDr}
C_c(r)\ \propto \ [G_{00}(r)]^2\ \propto \ \frac 1{r^2}
\exp\{-2m_Dr\}
  \ee
The correlator (\ref{CP}) can be attributed a physical
meaning. $-T \ln C(r)$ coincides with the change of free
energy of the system when putting there a pair of heavy
quark and antiquark at distance $r$, the averaging over
color spin orientations being assumed \cite{Larry}.

Also we shall see in the next section that the perturbative
corrections $\sim g^3T^4$ to a perfectly well defined and
physical quantity --- the free energy --- depend directly on
the value of $m_D$.

Thus the notion of Debye screening mass in $QGP$ is
physical, though, unfortunately, not to the same extent as
it is in the usual plasma. The correlator (\ref{CP}) with
the correlation length of fractions of  a fermi cannot be
directly measured even in a {\it gedanken} laboratory
experiment --- at least, we cannot contemplate such an
experiment. But it can be measured in lattice numerical
experiments which is almost as good. We shall see later that
a number of other characteristics of $QGP$ have a similar
semi-physical status.

\vspace{.3cm}

\centerline{\it Magnetic  screening.}

\vspace{.3cm}

Debye mass describes the screening of static electric
fields. In abelian plasma, magnetic fields are not screened
whatsoever. Such a screening could be provided by magnetic
monopoles, but they are not abundant in Nature. The absence
of monopoles is technically related to one of the Maxwell
equations $\partial_i B_i = 0$. In the nonabelian case, the
corresponding equation reads ${\cal D}_i B_i = 0$ where
$\vecg{\cal D}$ is a covariant derivative. Thus gluon field
configurations with local color magnetic charge density
$\rho_m \sim \partial_i B_i$ (with usual derivative) are
admissible. The presence of such configurations in the gluon
heat bath results in screening of chromomagnetic fields.

Let us see how it comes out in a perturbative calculation.
Consider the one-loop graph in Fig.\ref{Gl1loop}a which
contributes to
the polarization operator of the spatial components of the
gluon field $\Pi_{ij}(0, \vec{k}) \propto \Pi_t(0,
k)$. At high temperatures, it suffices to take into
account only the lowest Matsubara frequency (with $\omega
= 0$) in gluon propagators. We have
  \be
  \label{Pitr}
\Pi_t(0, k) \sim g^2 k^2 T \int \frac {d^3p}
{\vec{p}^2 (\vec{p} - \vec{k})^2}  \sim g^2 T k
  \ee
The numerical coefficient can be explicitly calculated, but
it depends on the gauge and makes as such a little sense.
The loop integral is determined by the low momentum region
$|\vec{p}|_{char} \sim k \ll T$ (as we are
interested in the large distance behavior of the gluon
Green's function, we keep $k$ small). Note that the
Feynman integral for $\Pi_l(0, k)$ has a completely
different behavior being saturated (in the leading order) by
the loop momenta $|\vec{p}| \sim T$ --- it is a so called
{\it hard thermal loop}.

The transverse part of the gluon Green's function is
\be
\label{Gtran}
 G_t(0, k) \sim \frac 1{\vec{k}^2 + \Pi_t(0, k)}
\ee
  We see that at $k \sim g^2 T$ the one-loop contribution to
the polarization operator is of the same order as the tree
term $k^2$. One can estimate a two-loop contribution to
$\Pi_t(0, k)$ which is of order $(g^2 T)^2$. The factor
$T^2$ here comes from two loops (we use the rule
(\ref{intT}) and take into account only the contribution of
the lowest Matsubara frequency in each loop).
Similarly, a three-loop contribution to $\Pi_t(0,
k)$ is of order $(g^2 T)^3/k$, a four-loop contribution is
of order $(g^2 T)^4/k^2$ etc. (The growing powers of $k$ in
the denominator are provided by infrared 3-dimensional loop
integrals. Infrared integrals  may also provide for a
logarithmic singularity in external momentum $k$ in the two-
loop
and higher loop contributions, but  weak logarithmic factors
 are of no concern for us here).
At $k \sim g^2T$, all contributions are of the same order
and
the perturbation theory breaks down \cite{Linde}.

There is an alternative way to see it. Let us write down the
expression for the partition function of the pure glue
system at finite temperature:
  \be
  \label{ZglT}
Z = \int \prod dA^a_\mu (\vec{x}, \tau) \exp \left\{
- \frac 1{4g^2} \int_0^\beta d\tau \int d^3x
(F^a_{\mu\nu})^2 \right\}
  \ee
When $T$ is large and $\beta$ small, the Euclidean time
dependence of the fields may be disregarded. Also the
effects due to $A_0(\vec{x}, \tau)$ can be disregarded ---
time components of the gluon field acquire the large mass
$m_D \sim gT \gg g^2T$ and decouple. We are left with the
expression
  \be
  \label{Z3}
Z = \int \prod dA^a_i (\vec{x}) \exp \left\{
- \frac 1{4g^2T} \int d^3x (F^a_{ij})^2 \right\}
  \ee
A theory with quarks is also reduced to (\ref{Z3}) in this
limit --- the fermions have high Matsubara frequencies $\sim
T$ and decouple. The partition function (\ref{Z3}) describes
a nonlinear 3D theory with the dimensional coupling
constant $g_3^2 \sim g^2T$
\footnote{One should be careful here. It would be wrong to
use the expression (\ref{Z3}) for calculating, say, the free
energy of $QGP$ at large $T$. The latter takes the
contributions from hard thermal loops (involving also
fermions !) with momenta of order $T$ which are not taken
into account in Eq. (\ref{Z3}). Eq. (\ref{Z3}) should be
understood as an {\it effective} theory describing soft
gluon modes with momenta of order $g^2T$.}.
No perturbative calculations in this theory are possible.

In $QED$, there is no magnetic screening (and the effective
3-dimensional theory is trivial). Presumably, perturbative
calculations in hot $QED$ can be carried out at any order,
though this question is not {\it absolutely} clear.

What is the physical meaning of this nonabelian screening ?
Can it actually be measured ?

The way we derived it, the magnetic screening shows up in
the large distance behavior of the spatial gluon propagator
$G_t(r) \propto \exp\{ik^* r\}$ where $k^*$ is the solution
to the  equation
  \be
  \label{disp}
k^2 + \Pi_t(0, k) = 0
  \ee
  There is no reason to expect that $k=0$ is a solution ---
the behavior of $\Pi_t(0, k)$ in the limit $k \to 0$ where
the
perturbation theory does not work is not known, but one can
tentatively guess that it tends to a constant $\sim
(g^2T)^2$ (As we have seen, $g^2T$ is the only relevant
scale in this limit). If so, $k^* \sim ig^2T$ and $G_t(r)$
falls down
exponentially $\propto \exp\{-Cg^2Tr\}$ at large distances.
Hence the term ``magnetic screening''.

Unfortunately, the gluon polarization operator is a gauge-
dependent
quantity and, would the God provide us with the
exact expression for $\Pi_t(0, k)$ in any gauge, even He
could not guarantee that the solution of the dispersive
equation (\ref{disp}) would be gauge-independent
\footnote{cf. Eq.(\ref{Siggauge}) and the related discussion
in sect. 6.4.}.

What one can do, however, is to consider the correlator of
chromomagnetic fields $C_{ab}(\vec{x}) \ =\
<\vec{B}^a(\vec{x}) \vec{B}^b(0)>_T$.
This correlator also depends on the gauge in a nonabelian
theory, but the gauge dependence amounts to rotation in
color
space: $C_{ab}(\vec{x}) \to \Omega_{a'a}(\vec{x})
C^{a'b'}(\vec{x}) \Omega_{b'b}(0)$ and cannot affect the
exponential behavior of the propagator. This is the way the
magnetic mass is usually measured on the lattices.

One can do even better considering gauge-invariant
correlators, the simplest one is \ \ $<G_{\mu\nu}^2(\vec{x})
G_{\mu\nu}^2(0)>_T$. Then the quantity
  \be
  \label{mudef}
\mu = - \lim_{r \to \infty} \frac 1r \ln
<G_{\mu\nu}^2(\vec{x})
G_{\mu\nu}^2(0)>_T
  \ee
provides an invariant definition of the magnetic screening
mass. The indefinite article is crucial here. Choosing other
gauge-invariant correlators, one would get other invariant
definitions. For example (and this will be important in the
following discussion), the true correlation length of the
correlator of Polyakov loops (\ref{CP}) is also of order
$(g^2T)^{-1}$. The asymptotics (\ref{CmDr}) is an {\it
intermediate} one and holds only in the region
  \be
\label{range}
  (gT)^{-1} \ll r \ll (g^2T)^{-1}
  \ee

\begin{figure}
\begin{center}
\SetScale{2}
\begin{picture}(200,60)(0,0)
\Vertex(50,30){1}
\Vertex(150,30){1}

\PhotonArc(65,30)(15,0,360){1}{10}
\PhotonArc(135,30)(15,0,360){1}{10}

\DashCArc(100,-4.4)(40,60,120){3}
\DashCArc(100,64.4)(40,240,300){3}
\end{picture}
\end{center}
\caption{Correlator of Polyakov loops at large distances.
Wavy lines are electric gluons and dashed lines are magnetic
gluons.}
\label{PolNadk}
\end{figure}
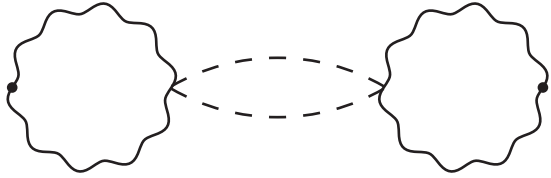

What was wrong with our previous derivation ? The matter is
we took into account earlier only the thermal corrections to
the gluon propagator and tacitly neglected corrections to
the vertices. Ab accurate analysis \cite{Nad} shows that
this is justified in the range (\ref{range}) but not beyond.
An example of the graph providing the leading contribution
in $<P(\vec{x}) P^*(0)>_T$ at $r \gg (g^2T)^{-1}$ is given
in Fig.\ref{PolNadk}.
\footnote{
The  asymptotics of the correlator of Polyakov loops at
large
enough $r$ is determined by the magnetic photon exchange
also in abelian plasma. Two magnetic photons can be coupled
to $P(x)$ via a fermion loop. In abelian case, magnetic
photons are massless and, as a result, the correlator has a
power asymptotics $\propto 1/r^6$ \cite{Arnold}. Physically,
it corresponds to Van-der-Vaals repulsion between the clouds
of virtual  electrons and positrons formed near two heavy
probe
charges.
Seemingly, the asymptotics $\propto 1/r^6$ for the Polyakov
lines correlator in $QGP$ found in \cite{Wong} has this
origin. But in nonabelian case, the asymptotics becomes
exponential when taking into account the magnetic screening
effects.}

\newpage

\newpage

Thus the ``experimental status'' of the magnetic mass
(\ref{mudef}) and of other similar quantities is roughly the
same and even better than for the Debye mass. The
exponential behavior $\propto \exp\{-\mu r\}$ of the
correlators is expected to hold at any large $r$ and the
value $\mu$ can in principle be determined in lattice
experiments with any desired accuracy. On the other hand,
the Debye screening mass cannot be determined with an
arbitrary accuracy due to the finite range (\ref{range})
where the
asymptotics (\ref{CmDr}) holds (being sophisticated enough,
an invariant
definition of Debye mass still can be suggested for $N_c
\geq
3$ --- see the
discussion in the following section). And, as was also the
case
for the Debye mass, we cannot invent any laboratory
experiment where the magnetic mass could be directly
measured.

Speaking of the lattice experiment, it can be done and
actually
{\it has} been done very recently  by the Bielefeld group.
Electric
and magnetic screening lengths were measured in the high
temperature
region up to $T \sim 10^6 {\rm GeV}$. Preliminary data
\cite{kar}
show a slow falloff of the quantities $m_D/T$ and $\mu_{\rm
magn}/T$
with temperature in a qualitative agreement with the
expected
behavior $m_D/T \sim g(T)$ and $\mu_{\rm magn}/T \sim
g^2(T)$.

\subsection{Static Properties of $QGP$: perturbative
Corrections.}

\vspace{.3cm}

\centerline{\it Debye mass.}

\vspace{.3cm}

To provide a smooth continuation of the discussion started
at the end of the previous section, we consider first
higher-loop effects in the Debye mass. Note first of all
that the
definition (\ref{mDdef}) is not suitable anymore.
Higher-loop corrections $\Delta m_D \sim g^2T$ to the Debye
mass as
defined in Eq.(\ref{mDdef}) depend on the gauge
\cite{Kaj1}. A better way is to define the Debye mass as
the solution of the dispersive equation $k^2 + \Pi_l(0,k) =
0$. In other words, we define \cite{Rebhan}
 \be
 \label{mDReb}
m_D^2 = \lim_{k^2 \to -m_D^2} \Pi_l(0,k)
  \ee
The longitudinal part of the gluon Green's function
$$
G_l(0,k) = \frac 1{k^2 + \Pi_l(0,k)}
$$
has then the pole at $k = im_D$. Pole
position has the ``tendency'' to remain  gauge-invariant
even though the Green's function
itself is not. We will see in the following
section that there {\it are} cases when formal arguments
displaying gauge-invariance of the pole position fail in
higher orders of perturbation theory due to severe infrared
singularities. Speaking of Debye mass as defined in
Eq.(\ref{mDReb}),
its gauge invariance can be shown in the order $\sim g^2T
\ln (C/g)$,
but the constant under the logarithm is not computable in
perturbation theory in principle, and it is difficult to
pose
a question whether it is gauge--invariant or not.
  In the next-to-leading order, the (gauge-invariant)
result is \cite{Rebhan}
  \be
  \label {mDcorr}
m_D \ =\ m_D^{(0)} + \frac{N_c g^2 T}{4\pi} \ln (C/g)
  \ee
with logarithmic accuracy. As was mentioned, the constant
$C$ is
not computable by perturbation theory.
We see that the correction is non-analytic in coupling
constant. The non-analyticity appears due to bad infrared
behavior of the loop integrals --- they involve a
logarithmic infrared divergence and depend on the low
momenta cutoff which is of order of magnetic mass scale
$g^2T$. Thus we have
$$
\Delta m_D^2 \propto \int_{\mu_{mag}}^{m_D} \frac {dp}p
= \ln {m_D \over \mu_{mag}} \sim \ln {1 \over g}
$$
The magnetic infrared cutoff is actually provided by higher-
order
graphs --- the orders of perturbation theory are mixed
up and a pure two-loop calculation is not self-consistent.
As we have seen, $\mu_{mag}$ cannon be determined
analytically which means that the correction $\sim g^2T$
without the logarithmic factor in the Debye mass cannot be
determined analytically.

Also the correction $\sim g^2T$  cannot be ``experimentally
observed''. Really, we have seen that the invariant physical
definition of $m_D$ refers to the correlator of Polyakov
loops (\ref{CP}). The latter displays the exponential
behavior (\ref{CmDr}) in the limited range (\ref{range}).
But that is tantamount to saying that the correction $\sim
g^2T$ in the Debye mass cannot be determined from the
correlator (\ref{CP}). To do this, one should probe the
distances $r \gg (g^2T)^{-1}$ where the correlator has a
completely different behavior being determined by the
magnetic scale. The correction (\ref{mDcorr}) is still
observable, however, due to a logarithmic enhancement
factor.

To be more precise, the correlator in the range
(\ref{range}) has the form \cite{Nad}
 \be
  \label{CPcorr}
C_c(r) \ =  \ \frac {N_c^2 - 1}{8N_c^2}\frac {g^4}{(4\pi r
T)^2} \exp \left\{
-2m_D^{(0)} r - \frac {N_c g^2 T r}{4\pi} \ln (m_D^{(0)}r)
\right\}
 \ee
where $m_D^{(0)}$ is the lowest order Debye mass
(\ref{mD}).

\begin{figure}
\begin{center}
\SetScale{2}
\begin{picture}(200,60)(0,0)
\Vertex(50,30){1}
\Vertex(80,30){1}
\put(65,5){a)}
\PhotonArc(65,30)(15,0,360){1}{10}
\DashCArc(65,45)(15,210,330){2}

\put(114,30){${\Large \frac 12}$}
\Vertex(150,30){1}
\Vertex(120,30){1}
\put(135,5){b)}
\DashLine(135,15)(135,45){2}
\PhotonArc(135,30)(15,0,360){1}{10}

\end{picture}
\end{center}
\caption{Correlator of Polyakov loops in next--to--leading
order.
Wavy lines are electric gluons with Debye mass
(\protect\ref{mD}) taken into account.
Dashed lines are magnetic gluons.}
\label{NadPP}
\end{figure}
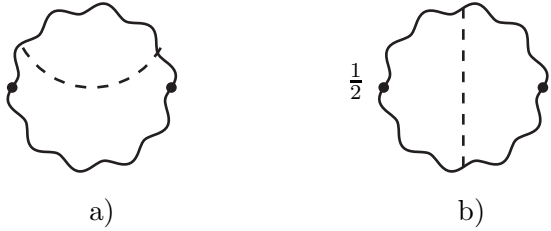

It is not just
  \be
  \label{corrnaiv}
C_c(r) \ \propto \ \exp\{-2m_D r\}
  \ee
, with $m_D$ being defined
in (\ref{mDcorr}), by two reasons.
First, at finite $r \ll (g^2T)^{-1}$, the infrared
cutoff in the loop integrals is provided by $r^{-1}$ rather
than $\mu_{mag}$ and we have
  \be
\label{recmD}
 \ln \frac 1g \sim \ln {m_D \over \mu_{mag}} \to \ln (m_Dr)
  \ee
Second, there are two kinds of graphs involving the
corrections both
in the gluon propagator and in the vertex (see
Fig.\ref{NadPP}).

\newpage

\newpage

 The graph in Fig.\ref{NadPP}a would give the correction
 \be
  \label{acor}
\frac{C_c(r)}{C_c^{(0)}(r)} \ \sim \ 1 - \frac {N_c g^2 T
r}{2\pi} \ln (m_D r)
  \ee
with logarithmic accuracy. It corresponds to
Eq.(\ref{corrnaiv}) where
the substitution (\ref{recmD}) is made. The second graph
also contributes
in this order and gives a positive correction with a twice
as small
coefficient \cite{Nad}. Exponentiating the one loop
correction
\footnote{ That the leading corrections
$\sim [g^2 T r \ln (m_D r)]^n$ exponentiate is an educated
guess.
To the best of our
knowledge, nobody has checked it so far explicitly.}, we
arrive at the result
(\ref{CPcorr}). Note that
the Fourier image of the correlator (\ref{CPcorr}) does {\it
not}
have a singularity at finite $k$ whatsoever. Thus the Debye
mass pole
in the gluon propagator does not show up as a pole in the
gauge-invariant correlator of Polyakov loops. Thereby, the
quantity
(\ref{mDcorr}) is not quite physical in spite of its gauge
invariance.

But, anyway, in the theoretical limit $\ln [1/g(T)] \to
\infty$, there is
a narrow range of $r$  where the condition $r \ll (g^2T)^{-1}$ is
fulfilled so that the intermediate asymptotic law
(\ref{CPcorr}) holds whereas the correction in the exponent
$\propto gm_D^{(0)} r \ln (m_D r)$ is of order 1
and can in principle  be singled out in numerical lattice
experiment.

There are two interesting recent proposals to define Debye
mass non-perturbatively in a gauge-invariant way
\cite{Arnold}. The first one is to consider the correlator
of imaginary parts of Polyakov lines in a complex
representation. If we are interested in a pure Yang--Mills
theory,
an additional requirement for the representation to be
invariant under the action of the
center of the group $Z_N$ (such is, say, the decouplet
representation
in $SU(3)$) should be imposed.
\footnote{The authors of Ref.\cite{Arnold} argued the
necessity to consider only the
$Z_N$ - invariant representations saying that otherwise the
correlators
give zero after averaging over different ``$Z_N$ - phases''.
Actually, as we have seen in Chapter 3, such
phases do not exist in Nature and a proper argumentation
should be that the
correlators which are not invariant under $Z_N$
transformations just do not have
a physical meaning \cite{bub}.}.
The point is that $P^*_{10} - P_{10}$ is odd
under Euclidean time reversion and cannot be coupled to only
magnetic gluons. That means that the correlator
 \be
 \label{CP10}
<P^*_{10}(\vec{x}) - P_{10}(\vec{x}),\ \ P^*_{10}(0) -
P_{10}(0)>_T
 \ee
exhibits the Debye screening falloff even at arbitrary large
$r$.
 In the intermediate region
(\ref{range}), the correlator (\ref{CP10}) is described by
the
graph with  exchange of 3 electric gluons and falls down  as
 $\propto \exp\{-3m_Dr\}$. At very large distances, the
dominant graph
corresponds to the exchange of {\it one} electric gluon
which is odd under
the time reversion + a cloud  of light magnetic gluons which
neutralize
the color. Debye mass extracted from the exponential falloff
of such a
correlator at large distances has a gauge invariant meaning,
but depends,
generally speaking, on the choice of the representation
(probably, not
in the order $g^2T \ln(1/g)$). Also this definition does not
work for $SU(2)$ where the Polyakov line is real
in any representation.

Another suggestion was to study the behavior of large
spatial
Wilson loops in {\it adjoint} color representation. At large
distances, the theory is effectively reduced to the 3-
dimensional
YM theory (\ref{Z3}). Adjoint color charges in
this theory are not confined but rather screened, and the
Wilson loop exhibits the perimeter law behavior
  \be
  \label{W3per}
W(C) \propto \exp \{ -m^* \times {\rm perimeter}(C)\}
\ee
One can show that $m^*$  is of order $g^2 \ln(1/g)T$  and
coincides with a
perturbative correction to the Debye mass in (\ref{mDcorr}).
The problem
here is that $m^*$ has no trace of the lowest order
contribution (\ref{mD}).
 Still, $m^*$ certainly has an invariant
meaning and is as such an interesting quantity to study.

\vspace{.3cm}

\centerline{\it Free energy.}

\vspace{.3cm}

This is the most basic and physical quantity of all. May be
this is the reason why perturbative corrections are known
here with record precision.

\newpage

\begin{figure}
\begin{center}
\SetScale{2}
\begin{picture}(200,60)(0,0)

\PhotonArc(15,30)(15,0,360){1}{10}
\PhotonArc(45,30)(15,0,360){1}{10}

\DashCArc(100,30)(25,0,360){4}
\Photon(100,5)(100,55){1}{7}

\CArc(170,30)(25,0,360)
\Photon(170,5)(170,55){1}{7}
\end{picture}
\end{center}
\caption{ Free energy in two loops.}
\label{FQGP2}
\end{figure}
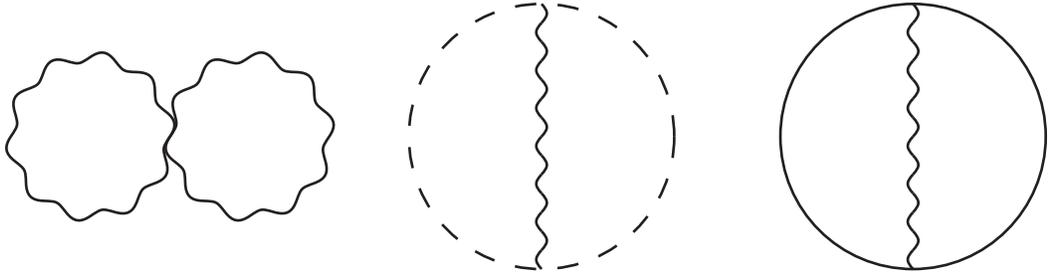

\newpage

The correction $\sim g^2 T^4$ has been found by Shuryak
\cite{ShurD}. To this end, one should calculate the two-loop
graphs depicted in Fig.\ref{FQGP2}. The behavior of Feynman
integrals
in this order is quite benign and no particular problem
arises.

\newpage

\begin{figure}
\begin{center}
\SetScale{2}
\begin{picture}(200,80)(0,0)

\PhotonArc(100,30)(30,0,360){1}{10}
\Photon(85,4)(85,56){1}{7}
\Photon(115,4)(115,56){1}{7}
\put(100,63){k}

\end{picture}
\end{center}

\caption{ An infrared-divergent 3-loop graph in free
energy.}
\label{FQGP3}
\end{figure}
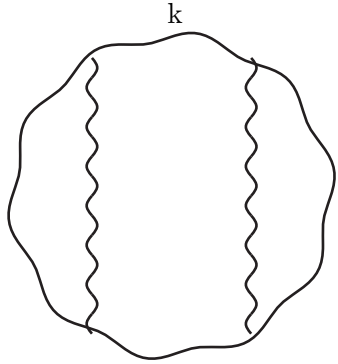

\newpage

\begin{figure}
\begin{center}
\SetScale{2}
\begin{picture}(200,100)(0,0)

\PhotonArc(50,50)(30,0,360){1}{10}
\GCirc(50,20){10}{0.5}
\GCirc(50,80){10}{0.5}

\PhotonArc(130,50)(30,0,360){1}{10}
\GCirc(130,20){10}{0.5}
\GCirc(104,65){10}{0.5}
\GCirc(156,65){10}{0.5}

\put(90,50){+}
\put(180,50){$+\ldots$}

\end{picture}

\end{center}
\caption{ Ring diagrams. Grey circles stand for a set of
one-loop graphs in Fig.\protect\ref{Gl1loop}}
\label{ring}
\end{figure}
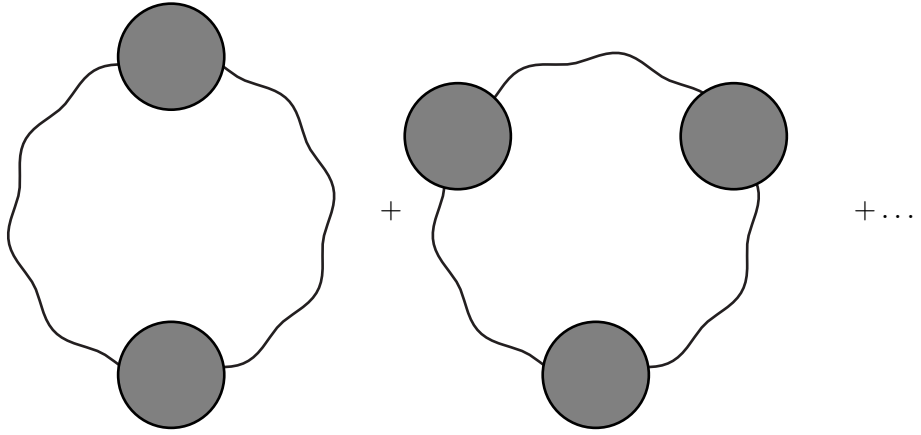

However, starting from the three-loop level, the problems
crop up. The
contribution of the graph depicted in Fig.\ref{FQGP3} is
infrared
divergent:
  \be
  \label{F5}
\Delta F^{Fig.\ref{FQGP3}} \sim g^4 T^5 \int \frac{dk}{k^2}
 \ee
That means that a pure 3-loop calculation is not self-
consistent and, to get a finite answer, one has to resum a
set of infrared-divergent graphs in all orders of
perturbation theory. This is, however, not a hopeless
problem, and it has been solved by Kapusta back in 1979
\cite{Kapusta}. The leading infrared singularity is due to
the so called ``ring diagrams'' depicted in Fig.\ref{ring}.

\newpage

\newpage

The sum of
the whole set of ring diagrams has the form
  \be
  \label{Fring}
F^{ring} \sim T \int d^3k \left\{ \ln\left[1 +\frac
{\Pi_l(0,k)}{k^2} \right] -  \frac {\Pi_l(0,k)}{k^2}
\right\}
\ee
The expansion of the integrand in $
{\Pi_l(0,k)}/{k^2}$ restores the original infrared-divergent
integrals. However, the whole integral is convergent being
saturated by momenta of order $k \sim [\Pi_l(0,0)]^{1/2}
\sim gT$. Neglecting the $k$-dependence in the polarization
operator, we obtain
   \be
  \label{F3}
F_{3} \sim T \int d^3k \left[ \ln\left(1 +\frac
{m_D^2}{k^2} \right) -  \frac {m_D^2}{k^2} \right] \sim
Tm_D^3 \sim g^3T^4
\ee
We see that the correction is non-analytic in the coupling
constant $\alpha_s$  which exactly reflects the fact that
the individual
graphs diverge and the orders of perturbation theory are
mixed up. Note, however, that infrared divergences here are
of comparatively benign variety --- the integral depends on
the scale $k^{char} \sim gT$ and we are far from the land of
no return $k \sim g^2T$. That is why an analytic
determination of the coefficient in (\ref{F3}) is possible.

The next correction has the order $\sim g^4 T^4 \ln(1/g)$.
It comes from the same ring graphs of Fig.\ref{ring} where
now the
term $\sim g^2Tk$ in $\Pi_l(0,k)$ should be taken into
account. This correction was determined by Toimela
\cite{Toimela}.

Presently, the correction $\sim g^4T^4$ (without the
logarithmic factor)
\cite{Zhai1} and the
correction $\sim g^5T^4$ \cite{Zhai2,Nieto} are known. This
is the absolute limit beyond which no perturbative
calculation is possible --- similar ring graphs as in
Fig.\ref{ring}
but with magnetic gluons would give the contribution
$\propto T\mu_{mag}^3 \sim g^6 T^4$ in the free energy. And,
as we have seen, $\mu_{mag}$ cannot be determined
perturbatively.

Collecting all the terms, the following result is obtained
\cite{Zhai2,Nieto}
  \be
 \label{Ffull}
F = - \frac{8\pi^2T^4}{45}
\left[ F_0 + F_2 \frac {\alpha_s(\mu)}{\pi} + F_3 \left(
\frac {\alpha_s(\mu)}{\pi} \right)^{3/2} +
\right. \nonumber \\
\left.
F_4 \left( \frac {\alpha_s}{\pi} \right)^{2} +
F_5 \left( \frac {\alpha_s}{\pi} \right)^{5/2} +
O(\alpha_s^3 \ln \alpha_s ) \right]
 \ee
where
\be
\label{F0c}
F_0 = 1 + \frac {21} {32} N_f
 \ee

\be
\label{F2c}
F_2 = -\frac{15}4 \left(1 + \frac {5} {12} N_f \right)
 \ee

\be
\label{F3c}
F_3 = 30 \left( 1 + \frac {N_f}6 \right)^{3/2}
 \ee

\be
\label{F4c}
F_4 = 237.2 + 15.97 N_f - 0.413 N_f^2 + \frac {135}2
\left( 1 + \frac {N_f}6 \right) \ln \left[ \frac
{\alpha_s}\pi \left( 1 + \frac {N_f}6 \right) \right]
\nonumber \\
- \frac {165}8 \left(1 + \frac {5} {12} N_f \right)
\left(1 -\frac {2} {33} N_f \right) \ln \frac \mu{2\pi T}
 \ee

 \be
 \label{F5c}
F_5 = \left(1 + \frac { N_f}6 \right)^{1/2} \left[
-799.2 - 21.96 N_f - 1.926 N_f^2 \right. \nonumber \\
\left. + \frac {485}2 \left(1 + \frac { N_f}6 \right)
\left(1 -\frac {2} {33} N_f \right) \ln \frac \mu{2\pi T}
\right]
 \ee
The expressions (\ref{F0c}) - (\ref{F5c}) are written for
$N_c =3$. The coefficients like 237.2 are not a result of
numerical integration but are expressed via certain special
functions.

A nice feature of the result (\ref{Ffull}) is its renorm-
invariance.
The coefficients $F_4$ and $F_5$ involve a
logarithmic $\mu$--dependence in such a way that the whole
sum does not depend on the renormalization scale $\mu$.

Let us choose $\mu = 2\pi T$ (this is a natural choice
\cite{Tsum},
$2\pi T$ being the lowest nonzero gluon Matsubara frequency
) and $N_f = 3$.
In that case, we have
 \be
 \label{series}
F = F_0 \left[1 - 0.9 \alpha_s + 3.3\alpha_s^{3/2} + (7.1 +
3.5
\ln \alpha_s ) \alpha_s^2 - 20.8 \alpha_s^{5/2} \right]
 \ee
Note a large numerical coefficient at $\alpha_s^{5/2}$. It
is rather troublesome because the correction $\sim
\alpha_s^{5/2}$ overshoots all previous terms up to very
high temperatures and, at temperatures which can  be
realistically ever reached at accelerators, makes    the
whole perturbative approach problematic.

Take $T \sim 0.5 \ {\rm GeV}$ (as was mentioned earlier,
this is the temperature one can hope to
achieve at RHIC \cite{RHICT}). Then $2\pi T \sim 3 \ {\rm
GeV}$
and $\alpha_s \sim 0.2$. (We use a conservative estimate for
$\alpha_s$ following from $\Upsilon$ physics \cite{Vol}.
Recent
measurements at LEP favor even larger values.). The series
(\ref{series}) takes the form
  \be
  \label{sernum}
 F = F_0[1 - .18 + .3 +.06 - .37 + \ldots]
 \ee
which is rather unsatisfactory. We emphasize that the
coefficient of $\alpha_s^{5/2}$ is rather trustworthy being
obtained independently by two different groups.

The estimate (\ref{sernum}) was obtained in the assumption
$N_f = 3$.
In pure Yang-Mills theory, the behavior of perturbative
series is even
worse
\footnote{I am indebted to F. Karsch who pointed my
attention to this fact.}:
 \be
 \label{serkar}
F = F_0 \left[1 - 1.19 \alpha_s + 5.39\alpha_s^{3/2} +
(16.20 + 6.84
\ln \alpha_s ) \alpha_s^2 - 45.69 \alpha_s^{5/2} \right]
 \ee
which means that perturbation theory cannot be trusted at
all at $\alpha_s$
$ \ ^>_\sim \ .15$. One could naively
expect that deviations of real free energy density from
Stephan--Boltzmann
value would be considerable up to very large temperatures.
Somewhat
surprisingly,
numerical lattice calculations show it is not the case.
According to recent
measurements   in pure $SU(3)$ Yang--Mills theory
\cite{karfree}, the free energy density at $T = 5T_c$
constitutes 85\%
of the Stephan--Boltzmann value. This suggests that the
coupling constant
in this region is small. Indeed,
the estimate for the coupling constant obtained in
\cite{karfree} from
measurement of another lattice quantity, the spatial string
tension,
is $\alpha_s \approx .12$ at $ T = 5T_c$. If substituting it
in the series
(\ref{serkar}), we find  $F/F_0 \approx .87$ in a perfect
agreement
with the data.
 Actually, the smallness of the coupling constant here
is an ``optical illusion''. When comparing the pure Yang--
Mills
theory with $QCD$ with quarks, one  should compare the scale
of perturbative
corrections which is of order $\sim \alpha_s c_V$ in  the
pure Yang--Mills
theory versus $\sim \alpha_s c_F$ in $QCD$. This scale is
roughly the same
in pure YM theory and in QCD for the same temperature
(measured in units
of string tension $\sigma$ or in units of $T_c$).

At lower temperatures, perturbation theory does not work
and non-perturbative effects (associated with instantons)
become important.
 The lattice values of the ratio
 $F/F_0$ rapidly decrease with temperature and are rather
small at $T \sim T_c$.

 The last remark is technical. The result (\ref{Ffull}) was
obtained in Euclidean technique. There is a real time
calculation which correctly reproduces the two-loop term
$\sim g^2T^4$ in free energy \cite{Landsman}, but nobody so
far succeeded in calculating in this way the terms $\sim g^3
T^4$ and higher. Certainly, real-time technique is not very
suitable for calculation of static quantities, and one way
to get the result is good enough, but, to my mind, it is an
interesting methodical problem.

\subsection{Collective excitations.}

One of striking and distinct physical phenomena
characteristic of usual plasma is a non-trivial dispersive
behavior of electromagnetic waves. In contrast to the vacuum
case where only transverse photons with the dispersive law
$\omega = |\vec{k}|$ propagate, two different branches
with different non-trivial dispersive laws $\omega_\bot(k)$
and $\omega_\|(k)$ appear in plasma. The value
$\omega_\bot(0) = \omega_\|(0) = \omega_{pl}$ characterizes
the
eigenfrequency of spatially homogeneous charge density
oscillations and is called the plasma frequency.

A similar phenomenon exists also in $QGP$. The spectrum of
$QGP$ involves collective excitations with quantum number of
quarks and gluons. Like in usual plasma, there are
transverse and longitudinal branches of gluon collective
excitations (alias, transverse and longitudinal {\it
plasmons}) and their properties {\it on the one-loop level}
are very similar to the properties of photon collective
excitations in usual plasma. A novel feature is the
appearance of non-trivial fermion collective excitations
({\it plasminos}).
But, again, they are not specific for a nonabelian theory
and appear also in ultrarelativistic $e^+e^-$ plasma (in the
limit when the electron mass can be neglected compared to
the temperature-induced gap in the electron spectrum
$\propto eT$).

\vspace{.3cm}

 \centerline{\it One-loop calculations.}

\vspace{.3cm}

The dispersive laws of quark and gluon collective
excitations can be obtained via solution of a nonabelian
analog of the Vlasov system involving the classical field
equations in the medium and Boltzmann kinetic equation
\cite{Blaizot}.
This way of derivation makes analogies with usual plasma
(where the Vlasov system is a standard technique) the most
transparent.

I will outline here another way of derivation which is more
conventional and more easy to understand for a field
theorist.
\footnote{We will meet Vlasov equations again and discuss
them
in a little more details in the end of this chapter.}
 This is actually the way the results were
originally derived \cite{Kal,Weldon} (Note, however, that
analogous results for {\it abelian} ultrarelativistic plasma
were first derived long time ago in Vlasov technique
\cite{Silin}).

Consider the gluon Green's function in a thermal medium. As
was mentioned in sect. 2, at $T \neq 0$, different kinds of
Green's function exist. To be precise, we are considering
now the {\it retarded} Green's function
  \be
  \label{Gret}
\left[ G_{\mu\nu}^{ab} (x) \right]^R\ = \ -i<\theta(t)
[A_\mu^a(x), A_\nu^b(0)] >_T
  \ee
which describes a response of the system on a small
perturbation applied at $t = 0$ at some later time $t > 0$.
\footnote{As in commonly used gauges $G^{ab} \propto
\delta^{ab} $, we shall suppress color indices in the
following. We have retained them here just to make clear
that we are dealing with a commutator of Heisenberg field
operators, not with a commutator of classical color fields.}
As was discussed in details in Sect. 2.2, the Fourier image
of
 (\ref{Gret}) is free of singularities
in the upper $\omega$ half--plane. The poles of
$G^R_{\mu\nu}(\omega, \vec{k})$ correspond to eigenmodes of
the system and exactly give us the desired spectrum of gluon
collective excitations. The dispersive equation
 \be
\label{disptens}
\det \| [G^R_{\mu\nu} (\omega, k)]^{-1} \| = 0
 \ee
splits up in two:
  \be
  \label{displt}
\omega^2 - k^2 - \Pi_t(\omega, k) = 0 \nonumber \\
k^2 + \Pi_l(\omega, k) = 0
\ee
with $\Pi_{l,t}(\omega, k)$ being defined in
Eq.(\ref{tens}). The solutions to the equations
(\ref{displt}) give two branches of the spectrum.

The explicit one-loop expressions for $\Pi_l(\omega, k)$ and
$\Pi_t(\omega, k)$ obtained by the calculation of the graphs
in Fig.\ref{Gl1loop} in the limit $\omega, k \ll T$ are
\cite{Kal,Weldon}
  \be
  \label{Pilt}
\Pi_l(\omega, k) = 3\omega_{pl}^2 [1 -  F(\omega/k)]
\nonumber \\
\Pi_t(\omega, k) = \frac 32 \omega_{pl}^2 \left[ \frac
{\omega^2}{k^2} + \frac{k^2 - \omega^2}{k^2} F(\omega/k)
\right]
 \ee
where
 \be
\label{ompl}
\omega_{pl} = \frac{gT}3 \sqrt{N_c + \frac {N_f}2}
 \ee
is the plasma frequency and
 \be
\label{Fx}
F(x) = \frac x2 \ln \frac{x+1}{x-1}
 \ee

\newpage

\begin{figure}
\grpicture{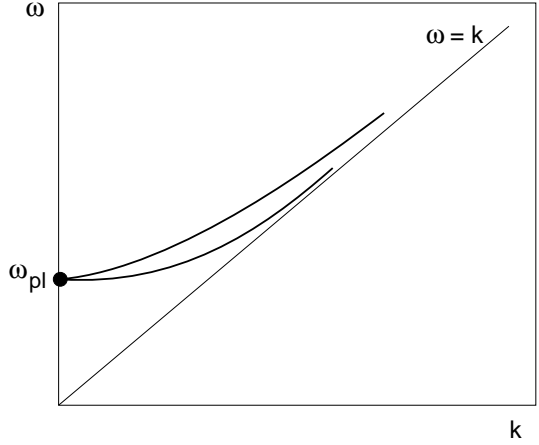}
\caption{ Plasmon spectrum in one loop.}
\label{plasmo}
\end{figure}

\newpage

 The behavior of dispersive curves is schematically shown in
Fig.\ref{plasmo}. At $k
= 0$, $\omega_\bot(0) = \omega_\|(0) = \omega_{pl}$. Then
the two branches diverge:
  \be
\label{lowk}
 \omega^2_\bot(k \ll gT) = \omega_{pl}^2 + \frac 65 k^2
\nonumber \\
 \omega^2_\|(k \ll gT) = \omega_{pl}^2 + \frac 35 k^2
\ee
At $k \gg gT$ both branches tend to the vacuum dispersive
law
  \be
\label{highk}
 \omega^2_\bot(k \gg gT) = k^2 + \frac 32 \omega_{pl}^2
\nonumber \\
 \omega^2_\|(k \gg gT) = k^2 \left[1 + 4\exp \left(
 - \frac{2k^2}{3\omega_{pl}^2} - 2 \right)
 \right]
\ee
We see that $\omega_\|(k)$ approaches the line $\omega = k$
exponentially fast.

The dispersive laws for quark collective excitations are
obtained from a similar analysis of the quark Green's
function.
One loop expression for the finite temperature contribution
in
the fermion polarization operator in the limit $\omega, k
\ll T$
is
  \be
 \label{Sigomk}
\Sigma(\omega, \vec{k}) \ =\ \gamma^0 \frac
{\omega_0^2}{\omega}
F(\omega/k) \ -\ \frac {\vecg{\gamma} \vec{k}}{k} \frac
{\omega_0^2}{k}
[F(\omega/k) - 1]
  \ee
where $\omega_0$ is the plasmino frequency at zero momentum:
\be
\label{om0}
\omega_0^2 = \frac{g^2T^2}8 c_F
 \ee
We see that the fermion polarization operator
involves two tensor structures $\omega \gamma^0 - \vec{k}
\vecg{\gamma}$ and $\omega \gamma^0 + \vec{k} \vecg{\gamma}$
which gives rise to two dispersive branches. We will call
the branch corresponding to the Lorentz-invariant structure
``transverse'' and the branch corresponding to the structure
$\omega \gamma^0 + \vec{k} \vecg{\gamma}$ ---
``longitudinal''. These terms may be misleading in the
fermion case because, in contrast to the plasmons with
photon or gluon quantum numbers, these branches are not
associated with transverse and longitudinal field
polarizations. Hence the quotation marks. But better names
were not invented, and using the words ``transverse'' and
``longitudinal''  still makes a certain sense because the
physical properties of ``transverse'' and ``longitudinal''
fermion branches are rather analogous to the physical
properties  of transverse and longitudinal gluon branches.

\newpage

\begin{figure}
\grpicture{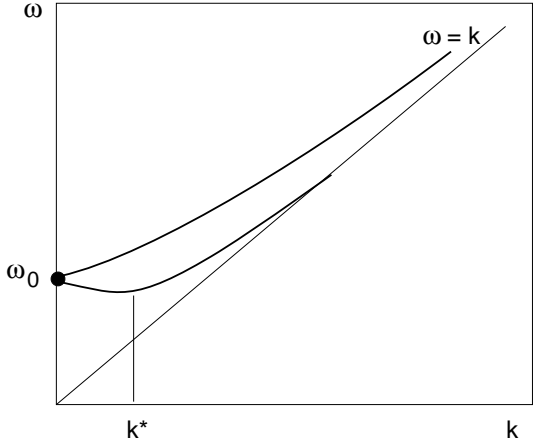}
\caption{ Plasmino spectrum in one loop.}
\label{plasmi}
\end{figure}

\newpage

The pattern of the quark spectrum is shown in
Fig.\ref{plasmi}. It is
similar to the gluon spectrum with one important distinction
--- at $k \sim 0$, $\omega_\bot(k)$ and  $\omega_\|(k)$
involve linear terms of opposite sign:
 \be
 \label{qlowk}
 \omega^q_\bot(k \ll gT) = \omega_0 + \frac k{3}
\nonumber \\
 \omega^q_\|(k \ll gT) = \omega_0 - \frac k{3}
 \ee

Thus  $\omega^q_\|(k)$ first goes down and reaches minimum
at
some $k^*$. The group velocity of the longitudinal plasmino
at
this point is zero. At large $k$,
 \be
\omega_\bot^{q2}(k \gg gT) \ =\ k^2 + 2\omega_0^2
 \ee
and $\omega_\|^{q}$ tends from above to the line $\omega =
k$ exponentially
fast.

\vspace{.3cm}

\centerline{\it Landau Damping.}

\vspace{.3cm}

The quoted one-loop results for the dispersive laws of
transverse plasmons
and plasminos are gauge-invariant and stable with respect to
higher-order
corrections. The latter is not true, however, for
longitudinal excitation
branches \cite{SilinU,spectr}. We have seen that
$\omega_\|^{1\ loop}(k)$ tends
to  the line $\omega=k$ exponentially fast at $k \gg gT$.
This can be easily
seen from the analysis of the dispersive equations for
longitudinal branches
which, in the limit $k \gg gT$, have the form
  \be
  \label{displk}
\omega_\| + k \sim \frac{g^2T^2}k \ln \frac k{\omega_\| -k}
 \ee
At $k \gg gT$, the solution exists when the logarithm is
large and
$\omega_\| - k$ is exponentially small. The logarithmic
factor
in Eq.(\ref{displk}) comes from the angular integral
 \be
 \label{ang}
\sim \int \frac{d\theta}{\omega - k \cos \theta}
 \ee
which diverges at $\omega = k$. This collinear divergence
appears due to
masslessness of quarks and gluons in the loop depicted in
Fig.\ref{Gl1loop}. But
 quarks and gluons in $QGP$ are {\it not} massless --- their
dispersive law
acquires the gap $\sim gT$ due to temperature effects. An
accurate calculation
requires substituting in the loops the {\it dressed}
propagators. As a result,
the logarithmic divergence in the integral (\ref{ang}) is
cut off and the
logarithmic factor in Eq.(\ref{displk}) is modified:
  \be
  \label{cutoff}
\ln \frac k{\omega - k} \to \frac 12 \ln \frac{k^2}{(\omega
- k)^2 + Cg^2T^2}
\ee
Dressing of propagators amounts to going beyond one-loop
approximation.
Strictly speaking, to be self-consistent, one should also
take into account
one-loop thermal corrections to the vertices (this procedure
is known as
{\it resummation of hard thermal loops} \cite{Pisar}), but
in this particular case these
corrections do not play an important role. What {\it is}
important is the
cutoff of the logarithmic collinear singularity due to
effective
temperature-induced masses.

Substituting (\ref{cutoff}) in (\ref{displk}), we see that
the new
dispersive equation does not at all have solutions with real
$\omega$ for
large enough $k$. This fact can be given a natural physical
explanation.
When $k$ is small compared to $gT$, there is no logarithmic
factor
in the dispersive equation, the modification
(\ref{cutoff}) is
irrelevant, and the dispersive law of longitudinal modes
does
not deviate
from the one-loop result. Then the logarithm appears, the
modification
(\ref{cutoff}) starts
playing a role, and, at some $k^{**} \sim gT$, the
longitudinal
dispersive
curve crosses the line $\omega = k$. At this point the
longitudinal
polarization operator acquires the imaginary part due to
{\it Landau
damping}.

In usual plasma, Landau damping is the process when
propagating
electromagnetic waves are ``absorbed'' by the electrons
moving
in plasma. In
the language of quantum field theory, it is a $2 \to 1$
process
  \be
\label{21}
\gamma^* + e \to e
  \ee
In real time technique, that corresponds to a contribution
to the imaginary
part of the polarization operator so that both internal
electron lines in
the loop are placed on mass shell. At $T = 0$ the standard
Cutkosky rules
imply positive energies of all particles in the direct
channel, and
the imaginary part appears only due to the decay $\gamma^*
\to
e^+ + e^-$. At
$T \neq 0$, Cutkosky rules are modified and both signs for
energy are
admissible.
\footnote{We already exploited this fact in Sect.4.3 while
calculating
the damping of pion collective excitations in low
temperature phase.}
 Physically, that corresponds to the presence of
real particles
in the heat bath so that the process (\ref{21}) may go.

Also in $QGP$ imaginary parts of polarization operators may
acquire
contributions due to Landau damping. The corresponding
processes are
 \be
  \label{12qg}
g^* + g \to g, \ \ g^* + q(\bar q) \to q(\bar q),\ \
q^*({\bar q}^*) +
g \to q(\bar q),\ \ q^*({\bar q}^*) +
\bar{q} ( q) \to g, \
 \ee
where $g^*, q^*, {\bar q}^*$ are plasmon and plasmino
collective
excitations and $g, q, {\bar q}$
are the excitations with characteristic momenta of order
temperature
(in this kinematic region, the dispersive laws are roughly
the same as for
free quarks and gluons and the star superscript is
redundant).

The kinematic condition for the scattering processes
(\ref{12qg}) to go is
that the frequency of collective excitations $g^*$ and $q^*$
would be less
than their momentum. We have seen that the condition $\omega
< k$ is
realized indeed for longitudinal plasmon and plasmino
excitations starting
from some  $k^{**} \sim gT$. At $k > k^{**}$, the Landau
damping switches on
and the dispersion law acquires an imaginary part. The
imaginary part
rapidly grows and very soon becomes of order of the real
part. From there
on, it makes no sense to talk about propagating longitudinal
modes anymore.
The situation is the same as in usual plasma where
longitudinal modes also
become overdamped at large enough momenta and disappear from
the physical
spectrum \cite{LP}.

\newpage

\begin{figure}
\grpicture{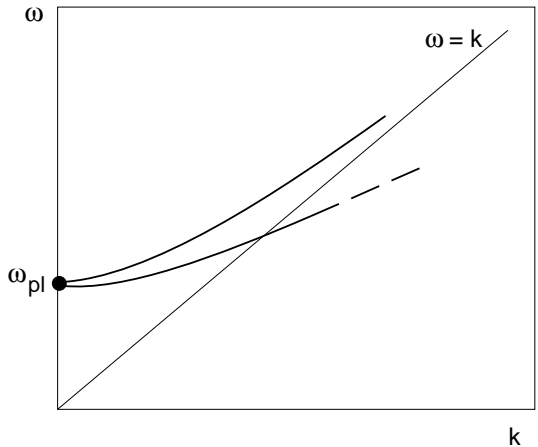}
\caption{ Plasmon spectrum.}
\label{landa}
\end{figure}

\newpage

The true pattern of plasmon collective modes is shown in
Fig.\ref{landa}. A similar
picture holds for plasminos.

\vspace{.3cm}

\centerline{\it Observability.}

\vspace{.3cm}

One-loop dispersive curves are gauge-invariant. However, the
question
whether these curves are physically observable is, again,
highly non-trivial. It is easy to measure explicitly photon
dispersive
curves in usual
plasma --- to this end, one should study  propagating
classical
electromagnetic waves, measure the electric charge density
(say, by laser
beams) as a function of time and spatial coordinates and
determine thereby
the frequency and the wave vector of the wave.

But there is no such thing in Nature as classical gluon
field due to
confinement, and no classical device which would measure the
color charge
density exists. Even more obviously, quark fields (which
have Grassmanian
nature) cannot be treated classically. Hence, one cannot
really measure the
energy and momentum of propagating colored waves in a
direct physical
experiment.

What one {\it can} measure are correlators of colorless (in
the first place, electromagnetic) currents. Modification of
dispersive laws affects these
correlators and that can be observed. However, a colorless
current always
couples to a {\it pair} of  colored particles and, as a
result, physical
correlators involve some integrals of quark and gluon
Green's functions
which are related to quark and gluon dispersive
characteristics only in an
indirect way. Also, thermal modification of vertices is
 as important here
as modification of the Green's functions.

However, there is one special point on the dispersive curves
which can in
principle be directly measured in experiment. This is the
point on the
longitudinal plasmino curve where its frequency acquires a
minimal value
and the group velocity turns to zero.
Consider the problem of
emission of relatively soft $e^+ e^-$ or $\mu^+\mu^-$ pairs
by $QGP$.
In a ``thermos bottle'' gedanken
experiment, one should make sure that the size of your
thermos bottle is
much less than the lepton mean free path. Otherwise, the
leptons are
thermalized and their spectrum is just Planckian. Btt in
heavy ion collisions  experiments, $QGP$ is produced in
small volume, the condition $L_{\rm char} \ll L^{em}_{\rm
free\ path}$
is satisfied,
and the spectrum of emitted leptons (and photons) can
provide a
non-trivial
information on dynamic characteristics of $QGP$.

The spectrum of soft dileptons was calculated in
\cite{Yuan}.
This is one
of very few  {\it physical} problems we know of
where the hard
thermal loop resummation
technique \cite{Pisar} should be used (and was used) at full
length. The spectrum feels the
effects due to quark and gluon interactions in the region
$E_{l^+l^-} \sim
P_{l^+l^-} \sim gT$ ---  the spectrum at larger energies and
momenta
is the same as for the gas
of free quarks. One particular source of soft dileptons is
the process
$q^*_\bot \to q^*_\| \ +\ l^+l^-$. The probability of this
process has a
``spike'' for the momentum  of $q_\bot^*$ and $q_\|^*$
coinciding with the
momentum $k^*$ on the longitudinal plasmino dispersive curve
with zero group
velocity. There are just many plasminos at the
vicinity of
this point and the phase space factor provides a singularity
at $E_{l^+l^-}
= \omega_\bot(k^*) - \omega_\|(k^*)$ in the spectrum.
Another spike comes
from the process when a longitudinal plasmino with momentum
$k^*$
annihilates with a longitudinal antiplasmino with the
opposite
 momentum to produce a lepton pair with zero net momentum
and
the energy $2\omega_\|(k^*)$.

Unfortunately, in the soft region, the main contribution in
the spectrum is
due to cuts. In other words, the most relevant elementary
kinetic
processes are
not  $q^* \to q^* + l^+l^-$ or $q^* + \bar{q}^* \to
l^+l^-$, but rather
$q^* + g^* \to q^* + g^* + l^+l^-$ etc. The spikes actually
have finite
width due to collisional damping of collective excitations
(the issue to be
discussed in the next section), and one can hope to see only
a tiny
resonance on a huge background. Still, such a resonance in
the spectrum
{\it is } an observable effect.

\subsection{ Damping mayhem and transport paradise.}

\vspace{.3cm}

\centerline{\it Direct decay.}

\vspace{.3cm}

In the previous section, we discussed the Landau damping
contribution to
the imaginary parts of polarization operators and,
correspondingly, to
imaginary parts of dispersion laws. It comes from the
kinematic region
$\omega < k$ and is physically related to absorption of
ingoing excitations
by thermal quanta like in (\ref{12qg}). However, we did not
say a word about the
contribution of direct decay processes $g^* \to q + \bar{q}$
etc. in the
timelike kinematic region $\omega > k$.

That was with a good reason. On the one-loop level, the
contribution of
decay processes in imaginary parts is nonzero and is of
order $\sim g^2T$.
Unfortunately, it depends on the gauge and, in some gauges,
has even the
wrong sign corresponding not to damping of excitations but
to instabilities
\cite{Zahed}. The point is that such one-loop calculation is
unstable with
respect to higher-order corrections. It is very clear
physically --- quarks
and gluons in $QGP$ cannot be treated as massless but
acquire dynamical
masses due to thermal effects. And the decay of a plasmon or
plasmino into
two other collective excitations is not kinematically
allowed. The only
exception is the process $g^* \to q^* + \bar{q}^*$ which in
principle may
go if \cite{spectr}
 \be
 \label{Nf6}
N_f > 9c_F - 2c_V = 6
 \ee
But there are at most three light flavors in real $QGP$ and
decay processes
can be safely forgotten.

\vspace{.3cm}

\centerline{\it Collisional damping.}

\vspace{.3cm}

Still, damping is there even in the timelike region $\omega
> k$ due to
collisions $g^* + q^* \to g^* + q^*$, $q^* + \bar{q}^* \to
g^* + g^*$ etc.
This is also the main source of damping of transverse
electromagnetic waves
in usual plasma \cite{LP}. A rough estimate for collisional
damping in $QGP$
can be  done very simply.

The meaning of damping is the inverse lifetime of
excitations.
We have
\be
\label{dampest}
\zeta \sim (\tau_{life})^{-1} \sim n \sigma^{tot}
  \ee
where $n \sim T^3$ is the density of the medium and
$\sigma^{tot}$ for
excitations which carry (color) charge has a Coulomb form
  \be
  \label{sigma}
\sigma^{tot} \sim g^4 \int \frac{dp_\bot^2}{p_\bot^4} \sim
\frac {g^2}{T^2}
 \ee
We took into account the fact that the power infrared
divergence for the
integral of Coulomb cross section is effectively cut off at
$p_\bot \sim
gT$ due to Debye screening
\footnote{and due to Landau damping effects --- see more
detailed discussion below.}. As a result, we obtain the
estimate
\be
\label{g2T}
\zeta^{q,g} \sim g^2T
 \ee

Note that this value for the damping is unusually  large. It
is much larger
than, say, the damping of photons in ultrarelativistic
$e^+e^-$ -- plasma.
The latter can also be estimated from the formula
(\ref{dampest}), but
$\sigma^{tot}$ is now not the Coulomb, but the Compton cross
section. The
integral has now the form $\int_{eT} dp_\bot^2/p_\bot^2 \sim
\ln (1/e)$ and
the estimate is
  \be
  \label{zetgam}
\zeta^\gamma \sim   e^4T \ln (1/e) \ll e^2T
 \ee
The question arises whether the new anomalously large scale
$\sim g^2T$ has a \ physical \ relevance \footnote{Do not
confuse this scale with the magnetic scale
which is also
of order $g^2T$. The former is related to kinetic properties
of the system
while the latter refers exclusively to static phenomena.}.
 We will return to discussion of this point a bit later.

\newpage

\begin{figure}
\begin{center}
\SetScale{2}
\begin{picture}(200,70)(0,0)
\Line(55,35)(70,35)
\Line(130,35)(145,35)

\PhotonArc(100,35)(30,0,180){1}{10}
\CArc(100,35)(30,180,360)

\Vertex(100,5){3}
\Vertex(100,65){3}

\put(55,38){k}

\end{picture}

\end{center}
\caption{}
\label{damLS}
\end{figure}

\newpage

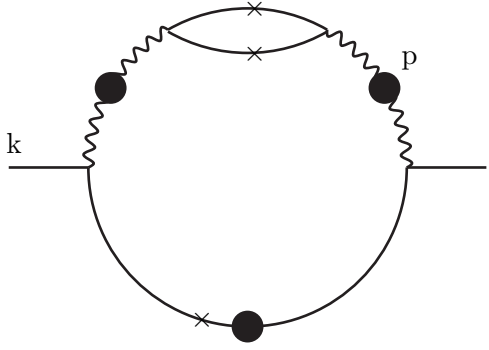
\begin{figure}
\begin{center}
\SetScale{2}
\begin{picture}(200,70)(0,0)
\Line(55,35)(70,35)
\Line(130,35)(145,35)

\PhotonArc(100,35)(30,0,60){1}{10}
\PhotonArc(100,35)(30,120,180){1}{10}
\CArc(100,35)(30,60,120)
\CArc(100,86.6)(30,240,300)

\CArc(100,35)(30,180,360)

\put(100,64){$\times$}
\put(100,55.5){$\times$}
\put(90,5){$\times$}

\Vertex(100,5){3}
\Vertex(125.8,50){3}
\Vertex(74.2,50){3}

\put(55,38){k}
\put(130,55){p}
\end{picture}

\end{center}
\caption{ A graph contributing to $ {\rm Im}\
\Sigma(k)$.  Crossed lines are put on mass shell. One of
the quarks in the internal loop has a negative energy.}
\label{coldam}
\end{figure}

 An accurate calculation of the damping of {\it fast moving}
($k \gg
gT$) quark and gluon excitations in
$QGP$ has been done in \cite{spectr,andam} (see also
\cite{Alt}).
Consider the graph in Fig.\ref{damLS}
for the quark polarization operator where the lines with
blobs
stand for quark and
gluon propagator dressed by thermal loops.
\footnote{It can be shown \cite{spectr} that when
calculating the leading
contribution in $\zeta$ in
the kinematic region $k \gg gT$ which is under discussion
now, vertex
corrections can be disregarded.}
 The imaginary part of the  loop in Fig.\ref{damLS} depends
on
the imaginary
parts of internal propagators. The imaginary part of the
gluon propagator
due to Landau damping turns out to be of paramount
importance. Physically,
this contribution just corresponds to the scattering
processes
$g^* + q^* \to g^* + q^*$ and $g^* + g^* \to g^* + g^*$
as can be
easily inferred if spelling out the exact gluon propagator
as in Fig.\ref{coldam} (there is also a similar graph with
internal gluon loop).

\newpage

\newpage

 Dressing  the quark propagator (indicated by the blob in
Fig.\ref{coldam}) is  important. If not
taking it
into account, the imaginary part of the propagator is
$\delta$ --
function and the result for ${\rm Im}\ \Sigma_R(\omega,
\vec{k})$
near the mass shell $\omega = |\vec{k}|$ has the
form
  \be
  \label{dkdk0}
 {\rm Im}\ \Sigma_R(\vec{k}) = - \frac 3{32\pi^2}
\omega_{pl}^2 g^2 c_F
T \gamma^0
\int_{-\infty}^\infty dp_0 \int_{-1}^1 d \cos
\theta
\int_0^\infty \frac {dp^2}{p^4 + \frac 9{16} \omega_{pl}^4
\frac
{p_0^2}{p^2}}\  \delta(p_0 - p \cos \theta)
  \ee
where $\theta$ is the angle between $\vec{k}$ and $\vec{p}$.
The integral has a power infrared behavior at $p \gg gT$,
but this divergence
is cut off due to Landau damping effects (the second term in
the denominator
in the integrand).
Still, the integral in (\ref{dkdk0}) diverges
logarithmically at $p
\ll gT$ and the result for the damping inferred from
Eq.(\ref{dkdk0})
is infinite.

The crucial observation is that the {\it dressed} quark
propagator does not
have singularities on the real $p_0$
axis. A self-consistent account of the collisional damping
for the quark
Green's function moves its singularities in the complex
plane.
As a result, $\delta$ -- function in the integrand is
replaced by a smooth
distribution with the width of order $\zeta \sim g^2T$:
 \be
\label{smooth}
  \delta(p_0 - p \cos \theta)\ \to \ \frac{\zeta/\pi}{(p_0 -
p \cos \theta)^2
+ \zeta^2}
 \ee
 This
smoothing cuts
off the logarithmic singularity in (\ref{dkdk0}) at $p \sim
g^2T$. The
other source for the cutoff could be provided by magnetic
screening effects
, but the latter is an essentially nonabelian phenomenon
whereas the
cutoff due to smearing out the $\delta$ -- function is a
universal effect
which occurs also in an abelian theory.

Note that the cutoff $p \ ^>_\sim \ g^2T$ corresponds to the
cutoff
 $p_0$ $ \ ^>_\sim \ g^4T$. Really, the integral in
(\ref{dkdk0}) is saturated
by the values
$p$ and $p_0$ such that two terms $p^4$ and $\sim
\omega_{\rm pl}^4 p_0^2/p^2$
in the denominator in the integrand are of the same order.
That means
that $p_0^{\rm char} \sim p^3/\omega_{\rm pl}^2$ which gives
the estimate
$p_0 \sim g^4T$ at the lower range of integration. Actually,
our calculation
corresponds to picking up the pole
corresponding to the so called {\it attenuating mode}
  \be
 \label{att}
 p_0 = -i\frac 43 \frac {p^3}{\omega_{\rm pl}^2}
  \ee
of the exact gluon propagator in the graph in
Fig.\ref{damLS}. This pole
presents an {\it extra} solution of the transverse
dispersive equation in
Eq.(\ref{displt}) when the imaginary part of the logarithm
in
$\Pi_t(p_0 , p)$ which is responsible for Landau damping is
taken into account, and we assumed $|p_0| \ll p$. The
attenuating mode
 (\ref{att}) was found in \cite{spectr}.
It exists in the region $g^2T \ ^<_\sim \ p \ ^<_\sim \ gT$
and the dispersion
law (\ref{att}) holds in the region  $g^2T \ \ll \ p \ \ll\
gT$.
When $p \ ^>_\sim \ gT$,
the assumption  $|p_0| \ll p$ is not fulfilled, and when $ p
\ ^<_\sim \ g^2T$,
perturbation theory breaks down.
A similar mode exists also in nonrelativistic plasma
(see Ref.\cite{LP}, p.135). We will meet the mode
(\ref{att}) and the
frequency scale $ \sim g^4T$ once again in the end of this
chapter when
discussing chirality non-conservation rate.

To find the dispersion law, one should add $i {\rm Im}\
\Sigma_R(\vec{k})$
from Eq.(\ref{dkdk0}) to the one-loop
result for $\Sigma_R(k)$ (which is real for $\omega \geq k$)
and solve the dispersive equation ${\rm Det} \|\gamma^0
\omega - \vecg{\gamma}
\vec{k} - \Sigma_R(\vec{k})\| \ =\ 0$.
The solution is
complex $\omega^{pole}(k) = \omega^{1\ loop}(k) - i\zeta
(k)$
  and the final result for $\zeta(k \gg gT)$ is very simple.
 \be
 \label{dampLS}
\zeta^q = \alpha_s c_F T \ln(C^q/g)  \nonumber \\
\zeta^g = \alpha_s c_V T \ln(C^g/g)
\ee
(damping of gluon fast moving excitations can be calculated
along
the same lines).
 Only the coefficient of logarithm can be calculated. The
constants $C^q,\ C^g$ under
the logarithm  cannot be determined. Actually, we will see
shortly that these
constants are gauge--dependent and cannot be {\it defined}
in a reasonable
way.

\newpage

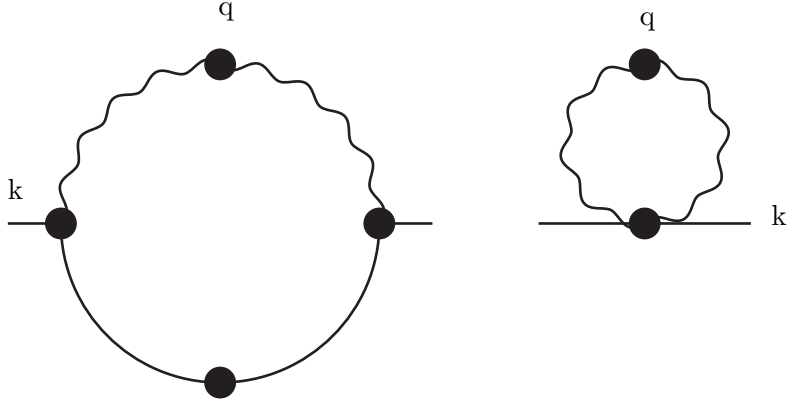
\begin{figure}
\begin{center}
\SetScale{2}
\begin{picture}(200,70)(0,0)

\put(0,40){k}
\Line(0,35)(10,35)
\Line(70,35)(80,35)

\put(40,75){q}
\PhotonArc(40,35)(30,0,180){1}{10}
\CArc(40,35)(30,180,360)

\Vertex(40,5){3}
\Vertex(40,65){3}
\Vertex(10,35){3}
\Vertex(70,35){3}

\put(145,35){k}
\Line(100,35)(140,35)
\PhotonArc(120,50)(15,0,360){1}{10}
\put(120,73){q}
\Vertex(120,65){3}
\Vertex(120,35){3}

\end{picture}

\end{center}
\caption{Polarization operator of soft plasmino.}
\label{soft}
\end{figure}

\newpage

The results (\ref{dampLS}) were obtained solving the
dispersive equation with
the polarization operator (\ref{dkdk0}) calculated at real
$\omega$.
A more refined analysis which takes into account the
modification
of ${\rm Im} \ \Sigma (k)$ for complex $\omega$ when one
starts to move
from the real $\omega$ axis towards a singularity
shows that the dispersive equation has actually no solutions
and the
singularity is no longer a pole, but a branching point
\cite{BNN,damp1,Pilon}. But this branching point is located
at the same distance from the real axis as the would-be pole
and brings about
the same damping behavior  of the gluon retarded Green's
function
$G^R(t) \sim \exp\{-\zeta t\}$ at large real times.

The modification of the fermion propagator due to higher
loops in the
graphs in Fig.\ref{damLS}, \ref{coldam} was taken into
account
 smearing out the $\delta$ -- function singularity
in the Green's function as in Eq.(\ref{smooth}).
In recent \cite{nonexp} it was found that if doing things
{\it quite}
carefully, namely resumming leading infrared
divergent pieces in the soft photon loops
contributing in the fermion Green's function
in any order in abelian theory, the infrared divergence in
(\ref{dkdk0})
eventually {\it survives}. 
That essentially means that $G^R(\omega)$ does not involve
singularities at all at finite distance from the real axis.
This results in a non-exponential
decay of  $G^R(t)$ at large real time:
  \be
\label{BlIanc}
G^R(t) \sim
\exp\{- \alpha
 T t\ \ln(eTt)\}
  \ee
 In spite of 
the fact that this result is
rather unexpected, it seems to be correct. Non-standard logarithmic factor
in (\ref{BlIanc})  resembles the logarithmic factor in the exponent
in (\ref{CPcorr}) . An analogy can also be drawn with the well-known 
logarithmic
divergence in the phases of Coulomb scattering \cite{Ianpriv}.
 A careful   independent
study of this question is, however, highly 
desirable. 

Damping of excitations in another kinematic region $\vec{k}
= 0$ (``standing''
plasmons and plasminos) was studied in \cite{Pisar}. Here
the vertex
corrections are as important as corrections to the
propagators, however an
accurate analysis of \cite{Pisar} shows that it suffices to
take into
account only one (hard thermal) loop corrections both in
polarization operators
and  vertices. Consider for definiteness the damping of
standing plasminos.
The graphs contributing to the soft plasmino polarization
operator are
shown in Fig.\ref{soft}. Using Keldysh technique one can
derive in
the limit of
soft external momentum
  \be
  \label{SigmaR}
\Sigma^R(k) = -2ig^2 \int \frac {d^4q}{(2\pi)^4} \frac T q_0
{\rm Im}\
D_{\mu\nu}^R(q) \left[ \frac 12 \Gamma^R_{\mu\nu}(k;q) +
\right.
\nonumber \\
\left. \Gamma^R_\mu (k,k+q; q) G^R(k+q) \Gamma_\nu^R(k+q,k;-
q) \right]
\ee
Here $\Gamma^R$ are the retarded
vertices presenting certain combinations of the Keldysh
components
$\Gamma_\mu^{abc}$ and $\Gamma_{\mu\nu}^{abcd}$:
  \be
\label{GamR}
\Gamma_\mu^R \ =\ \sum_{b,c = 1}^2 \Gamma_\mu^{1bc}, \ \ \ \
\ \
\Gamma_{\mu\nu}^R \ =\  \sum_{b,c,d = 1}^2
\Gamma_{\mu\nu}^{1bcd}
  \ee
where the index 1 is put on the outgoing fermion line.
Like retarded propagators,
$\Gamma^R$  are analytic in the upper $\omega$ half--plane.
Substituting here transverse and longitudinal parts of the
gluon Green's
function and one-loop vertices $\Gamma_{\mu\nu}$ and
$\Gamma_\mu$,
calculating the integral (which in this case can be done
only numerically),
and solving the dispersive equation, one arrives at the
result
 \be
 \label{dampPis}
\zeta^q(k=0) = 1.43 c_F \frac {g^2T}{4\pi}
 \ee
However, the gluon Green's function involves also a
gauge-dependent part
  \be
  \label{Dgauge}
D_{\mu\nu}^{R(\alpha)}(q) = (\alpha -1) \frac{q_\mu
q_\nu}{[(q_0 +
i\epsilon)^2 - \vec{q}^2]^2}
 \ee
where $\alpha$ is a gauge parameter and an infinitesimal
$i\epsilon$ is
introduced to provide for the right analytical properties.
At  first
sight, this gauge-dependent piece should not affect the
position of the
pole. Really, one can use the Ward identities
  \be
  \label{Ward}
q_\mu \Gamma_\mu^R(k, k+q;q) \ =\ t^a \left\{ [G^R(k+q)]^{-
1}
-  [G^R(k)]^{-1} \right\} \nonumber \\
q_\mu \Gamma_{\mu\nu}^R(k; q) \ =\ t^a \left\{
\Gamma_\nu^R(k,k-q;-q) -
\Gamma_\nu^R(k+q,k;-q) \right\}
  \ee
 which hold also at finite
temperature for {\it retarded} propagators and vertices
\cite{Taylor,andam}
to derive
  \be
\label{SigWard}
\Sigma^{R(\alpha)} (k) = -ig^2 T c_F (\alpha -1) [G^R(k)]^{-
1} \times \nonumber \\
\int \frac {d^4q}{(2\pi)^4} \frac 1  q_0 \left\{ \frac
{1}{[(q_0 +
i\epsilon)^2 - \vec{q}^2]^2} - \frac{1}{[(q_0 - i\epsilon)^2
- \vec{q}^2]^2}
\right\}
\left[ G^R(k+q) - G^R(k) \right] [G^R(k)]^{-1}
  \ee
We see  the presence of factors $[G^R(k)]^{-1}$ both on the
right and on the left. $G^R(k)$ is singular at the pole and
$[G^R(k)]^{-1}$ is zero. One might infer from this that a
gauge-dependent contribution to $\Sigma^R({\rm  pole})$ and
hence to the corresponding solution of the dispersive
equation determining the pole position is also zero.

However, this is wrong \cite{BKS,damp1}. The point is that
the integral in (\ref{SigWard}) involves a severe power
infrared divergence and is infinite at the pole. We have
thereby a $0 \times \infty$ uncertainty. This uncertainty
can
be resolved by
 choosing $k$ not exactly at the pole but slightly off mass
shell.
 Then $[G^R(k)]^{-1}$ is not exactly zero and also the
divergence in the integral is cut off  by an
off-mass-shellness. When the distance from the mass shell is
small, the
final result for $\Sigma^{R(\alpha)}(k \to {\rm pole})$ does
not depend on this distance and is just finite. In the soft
momenta region
 \be
 \label{Siggauge}
\Sigma^{R(\alpha)}_{\rm soft}(k) = - \frac
{i(\alpha -1) g^2 T c_F}{4\pi} \gamma^0
 \ee
This brings about a gauge-dependent part $\sim g^2T$ in the
damping of soft plasminos. A similar analysis with the same
conclusion can be carried out for plasmons.

In Ref.\cite{Rebmu}, another regularization procedure was
suggested
where the external momentum $k$ was always kept strictly on
mass
shell, but the divergent integral in Eq.(\ref{SigWard}) was
regularized
introducing an explicit infrared cutoff $\mu$. Then, of
course,
a gauge--dependent piece is absent even after the cutoff is
eventually
sent to zero. It does not solve a problem, however. The {\it
physical} definition of damping (if any) should be related
to
 the exponential asymptotics of
$G^R(t)$ at large real times. $G^R(t)$ is given by the
Fourier integral
 over the {\it real} $\omega$ axis
$$G^R(t) \ =\ \int_{-\infty}^\infty G_R(\omega) \ e^{-
i\omega t}\ d\omega$$
Thus it is determined by the behavior of $G_R(\omega)$ at
real $\omega$,
i.e.  {\it off mass shell}. The damping defined as
$$ \zeta \ =\ -\lim_{t \to \infty} \ \frac {G_R(t)}{t}$$
{\it is} gauge--dependent  \cite{damp1}.

The same
gauge-dependence shows up in the damping of energetic
plasmons and
plasminos, but in the latter case, this gauge-dependence is
parametrically overwhelmed by
the leading gauge-independent contribution (\ref{dampLS})
involving the factor $\ln (1/g)$.

\vspace{.3cm}

\centerline{\it Observability.}

\vspace{.3cm}

The observed gauge-dependence of the damping obviously
indicates that it is not a physical quantity. This is
definitely true at least for soft plasmons and plasminos
where the gauge-dependent part and the gauge-independent
part
(\ref{dampPis}) are of the same order $\sim g^2T$.

Indeed, it is not possible to contemplate a physical
experiment where this quantity could be measured.
That should be confronted with the case of abelian plasma
where damping of electromagnetic waves is a perfectly
physical quantity and can be directly observed by measuring
the attenuation of the amplitude of a classical wave with
time. But as we already noted, no classical gluon or quark
waves exist. This observation refers also for the damping of
{\it electron and positron} collective excitations in the
ultrarelativistic  abelian plasma. It also has an
anomalously large scale $\sim e^2T$ [with the extra
logarithmic factor \ $\ln (1/e)$ for $k \gg eT$] and it also
cannot be directly measured.

One could try to observe the effects due to damping in
gauge-invariant quantities like the polarization operator of
electromagnetic currents. An accurate analysis which goes
beyond the conventional hard thermal loop resummation
technique and effectively resums a set of ladder graphs
shows, however, that a self-consistent account of the
corrections due to damping in the quark Green's functions
and in
the vertices results in the exact cancelation of the
anomalously large scale $\sim g^2T$ in the final answer
\cite{andam}.
(for a similar analysis with a similar conclusion in scalar
QED see \cite{Rebscal}).

However, there is a physical problem where the  scale
$\sim g^2T$ can in principle show up.
This is the already discussed problem of lepton pair
production in $QGP$. We have seen that the spectrum of
leptons pairs involves spikes associated with a special
point
with zero group velocity on the longitudinal plasmino
dispersive curve.
Going beyond the hard thermal loop approximation and taking
into account the effects due to collisional damping in the
Green's functions {\it and} in the vertices would bring
about a finite width for these spikes of order $g^2T$ and
there is a principle possibility to measure this width. This
problem has not been studied, and it is not clear by now
whether the width of the spike can be calculated
analytically and whether one can single out this spike out
of the background.

What one can say quite definitely is that this width should
crucially depend {\it both}  on modification of propagators
 due to
collisional effects {\it and} on a similar modification of
vertices and has little to do with the
(gauge-dependent) position of the pole (or whatever  the
real singularity is \cite{BNN,damp1,Pilon}) of the quark and
gluon Green's function.

Thus we are convinced that the latter is not a physical
quantity probably even for energetic plasmons in spite
of the fact that the
leading contribution (\ref{dampLS}) is gauge-independent
there.
We just do not know how on earth this quantity could be
measured.

\vspace{.3cm}

\centerline{\it Transport phenomena.}

\vspace{.3cm}

 There is a lot of kinetic phenomena in $QGP$ which are
physical and measurable.  Indeed, nothing {\it in principle}
prevents measuring the electric resistance of a vessel with
$QGP$ or studying the flow of $QGP$ through narrow tubes.
They do not depend, however, on the anomalous damping scale
$g^2T$, but rather on a much smaller scale
  \be
 \label{char}
(\tau_{char})^{-1} \sim g^4T \ln(1/g)
  \ee
 This scale already appeared
in (\ref{zetgam}) determining the damping of electromagnetic
waves in $e^+e^-$ -- plasma. And it is also the scale which
determines the
mentioned physical effects of viscosity and electric
conductivity, and  many others --- heat conductivity,
energy losses of a heavy particle moving through plasma,
etc.

The appearance of the scale $g^4T \ln(1/g)$ has a clear
physical origin. All the mentioned effects are inherently
 related to the rate of relaxation of the system to thermal
equilibrium. The latter can be estimated as
  \be
  \label{relax}
(\tau_{rel})^{-1} \sim n \sigma^{trans}
 \ee
It looks the same as the estimate for lifetime
(\ref{dampest}) but with an essential difference --- in
contrast to (\ref{dampest}), the estimate (\ref{relax})
involves the
{\it transport} rather than the total cross section. The
transport cross section is defined as
  \be
  \label{trans}
   \sigma^{trans} = \int d \sigma (1 - \cos \theta)
  \ee
where $\theta$ is the scattering angle. The factor $(1 -
\cos \theta)$ takes care of the fact that small--angle
scattering though contributes to the total cross section,
does not essentially affect the distribution functions
$n_g(\vec{p})$, $n_q(\vec{p})$ and is not effective in
relaxation processes. For the Coulomb scattering in
ultrarelativistic plasma, the transport cross section is
    \be
  \label{sigtr}
\sigma^{trans} \sim g^4 \int_{(gT)^2}
\frac{dp_\bot^2}{p_\bot^4}
\frac {p_\bot^2}{T^2} \sim \frac {g^4}{T^2} \ln \frac 1g
 \ee
Multiplying it by $n \sim T^3$ and substituting it in
(\ref{relax}), the estimate (\ref{char}) is reproduced.

Viscosity and all other similar quantities can be
calculated analytically  in the leading order (probably,
magnetic infrared divergences prevent an analytic evaluation
of these quantities in next orders in $g$, but this question
is not yet well studied). It is interesting that Feynman
diagram technique proves to be technically inconvenient here
, and the good old Boltzmann kinetic equation is the tool
people usually use (see e.g. \cite{Baym}).

  Let us make two illustrative estimates which make clear
how the relaxation scale (\ref{char}) depending on the
transport cross  section (\ref{sigtr}) arises.

First, let us estimate the electric conductivity of $QGP$
(it is the quite conventional conductivity,  not the
``color conductivity''
which is sometimes discussed in the literature, depends on
the
anomalous damping scale $g^2T$, and  is {\it not} a
physically
observable quantity --- we do not have batteries with color
charge at our disposal). Suppose at $t=0$ the system was at
thermal equilibrium so that the quark distribution functions
are $n_0(\vec{p}, \vec{x}, t=0) = n_F(|\vec{p}|)$.
When we switch on the electric field, the distribution
function
starts to evolve according to the kinetic equation
  \be
  \label{kineq}
  \frac {\partial n(\vec{p}, \vec{x}, t)}{\partial t}  +
\vec{v}
  \frac {\partial n(\vec{p}, \vec{x}, t)}{\partial \vec{x}}
  = e\vec{E} \frac{\partial n
  (\vec{p}, \vec{x}, t)}{\partial \vec{p}} + \ldots
  \ee
with $\vec{v} = \vec{p}/|\vec{p}|$.
 Dots in RHS of Eq.(\ref{kineq}) stand for the collision
term which becomes
  relevant at $t \sim x \sim \tau_{\rm rel}  \sim \tau_{\rm
free\ path}
  \sim [g^4 T \ln(1/g)]^{-1}$.
  Thus the electric field brings about distortions of the
distribution
  function which grow up to the characteristic value
       $$\delta n \sim e  \vec{E} \frac {\vec{v}}T n_0
\tau_{\rm free \ path}
  \sim \frac {eT  \vec{E} \vec{v}  }{g^4 \ln(1/g)}   $$
  At this point, collisional effects stop the growth (a
particle drifting
  in external electric field collides with a particle in the
medium, forgets
  what happened before, and starts drifting anew).
  The density of electric current in the medium is
   \be
    \label{j}
   \vec{j}  = e  \int \vec{v} \delta n \ d^3p
   \sim \vec{E} \frac {e^2 T}{g^4 \ln(1/g)}
   \ee
   The coefficient between $\vec{j}$ and $\vec{E}$ gives the
conductivity.

Let us estimate now the energy losses of a heavy energetic
quark in $QGP$. Of course, free quarks do not exist, but a
physical experimental setup would be sending into the bottle
with $QGP$ a heavy meson ${Q}\bar{q}$ with open beauty or
top. In $QGP$, the meson dissociates, and a naked
heavy quark propagates losing its energy due to interaction
with the medium. It goes out then   on the other side of the
bottle dressed again with light quarks, but not necessarily
in the same way as before.  When $M_Q \gg \Lambda_{QCD}$,
this dressing does not essentially affect its energy. A
heavy particle containing $Q$ can be detected and its energy
can be measured.

Suppose a heavy quark is ultrarelativistic, but its energy
is not high enough for the Cerenkov radiation processes to
be important. Then the energy would be lost mainly due to
individual incoherent scatterings.
  The mean energy loss in each scattering is $\Delta E \sim
T$ ($T$ --- is a characteristic energy of the particles in
heat bath on which our heavy quark scatters). The mean time
interval between scatterings is $\tau_{\rm free\ path} \sim
 [g^4T \ln (1/g)]^{-1}$. We obtain
  \be
  \label{dEdx}
-  \frac {dE}{dx} \sim \frac {\Delta E}{\tau_{\rm free\
path}} = C \alpha_s^2 T^2 \ln(1/g)
  \ee
This estimate turns out to be  correct up to the argument of
the
logarithm which in reality is energy--dependent
\cite{Thoma}.
 The  numerical coefficient $C = 4\pi c_F/3$ was
determined by Bjorken \cite{Bjorken}

\subsection{Chirality drift.}

Up to now, we were concentrated mainly on the perturbative
calculations
in $QGP$. Also we discussed the magnetic screening
phenomenon which,
as was explained in Sect. 6.1, is essentially non-
perturbative. Magnetic
screening scale $\sim g^2T$ shows up in static correlators
of different
quantities at large distances. Magnetic screening also
affects kinetic
characteristics of $QGP$ such as plasmon and plasmino
dispersion
laws, transport coefficients etc. The latter are well--
defined
and analytically calculable in the lowest perturbative
order. The
problems appear (and non--perturbative physics come into
play)
only when trying to determine higher--order corrections.
There is, however, a beautiful {\it kinetic} effect which is
{\it
purely} non--perturbative. This is the axial charge non--
conservation
in hot quark--gluon plasma, or the ``chirality drift''
presenting
the subject of this conclusive section.

\vspace{.3cm}

\centerline{\it Theoretical picture.}

\vspace{.3cm}

 The primary
practical interest of this quantity is related not to $QCD$,
but
to the dynamics of electroweak theory in early hot Universe.
Long
time ago, `t Hooft observed \cite{Hooft} that the baryon
charge is conserved
in standard model only on the classical lagrangian level. It
{\it is} not
conserved in the full quantum theory due to anomaly
  \be
 \label{Banom}
\partial_\mu j_\mu^B \ = \ \frac{g^2}{32\pi^2} G_{\mu\nu}^a
\tilde{G}_{\mu\nu}^a
  \ee
where $G_{\mu\nu}^a$ is the $SU(2)$ electroweak field.
Baryon number
non-conservation is precipitated by the gauge field
configurations with
nonzero Pontryagin number, the instantons. At zero
temperature, the
 action of electroweak instantons is very large $\sim
2\pi/\alpha_W$ and the
rate of baryon number non-conservation involves the
exponential factor
$\sim \exp\{- 2\pi/\alpha_W\}$ which is zero in all
practical sense.

\begin{figure}
\begin{center}
\grpicture{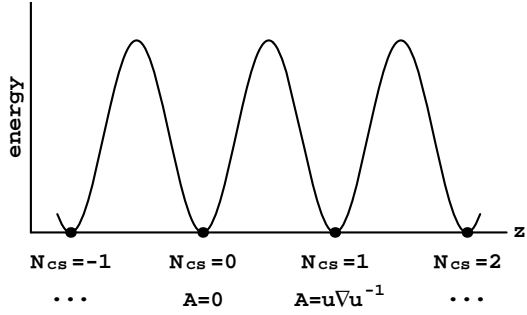}
\end{center}
\caption{A schematic representation of the (bosonic)
potential energy
along a particular direction (labeled $z$) in field space,
corresponding to
 topologically non-trivial transitions between vacua.}
\label{Yaf}
\end{figure}

Physically, instantons present tunneling trajectories
through a barrier
in the functional space. This barrier is associated with a
collective degree
of freedom called Chern--Simons number
  \be
 \label{Chern}
Q \ =\ \frac{g^2}{32\pi^2} \epsilon_{ijk} \int d^3x \left[
G_{ij}^a A_k^a
\ -\ \frac g3 \epsilon^{abc} A_i^a A_j^b A_k^c \right]
 \ee
 The integrand in (\ref{Chern}) is not gauge invariant, but
$Q(t)$ itself
is invariant under topologically trivial gauge
transformations. The
classical energy functional in Yang--Mills theory has
degenerate
minima at integer $Q$ (see Fig.\ref{Yaf}). These minima are
related to each
other by topologically
non-trivial gauge transformations. In electroweak theory
where gauge symmetry
is broken spontaneously
 by the Higgs mechanism
\footnote{ May be, this generally adopted terminology is not
the best. In some
sense, gauge symmetry is never broken as the generator of
gauge transformations
, the Gauss law constraint, still gives zero acting on the
vacuum state,
and there are no massless Goldstone bosons. However, better
words were not
invented and, anyway, everybody understands what the Higgs
mechanism is.}
, high barrier between
adjacent minima is created. The top of the barrier between,
say,
the minima with $Q= 0$ and $Q = 1$ (actually, it is a saddle
point in
the functional space with only one unstable mode)
is called the sphaleron \cite{Klink} and
has the energy $E_{\rm sph} \ \sim \ m_W/\alpha_W$.
 Chern--Simons number of the sphaleron configuration
is $Q_{\rm sph} = 1/2$.

\newpage

\newpage

It was observed in \cite{Kuz} that, when the temperature
grows, the height
of the barrier decreases.
Baryon number non-conservation processes are thus greatly
facilitated.
Actually,
as soon as $T \gg m_W$, the leading contribution in the rate
comes not from
quantum tunneling processes ``through'' the barrier, but
from the classical
thermal jumps ``over'' the barrier. These classical
transitions
are stochastic in nature and can occur both with increasing
and  decreasing
of baryon number. We have the physical picture of Brownian
motion.
Baryon charge ``drifts'' so that the average
square of the fluctuation of the baryon charge $[B(t) -
B(0)]^2$ grows
linearly with time. Due to the algebraic identity

  \be
  \label{Qdot}
\dot{Q} \ =\ \frac{g^2}{32\pi^2} \int d^3x  G_{\mu\nu}^a
\tilde{G}_{\mu\nu}^a
\ =\ \dot{B},
 \ee
one can study alternatively the fluctuations of Chern--
Simons number $Q(t)$.
 The rate of baryon number non-conservation $\Gamma^{\Delta
B \neq 0}(T)$ is
defined as the coefficient of the large time asymptotics of
the
thermal correlator
  \be
\label{CQt}
C(t) \ =\ <[Q(t) - Q(0)]^2>_T \ \stackrel {{\rm large}\
t}{\longrightarrow}\
 \Gamma^{\Delta B \neq 0}(T) Vt,
 \ee
where $V$ is the spatial volume of the system (naturally,
the larger
is the volume, the larger is the net rate), and we
subtracted the
time--independent part.
In the temperature region $m_W(T) \ll T \ll E_{\rm sph}(T)$,
the estimate for the rate is $\Gamma^{\Delta B \neq 0}(T)
 \propto \exp\{-E_{\rm sph}(T)/T\}$.
At some temperature (which is close to the critical
temperature $T_c$ where the spontaneously broken gauge
symmetry
$SU(2) \otimes
U(1)$ is restored), the characteristic energy of the
excitations is of the
same order as the barrier height at which point the
exponential suppression
disappears. An accurate calculation of the rate in the
temperature region
where the suppression is still there was performed in
\cite{LarArn}.

The question arises what happens at large temperatures $T
\gg T_c$ where
no barrier exists and the quasi-classical approximation
breaks down. It was
argued in \cite{LarArn} that the rate behaves as
  \be
 \label{Best}
 \Gamma^{\Delta B \neq 0}(T) \ =\ \kappa (\alpha_W T)^4
  \ee
where $\kappa$ is a constant which cannot be calculated
analytically.
The rate (\ref{Best}) is a purely non-perturbative quantity.
One
can show that the rate $\Gamma^{\Delta B \neq 0}(T) $
defined in
(\ref{CQt}) is zero in any order of perturbation theory.
\footnote{That does not mean that the correlator $C(t)$
itself is zero
in perturbation theory. It is not zero, diverges in
ultraviolet as a power,
and not only in nonabelian theory, but also in standard
$QED$. Already the
simplest graph with a photon loop gives a nonzero
contribution
$\sim \alpha^2 T^2 \Lambda_{UV} V$ in $C(t)$
 which is
physically related to the fact that a circularly polarized
photon has nonzero
Chern--Simons number density. These perturbative
contributions do not
grow, however, at large $t$ and do not affect the rate
\cite{4shap,Krasn}.}
The rate (\ref{Best}) is large, it leads to the dissipation
of
any primordial baryon charge density (if $B - L$ is exactly
conserved
as it does in most models of grand unification under
discussion) and
makes our very existence problematic. I will not pursue the
discussion of
this issue further here and refer the reader to a
comprehensive review
\cite{shap}.

The estimate (\ref{Best}) can be obtained as follows. As was
mentioned,
the barrier does not exist at high $T$. There are field
configurations
$\vec{A}(\vec{x})$
with arbitrary small energy which have $Q = 1/2$ and
interpolate between
adjacent minima. Such configurations should have, however,
large spatial
size: the larger is the size, the smaller are characteristic
field
gradients and the smaller is the energy $E \sim \int
d\vec{x} (\partial A)^2$.
\footnote{In other words, conformal symmetry of the
classical Yang--Mills
lagrangian does not allow soliton and sphaleron solutions.
The latter exist
when conformal symmetry is broken and a characteristic mass
scale
appears. In electroweak theory, this scale is provided by
the Higgs expectation
value. An interesting model is Yang--Mills theory on a
spatial sphere of finite
radius where the latter brings about the scale and sphaleron
solutions
appear \cite{S3}.}
One can visualize these configurations as the slices $\tau =
0$ of the
standard BPST instantons in the hamiltonian gauge $A_0  =
0$. It is
easy to see that the energy of such a slice for the
instanton of size $\rho$
is $E(\rho) \sim 1/(\alpha_W \rho)$. Writing the integral
over the collective
coordinate $\rho$ with the standard zero temperature
instanton measure
 $\sim \int d\rho/\rho^5$ (see Ref.\cite{measHoft}), we
obtain
  \be
 \label{rho5}
 \Gamma^{\Delta B \neq 0}(T) \ \sim\ \int
\frac{d\rho}{\rho^5} \exp \left\{
- \frac 1{\alpha_W \rho T} \right \} \ \sim \ (\alpha_W T)^4
 \ee
The characteristic $\rho$ where the integral is saturated is
nothing else
as our old friend, the magnetic screening scale. Actually,
one can forget  at this point, if one wishes, about
electroweak theory
altogether. In the
high temperature region, it presents ``electroweak plasma'',
and its
 physics is basically the same as the physics
of $QGP$ which is of main interest for us here. Instead of
baryon number
non-conservation, we have here the axial
charge non-conservation. Like baryon charge, it is conserved
on the classical
lagrangian level in the theory with massless quarks and {\it
is} not
conserved in the full quantum theory due to $U_A(1)$
anomaly. The rate
of axial charge non-conservation in hot $QCD$ can be
determined from the
 asymptotics of the Chern--Simons correlator (\ref{CQt}) at
large Minkowski times.

The result (\ref{Best}) means that roughly one transition
occurs in the
volume $\sim (g^2T)^{-3}$ by the time $\sim (g^2T)^{-1}$.
The estimate
for the characteristic volume seems to be rather solid:
indeed, the
barrier becomes irrelevant and also perturbation theory
breaks down at
the scale $\rho \sim (g^2 T)^{-1}$. However, the estimate
for the
characteristic {\it time} of the transition $t_{\rm char}
\sim (g^2 T)^{-1}$
was recently criticized in \cite{Son}. The time $t \sim (g^2
T)^{-1}$ is
the characteristic time of smearing out the wave packet with
the size
$\rho \sim (g^2 T)^{-1}$ in {\it empty} space. But our space
is not empty,
it presents a heat bath densely packed with gluon and quark
excitations.
It was argued in Ref.\cite{Son} that the time it takes for a
quasi-sphaleron
to decay in quark-gluon plasma plasma is actually suppressed
by two powers
of $g$ and is estimated as
\be
\label{Sontime}
t_{\rm decay} \ \sim \ (g^4T)^{-1}
\ee
 Correspondingly, the new estimate
for the rate is
  \be
 \label{estSon}
\Gamma \ \sim \ \frac 1{\rho_{\rm char}^3} \frac 1 {t_{\rm
decay}} \ \sim
\ \alpha^5 T^4
  \ee
The estimate (\ref{Sontime}) and its corrolary
(\ref{estSon})
 can be obtained as follows. Suppose, we have at
$t = 0$ a wave packet with the size $\sim (g^2T)^{-1}$.
Expand it in Fourier
series and solve a classical equation of motion (which
takes into account the heat bath effects) for an individual
mode
$\vec{A}_k$ with initial condition
  \be
 \label{ini}
\dot{\vec{A}}_k = 0
  \ee
The solution presents a sum of three exponentials
  \be
  \label{Akt}
 \vec{A}_k(t) \ =\ \alpha_1 \exp\{-C_1 g^2T t -
i\omega_\bot(k)t)\}\ +\
  \alpha_2 \exp\{-C_2 g^2T t - i\omega_\|(k)t)\} \nonumber
\\
+ \beta \exp\{-C_3 g^4T t\}
  \ee
 The first two terms  correspond
to  standard transverse and longitudinal plasmons and
involve
 a plasmon decay rate factor with
$\zeta \sim g^2T $ while the third one corresponds to
attenuating
mode (\ref{att}) with the decay rate $\sim g^4T$ at $k \sim
g^2T$. The
first and the second exponential come with small coefficient
$\sim g^2$
due to the boundary conditions (\ref{ini})
and will die away at the
time scale $\sim (g^2T)^{-1}$ while the second exponential
is still
alive. It would eventually determine the quasi-sphaleron
decay rate.

In our own opinion, these arguments are not conclusive
enough.
Let us first modify the initial conditions and put
$\dot{\vec{A}}_k \sim g^2T \vec{A}_k $. Such initial
conditions
seem to be as reasonable as (\ref{ini}). But then the
coefficients
$\alpha_{1,2}$
and $\beta$ would be of the same order and the Chern--Simons
number
of the field configuration (\ref{Akt}) would be essentially
modified
on the time scale $(g^2T)^{-1}$ rather than $(g^4T)^{-1}$.
Second, the
amplitude of Fourier harmonics in quasi--sphaleron wave
packet is large
and their nonlinear interaction is essential. Even if we
started up
with only attenuating mode at the initial moment, normal
plasmon modes
would be excited due to nonlinear effects and would
eventually determine
the time decay scale.
\footnote{As B. Muller pointed out to me, there is also
another effect of
losing of coherence (or spread) of the quasi-sphaleron wave
pocket due
 to dispersion.
But a characteristic time scale of this effect is estimated
as
$(\tau_{\rm spread})^{-1}\  \sim \ (d\omega_{\rm
pl}/dk)k_{\rm char}
\ \sim \ g^3T$. This time is larger than $(g^2T)^{-1}$ so that
the all
harmonics in the wave packet would die away before the
coherence
is lost.}

 Another question which can be asked here is why we assumed
that
the plasmon harmonics die away with the anomalous damping
scale $\sim g^2T$
rather than with the scale $\sim g^4T$ which is
characteristic for
transport phenomena in plasma and which, as we repeatedly
argued
in the previous sections, is more physical. The point is
that the latter scale
depends  on the transport cross section (\ref{trans}) where
the region of small
momentum transfer $p_\bot \sim gT$ is suppressed. In this
particular problem,
we do not see a reason to include such a suppression,
however. The scale
$gT$ is {\it large} compared to the characteristic momentum
scale $g^2T$
determining chirality drift. When a mode acquires the
momentum $\sim gT$, it
is lost as far as the coherent quasi--sphaleron wave packet
is concerned.

But all these arguments are extremely heuristic. To solve
the problem
accurately, it does {\it not} suffice to play around with
small
fluctuations on the vacuum background. Actually, one has to
expand the field near the quasi--sphaleron background
$\vec{A} \ =\ \vec{A}_b
+ \delta \vec{A}$ and to solve the linearized equations for
$\delta \vec{A}$
in the heat bath.

Let us first see what happens in the absence of heat bath.
Vast majority
of the modes are stable. There are two directions in the
functional
space decreasing the energy of the field configuration. One
of them corresponds
to rescaling the quasi-sphaleron configuration and can be
handled introducing
the collective coordinate $\rho$ and integrating over it as
in (\ref{rho5}).
The mode of main interest for us is the intrinsic sphaleron
unstable mode
$\delta \vec{A}_{unst}$ which corresponds to rolling down
the slope away from
the quasi-sphaleron ridge. In vacuum, the corresponding
linearized equation
is $\delta \ddot{\vec{A}} \ \sim \ [k_{\rm char}^2 ]
\delta {\vec{A}}$ with $k_{\rm char} \sim g^2T$. It
describes the decay
of quasi--sphaleron by the time
$\sim (g^2T)^{-1}$.

Let us now switch on the medium effects. The problem has not
been accurately
solved yet, but, taking insight from the dispersive equation
(\ref{displt}) on the vacuum background, we can tentatively
{\it add} to
$k_{\rm char}^2$ the transverse polarization operator
$\Pi_t(\omega, k_{\rm char})$. The new dispersive equation
looks something
like
  \be
  \label{newdisp}
-\omega^2 \ =\  k_{\rm char}^2 + \Pi_t(\omega, k_{\rm char})
  \ee
It differs from Eq.(\ref{displt}) only by the sign of
$\omega^2$ term.
Substitute now here $ \Pi_t(\omega, k_{\rm char}) \sim -
i\omega_{\rm pl}^2
 \omega /{k_{\rm char}}$  which is true at small $\omega \ll
k_{\rm char}$.
The solution to the dispersive equation (\ref{newdisp}) is
  \be
 \label{dampg4T}
\omega \ \sim \ - \frac {ik_{\rm char}^3}{\omega_{\rm pl}^2}
\ \sim \ -ig^4T
  \ee
This mode corresponds to {\it damping} of the fluctuations
$\delta \vec{A}$
with time so that a perturbed field tends asymptotically
back
to the quasi--sphaleron
configuration $\vec{A}_b$. This mode has nothing to do with
instability.

The dispersive equation (\ref{newdisp}) has also an unstable
solution, but
it corresponds to large $\omega \gg k_{\rm char}$ where
$ \Pi_t(\omega, k_{\rm char}) \ \sim \ \omega_{\rm pl}^2$.
The corresponding
unstable mode is
  \be
  \label{instgT}
\omega_{\rm unst} \ \sim \ i\omega_{\rm pl} \ \sim \ igT
  \ee
Thus medium {\it increases} the rate of the quasi-sphaleron
decay rather
than decreases it. The estimate is
  \be
 \label{g7T4}
\Gamma \ \sim \ g^7T^4
  \ee
This reasoning is quite parallel to the reasoning of
Ref.\cite{Son}, only the
sign of $ \Pi_t(\omega, k_{\rm char})$ in the dispersive
equation
(\ref{newdisp}) is chosen positive rather than negative.
Negative sign would
transform the damping mode (\ref{dampg4T}) into an unstable
one, and the
estimate (\ref{estSon}) would be reproduced.

By no means we  want to insist here that it is the estimate
(\ref{g7T4}) rather
than (\ref{estSon}) which {\it is} correct. Choosing the
positive sign
for $\Pi_t(\omega, k)$ in (\ref{newdisp}) looks more
natural, but the
thermal
polarization operator corresponding to fluctuations on a
quasi--sphaleron
background (with the Green's functions on a quasi--sphaleron
background in the loops) could essentially differ from the
vacuum one,
and only its accurate calculation   would solve the problem.
It is
possible that the
the rate is suppressed compared to the estimate
(\ref{Best}), and it is
also possible that it is enhanced.
\footnote{We should note, however, that the enhancement of
the rate as
is suggested by the estimate (\ref{g7T4}) does not look
physically
appealing.
It implies that the object with the size $\sim (g^2T)^{-1}$
dissipates
faster than it takes light to travel it across. Basically,
we are
presenting
this estimate just as an illustration that the same
reasoning as in
\cite{Son} with only a little natural modification leads to
completely
different results.}
The third possibility (which is actually the most
attractive)
is that plasma effects neither
increase nor decrease the rate of decay, but leave it
basically intact.

\vspace{.3cm}

\centerline{\it Numerical experiment. Kinetic equations.}

\vspace{.3cm}

Let us discuss now experimental (lattice) data. A very
serious difficulty here
is
that the rate of baryon number (chiral charge) non-
conservation is a kinetic
quantity defined in real time. We are not able currently to
calculate
 numerically field theory path integrals in Minkowski space-
-time and have
to look for some other way to solve the problem.

As was mentioned earlier, the physical mechanism of
chirality non--conservation
is the classical drift over the barrier. The idea arises to
obtain the result
just solving the {\it classical} equations of motion and
looking at how the
classical Chern--Simons charge (\ref{Chern}) is changed with
time.
Performing an average over initial conditions with the
classical weight
$\sim \exp\{-\beta H^{\rm cl}\}$, one estimates the
rate  of the drift in hot field theory. This algorithm was
suggested in
\cite{Grig} and successfully tested in some 2--dimensional
models
\cite{Grig,Grigfol}. In \cite{4shap} and in recent
\cite{Krasn}, the
 same algorithm was applied to  4--dimensional gauge
theories. To discretize
the classical equations of motion, the system was put on a
3--dimensional
hamiltonian lattice, the lattice spacing $a$ playing the
role of the ultraviolet
cutoff.

The question whether such a classical algorithm is justified
or not is under
discussion now. The
main problem here are severe power ultraviolet divergences
inherent
for a classical field theory at finite $T$. A classical
Rayleigh--Jeans
ultraviolet divergence in the  energy density of the photon
gas
  \be
 \label{RalJean}
E^{cl} \ \sim \ \int d^3p |\vec{p}|\ n^{cl}_B(|\vec{p}|) \
\sim \
T\int d^3p \ \sim \ T\Lambda_{UV}^3
  \ee
[where the classical limit of the Bose distribution is
$n_B^{cl}(\epsilon)
= 1/(\beta \epsilon)$] was the reason to develop the quantum
theory in
the first place. Classical contribution to the plasma
frequency
  \be
 \label{omcl}
(\omega_{\rm pl}^{\rm cl})^ 2 \ \sim \ g^2 \int \frac
{d^3p}{|\vec{p}|}
\frac 1{\beta |\vec{p}|} \ \sim \ g^2 T \Lambda_{UV}
  \ee
is also ultraviolet divergent.

In the full theory, the divergence is cut off at $|\vec{p}|
\sim T$ (where
quantum corrections in the Bose distribution function come
into play). For
the classical approximation to be justified,
 one should always keep $a^{-1} \sim \Lambda_{UV} \ll
T$.

In principle, chirality non-conservation rate calculated by
the classical
algorithm could also be cutoff--dependent. Moreover, if the
estimate
(\ref{estSon}) of Ref.\cite{Son} or the estimate
(\ref{g7T4}) is correct,
one {\it should} expect cutoff--dependence of $\Gamma$. In
both cases,
$\Gamma$ depends explicitly on the plasma frequency scale
which is
cutoff--dependent when calculated classically. One easily
estimates
  \be
\label{Gamcl}
\Gamma_{\rm ASY} \ \sim \ g^{10} T^4 \Longrightarrow
\Gamma_{\rm ASY}^{\rm cl} \ \sim \frac {k_{\rm
char}^6}{(\omega_{\rm pl}^
{\rm cl})^2} \ \sim \ \frac {g^{10} T^5}{\Lambda_{UV}}
\nonumber \\
\Gamma_{(\ref{g7T4})} \ \sim \ g^{7} T^4 \Longrightarrow
\Gamma_{(\ref{g7T4})}^{\rm cl} \  \sim  {k_{\rm
char}^3}{\omega_{\rm pl}^
{\rm cl}} \ \sim \  g^{7} T^{7/2}\Lambda_{UV}^{1/2}
 \ee
However, if the original estimate (\ref{Best}) is {\it not}
affected by
the medium effects, $\Gamma$ does not depend on the plasma
frequency scale,
but only on the magnetic mass scale $g^2T$. In this case,
one could expect
that classical numerical results for $\Gamma$ do not depend
on cutoff,
and the whole algorithm is trustworthy if the condition
$$g^2T \ \ll \ \Lambda_{UV} \ \ll T$$
is satisfied.

Following the suggestion of \cite{Bodeker}, the question of
whether classical
rate depends on cutoff or not, was studied in \cite{Krasn}.
Ambjorn and Krasnitz
find that the rate depends neither on cutoff nor on
$\omega_{\rm pl}$.
Their answer for the rate is
  \be
  \label{AKrate}
 \Gamma^{\Delta B \neq 0}(T) \ \approx\ 1.1 (\alpha_W T)^4
 \ee
This result directly contradicts the estimate (\ref{estSon})
and also
(\ref{g7T4}). It was suggested in \cite{Son} that the result
(\ref{AKrate})
is actually a lattice artifact related to an imperfect
lattice definition
of the Chern--Simons number. It may be, though it is not
quite clear
why an alleged
lattice artifact (\ref{AKrate}) does not depend on the
lattice spacing.

On pure esthetic grounds, it would be much nicer if the
plasma
effects would not eventually influence the quasi-sphaleron
decay rate
and the original result (\ref{Best}) and the numerical
calculations
of \cite{Krasn} were correct. Our heuristic arguments which
take into
account not only the attenuating but also the plasmon modes
and nonlinear
interaction between them [see the discussion after
Eq.(\ref{Akt})]
also seem to indicate that.
We would rather lay our bets on this
possibility, but a {\it scientific} answer to this question
is not yet
found. Thus we are in a position to discuss  what happens if
an
unfortunate possibility
 that the rate depends on the scale $\omega_{\rm pl}$ and
thereby on the
ultraviolet cutoff, when calculated classically,
is realized. Can one still hope to suggest
a numerical procedure where the rate of the chirality drift
in the full
quantum theory could be evaluated quantitatively ?

This question was studied in \cite{Bodeker} (see also recent
\cite{Hu}).
An algorithm based on the
separation of low momentum  and high momentum scale was
suggested. The idea
was to take into account explicitly the contribution of high
momentum modes
in the effective hamiltonian and then to solve numerically
the classical
equations of motion of this effective hamiltonian for soft
modes.

\begin{figure}
\begin{center}
\begin{picture}(90,90)
\SetScale{2.845}
\PhotonArc(45,30)(25,0,360){2}{15}
\Photon(0,30)(20,30){1.5}{5}
\Photon(70,30)(90,30){1.5}{5}
\Photon(45,55)(45,75){1.5}{5}

\put(45,0){ P}

\end{picture}
\end{center}
\caption{An example of HTL graph contributing to the
effective 3--gluon
vertex. The loop momentum is hard $p \sim T$. External
momenta are soft.}
\label{HTL}
\end{figure}
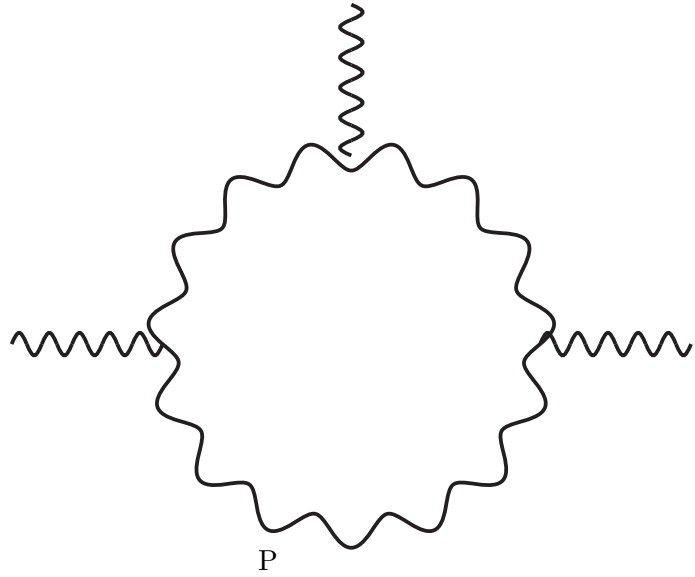

The effective theory for soft modes (with momenta of order
$gT$) was first
formulated in \cite{Pisar,Taylor} in lagrangian form. This
lagrangian involves
multiple quark and gluon effective vertices of the kind
shown in Fig.\ref{HTL}.
Characteristic momenta flowing in the hard thermal loop in
Fig.\ref{HTL}
are of order $T$. We already discussed at length two--point
quark and gluon
Green's functions which are contained in this lagrangian
[see Eqs.(\ref{Pilt}),
 (\ref{Sigomk})]. Both 2--point and also 3--point, 4--point
etc. effective
vertices are nonlocal.

\newpage

\ \ \ 

\newpage
A nonlocal lagrangian is very inconvenient for a numerical
study. Fortunately,
it can be presented in a local form \cite{Nair,Blaizot}. The
price is
introducing new variables depending on $x$ {\it and} on an
extra
3--dimensional unit vector $\vec{v}$.
\footnote{Then the effective action acquires an elegant
mathematical structure,
 being closely related to the 2D Chern--Simons theory in
light cone variables
 $x_\pm = t \pm \vec{x} \vec{v}$ \cite{NairCS} }
The effective HTL hamiltonian of pure gluodynamics
can be written in the form
  \be
  \label{HNair}
H \ =\ \int d^3x \ {\rm Tr} \left\{ \vec{E} \cdot \vec{E} +
\vec{B} \cdot \vec{B} + 3\omega_{\rm pl}^2 \int \frac
{d\vec{v}}{4\pi}
w(x,\vec{v}) w(x,\vec{v}) \right\} \nonumber \\
 \ee
where $w(x, \vec{v}) \equiv w^a(x,\vec{v}) t^a$ and
$\omega_{\rm pl}
= gT\sqrt{N_c}/3$.
The variables $w^a(x, \vec{v})$ talk to gauge fields by
virtue of nontrivial
equal time commutators
  \be
\label{Naircom}
\left[E^a_i(\vec{x}), w^b(\vec{x}', \vec{v})\right] \ =\
i \delta^{ab} v_i \delta(\vec{x} - \vec{x}') \nonumber \\
 \left[ w^a(\vec{x}, \vec{v}), w^b(\vec{x}',
\vec{v}')\right] \ =\
\frac{8\pi}{3\omega_{\rm pl}^2} \left( i f^{abc} w^c - v_i
{\cal D}_i^{ab}
\right) \delta(\vec{x} - \vec{x}') \delta(\vec{v} -
\vec{v}')
  \ee
(${\cal D}_i^{ab}$ is the covariant derivative)
and of the Gauss law constraint. Integration over
$\prod dw(x,\vec{v})$ restores original nonlocalities.

Actually, as was shown in \cite{Blaizot}, the variables $w$
have a transparent
physical meaning. They can be related to the deviation of
the phase space
distribution function of hard gluons $\delta n(x,\vec{p})
\equiv
\delta n^a(x,\vec{p})t^a$ from the thermal equilibrium Bose
distribution
$n_0(|\vec{p}|)$:
  \be
  \label{wdeln}
w\left(x,\frac {\vec{p}}{|\vec{p}|}\right) \ =\ \frac
3{g\pi^2 T^2} \int
\delta n(x, \vec{p}) |\vec{p}|^2 \ d|\vec{p}|
  \ee
The equations of motion corresponding to the hamiltonian
(\ref{HNair}) are
equivalent to a nonabelian analog of the Vlasov equation
system
  \be
  \label{Vl1}
{\cal D}_\mu F^{\mu\nu} \ =\ 2gN_c \int \frac
{d^3p}{(2\pi)^3}\ v^\nu
\ \delta n(x, \vec{p})
  \ee
  \be
  \label{Vl2}
v^\mu {\cal D}_\mu \delta n(x, \vec{p})\ =\ -g\vec{v}
\vec{E}
\frac {dn_0 (|\vec{p}|)}{d|\vec{p}|}
  \ee
where $v^\mu \equiv (1, \vec{v})$, $F^{\mu\nu}$ is the
strength of mean gluon
field (which is soft),
and $\delta n(x, \vec{p})$ are  degrees of freedom
associated with
hard particles.
We already met the abelian analog of Eq.(\ref{Vl2}) in Sect.
6.4, but
note the appearance of the covariant derivative ${\cal
D}_\mu$ :
in contrast to the abelian case, $\delta n(x, \vec{p})$
carries here color
charge. The RHS of Eq. (\ref{Vl1}) is the induced color
current of hard
particles.

When also fermions are present, the effective theory can
also be formulated
in kinetic equation language \cite{BlQED,Blaizot}.
To this end, one has to introduce mean fermion soft fields
and also mixed
boson--fermion densities $\sim <A_\mu \psi>$.
Such densities and the
corresponding kinetic equations were first introduced and
analyzed
in \cite{sound} where they were used to
study the problem of goldstino ($\equiv$ phonino) dispersion
law in a
supersymmetric thermal medium.

However, as far as the problem of constructing a numerical
algorithm to
 calculate
diffusion rate in a possible case when classical equations
of motion for the
tree hamiltonian produce an ultraviolet cutoff dependent
result is
concerned, a simple
use of the kinetic equations system (\ref{Vl1}), (\ref{Vl2})
instead
of the classical field equations is not yet a
solution.

The hamiltonian (\ref{HNair}) and the equation system
(\ref{Vl1}), (\ref{Vl2})
were derived in the assumption that integration in the hard
thermal loop
in Fig.\ref{HTL} etc. goes over {\it all} momenta, both hard
and soft.
If we just put the system (\ref{Vl1}), (\ref{Vl2}) on the
lattice with an
explicit ultraviolet cutoff $\Lambda_{\rm UV} = a^{-1}$,
soft modes would
be counted twice: once in the loop and once again in the
classical
dynamics. {\it If} the results of the simplistic classical
calculation depend
on the lattice spacing, so the results of the calculation
with the kinetic
equations (\ref{Vl1}), (\ref{Vl2}) would.
\footnote{On the other hand, {\it if} the classical
calculations in the
spirit of \cite{4shap,Krasn} eventually prove to be cutoff--
independent,
{\it they} give the correct result, and all the stuff of
effective
lagrangian for soft modes and Vlasov equations would play
here a role of
``purple hands on
enamelled wall'' \cite{Brusov} --- very beautiful, but
practically irrelevant.}

One can hope to obtain a cutoff--independent result only
when separating
carefully soft and hard modes and including only hard ones
(with
 momenta exceeding the separation scale) in the quantum
loops \cite{Bodeker}.
Unfortunately,
we were not able to present an explicit and consistent
procedure of such a
scale separation. A  simplistic momentum cutoff  breaks
gauge--invariance.
The lattice cutoff preserves gauge invariance but
 brings about a lot of new
rotationally--noninvariant structures in the effective
hamiltonian which we
could not handle.
\footnote{The last technical comment is that if one
eventually succeeds
 in constructing
a self--consistent algorithm of scale separation, he would
simultaneously
solve another problem which we did not  mention yet. HTL
approximation
for the effective lagrangian is justified when a
characteristic momenta
of soft modes is of order $gT$. In this case, however, we
need to study the
dynamics of modes with momenta of order $g^2T$. In this
region, HTL
approximation breaks down, and one has to resum higher--loop
ladder graphs
which are all of the same order. In a simple case (for the
polarization
operator of electromagnetic current in abelian theory), such
a resummation
was done in
\cite{andam}, but to do it for the full effective lagrangian
is a
formidable problem which is far from being solved now.
However, as was noted
in \cite{Bodeker}, the dominance of one loop HTL graphs is
restored if
taking into account only the momenta $p \gg gT$ in the
loops. Thus, if
the separation scale lies in the region $gT \ll \mu \ll T$,
HTL approximation
is justified.}

It is quite obvious that more theoretical and numerical
studies of this
question are highly desirable.

\section{Acknowlegdements}

This review presents an updated and considerably expanded
combination
of  two my lectures at E. Fermi Int. School at Varenna
\cite{Varenna} and
at XXIV ITEP Winter School at Snegiri \cite{ITEP} devoted to
the physics
of $QCD$
at low and, correspondingly, at high temperature. I take the
opportunity
to thank again the organizers of these Schools
for  vivid and inspired scientific atmosphere of
these meetings.

I have benefited a lot from illuminating discussions and
correspondence
with  J.P. Blaizot, V.L. Eletsky, E. Iancu, J. Kapusta, F. Karsch, H.
Leutwyler,
 B. Muller, S. Peigne, E. Pilon, A. Rebhan, D. Schiff,
E. Shuryak, A.I. Vainshtein,  and M.B. Voloshin. It is a  pleasure for me to
acknowledge  kind hospitality  extended to me at TPI at
University of
Minnesota where  the review was written. This work
has been done under the partial support of the INTAS Grants
CRNS--CT93--0023, 93--283, and 94--2851, and the U.S.
Civilian
Research and Develpoment Foundation Grant RP2--132.

\end{document}